\documentclass[pre,twocolumn,superscriptaddress,tightenlines,showpacs,longbibliography,floatfix,footinbib]{revtex4-1}

\usepackage{latexsym,amsmath,amsfonts,tabularx,amssymb,bm,empheq}
\usepackage{subfigure}

\usepackage{color}
\usepackage[dvipsnames]{xcolor}
\definecolor{nblue}{RGB}{28,130,185}
\definecolor{cgreen}{RGB}{76,153,0}
\definecolor{myorange}{RGB}{245,156,74}
\DeclareMathOperator{\sgn}{sgn}
\usepackage{ulem}

\usepackage{hyperref}
\hypersetup{
  colorlinks=true,
  citecolor=magenta,
  urlcolor=-myorange
}

\usepackage{tensor}

\usepackage{blkarray}
\usepackage{mathtools}
\usepackage{multirow}

\begin{document}

\title{Scaling Description of Dynamical Heterogeneity and Avalanches of Relaxation in Glass-Forming Liquids }

\author{Ali Tahaei}
\affiliation{Max Planck Institute for the Physics of Complex Systems, 	N\"{o}thnitzer Str.38, 01187 Dresden, Germany}

\author{Giulio Biroli}
\affiliation{Laboratoire de Physique de l'Ecole Normale Sup{\'e}rieure, ENS,
Universit\'e PSL, CNRS, Sorbonne Universit\'e, Universit\'e Paris Cit\'e, F-75005 Paris, France}

\author{Misaki Ozawa}
\affiliation{Univ. Grenoble Alpes, CNRS, LIPhy, 38000 Grenoble, France}

\author{Marko Popović}
\affiliation{Max Planck Institute for the Physics of Complex Systems, 	N\"{o}thnitzer Str.38, 01187 Dresden, Germany}
\affiliation{Center for Systems Biology Dresden, Pfotenhauerstrasse 108, 01307 Dresden, Germany}

\author{Matthieu Wyart}
\affiliation{Institute of Physics, EPFL, Lausanne, Switzerland}


\begin{abstract}
We provide a theoretical description of dynamical heterogeneities in glass-forming liquids, based on the premise that relaxation occurs via local rearrangements coupled by elasticity. In our framework, the growth of the dynamical correlation length $\xi$ and of the correlation volume $\chi_4$ are controlled by a zero-temperature fixed point. We connect this critical behavior to the properties of the distribution of local energy barriers at zero temperature. Our description makes a direct connection between dynamical heterogeneities and avalanche-type relaxation associated to dynamic facilitation, allowing us to relate the size distribution of heterogeneities to their time evolution. Within an avalanche, a local region relaxes multiple times, the more the larger is the avalanche. This property, related to the nature of the zero-temperature fixed point, directly leads to decoupling of particle diffusion and relaxation time (the so-called Stokes-Einstein violation).
Our most salient predictions are tested and confirmed by numerical simulations of  scalar and tensorial thermal elasto-plastic models.
\end{abstract}

\maketitle

\section{Introduction}


As a liquid is cooled,  the relaxation time $\tau_\alpha$ below which it acts as a solid -- before displaying flow --  grows from picoseconds at high temperatures up to minutes at the glass transition temperature $T_g$ \cite{Anderson95,ediger1996supercooled,Debenedetti01,berthier2011theoretical}. In this regime the effective activation energy associated to $\tau_\alpha$ grows for many liquids, leading to a super-Arrhenius behavior. 
Approaching $T_g$, dynamics also becomes heterogeneous on a growing correlation length scale $\xi$ ~\cite{kob1997dynamical,yamamoto1998dynamics,dalle2007spatial,karmakar2014growing}. The underlying causes for these observations are still debated. In some views, activation is cooperative: the slowing down of the dynamics is governed by complex motion taking place on an increasingly large static length-scale. 
In particular, cooperativity is central in the Random First Order Theory  \cite{kirkpatrick1989scaling,lubchenko2007theory,Biroli12} and due to the growth of amorphous order. Another approach focuses on dynamical facilitation, the phenomenon by which a region's relaxation is made much more likely by a relaxation nearby. In  kinetically constrained models, such as the East model, kinetic rules induce dynamic facilitation and growth of dynamical correlations~\cite{ritort2003glassy,berthier2011dynamical}. This lead to a theory \cite{garrahan2002geometrical,garrahan2007dynamical,hedges2009dynamic} in which thermodynamics plays almost no role, but dynamics is heterogeneous (due to kinetic constraints) and the super-Arrhenius behavior is due to non-local rearrangements taking place over $\xi$. 
Free volume \cite{turnbull1961free} or elastic \cite{dyre2006colloquium,rainone2020pinching,JeppeEdan} models assume that the activation energy is not controlled by a static growing length scale: it is governed by the energy barrier of elementary rearrangements, or excitations. Recently,  measurements indicated that the distribution of local energy barriers shifts to higher energy under cooling, opening up a gap at low energies where excitations are nearly absent \cite{massimo}. 
This shift accounts quantitatively for the dynamical slowing down of the liquid,  supporting that local energy barriers may indeed control the dynamics. 
Yet in these views, what causes the existence of a growing length scale is unclear. 
An intuitive resolution of this paradox stems from the fact that on relatively short time scales, a super-cooled liquid acts as a solid~\cite{dyre2006colloquium,schroder2020solid}: a rearrangement corresponds to a local (plastic) strain, that  affects stress away from it. Lemaitre \cite{Lemaitre14} was the first in stressing the possible relevance of long-range stress correlations in the dynamics of super-cooled liquids, in line with various theoretical and numerical studies~\cite{chowdhury2016long,tong2020emergent,wu2015anisotropic,maier2017emergence,steffen2022molecular,flenner2015long,klochko2022theory}.  Recently, it was shown that indeed elastic interactions play an important role in dynamical facilitation~\cite{chacko2021elastoplasticity}.
Moreover, Ref.~\cite{lerbinger2022relevance} showed that heterogeneous relaxations take place in the form of plastic rearrangements that are called shear transformations~\cite{falk1998dynamics}. Following these molecular simulations studies, Ref.~\cite{ozawa2023elasticity} demonstrated with elasto-plastic models  (traditionally used to study the plasticity of amorphous solids under loading \cite{picard2004elastic,Nicolas2018,Vandembroucq2004,Lin2014b,rossi2022finite}) that while being controlled by local energy barriers, the dynamics can at the same time display growing dynamical correlations similar to observations in experiments and molecular simulations~\cite{berthier2005direct,berthier2011dynamical}. Furthermore, such models also capture the emergence of a gap in the distribution of local barriers under cooling, which controls the dynamics \cite{popovic2021thermally,ozawa2023elasticity}.
We expect that the elasto-plastic description discussed above becomes more and more relevant with decreasing temperature. This paper focuses on such a lower temperature regime.

In this work we propose a scaling description of dynamical heterogeneities in glass-forming liquids, modeled as undergoing local irreversible rearrangements coupled by elasticity. 
We show that the dynamical correlations observed in elasto-plastic models of equilibrium glassy dynamics \cite{ozawa2023elasticity} are due to "thermal avalanches", where rare nucleated events are followed by a pulse of faster (or facilitated) events. These avalanches are very reminiscent of the ones found in molecular simulations of super-cooled liquids~\cite{candelier2010spatiotemporal,keys2011excitations,yanagishima2017common}. 
Our analysis builds a link with systems that crackle such as disordered magnets, granular materials, and earthquakes~\cite{sethna2001crackling,rosso2022avalanches} even at finite temperatures~\cite{popovic2021thermally,purrello2017creep,yao2023thermal,korchinski2022dynamic}, revealing that dynamical heterogeneities are controlled by a critical point at zero temperature, with a diverging length scale $\xi \sim T^{-\nu}$ and correlation volume $\chi_4^* \sim T^{-\gamma}$. We provide a scaling argument expressing $\nu$ and $\gamma$ in terms of the distribution of energy barriers at $T=0$. Our analysis makes predictions on the power-law distribution of the size of dynamical heterogeneities, which could be tested in numerical simulations of super-cooled liquids thanks to the recent advances in the characterization of dynamical correlations~\cite{scalliet2021excess,scalliet2022thirty}. 
Based on the properties of the zero-temperature fixed point governing thermal avalanches, we also show that the decoupling $D \tau_\alpha$ between particle diffusion $D$ and relaxation time $\tau_\alpha$ (the so-called Stoke-Einstein violation|~\cite{tarjus1995breakdown,ediger2000spatially,sengupta2013breakdown,charbonneau2014hopping,kawasaki2017identifying}) diverges as $D \tau_\alpha \sim T^{-h}$, where $h$ can be expressed in terms of avalanche properties at vanishing temperature. 
The key physical mechanism behind the Stoke-Einstein violation is the intensive accumulation of rearrangements in the mobile region that quantifies a larger diffusion constant $D$, relative to the structural relaxation time $\tau_\alpha$~\cite{jung2004excitation,berthier2004length,hedges2007decoupling,chaudhuri2007universal,pastore2021breakdown}. 
We perform numerical simulations and analyze the results by finite size scaling 
in order to test our salient predictions. Our findings agree well with the scaling theory both for scalar and tensorial thermal elasto-plastic models.



\section{Thermal Elasto-plastic models}

To study dynamical heterogeneities in low-temperature glass-forming liquids, we employ a generalization of elasto-plastic models~\cite{picard2004elastic,Nicolas2018} which are a class of mesoscopic models designed to capture the essential features of localized plastic events (shear transformations) with stress relaxation accompanied by long-range elastic interactions. An elasto-plastic model contains $N=L^d$ mesoscopic sites that are arranged on a regular grid whose linear size is $L$, and $d$ is the spacial dimension. Each site exhibits a plastic event when the magnitude of the local shear stress becomes sufficiently large, which leads to local stress relaxation and stress redistribution in the rest of the system via the form of Eshelby fields. Elasto-plastic models were originally introduced to study the flow of amorphous solids under external loading~\cite{picard2004elastic,Lin2014b,Nicolas2018,rossi2022finite} (some exceptions e.g., Ref.~\cite{bulatov1994stochastic}) and they were generalized to take into account thermal fluctuations~\cite{ferrero2014relaxation} and then to study glass-forming liquids~\cite{ozawa2023elasticity}. Here, we study their low-temperature relaxation dynamics in the absence of loading. Following the idea that  supercooled liquids can be viewed as solids that flow~\cite{dyre2006colloquium,Lemaitre14}, we develop a physical scenario based on the assumption that local rearrangements are elastically 
coupled, and we model this physical mechanism by elasto-plastic models.

In this paper, we use a tensorial elasto-plastic model that accounts for the shear stress tensor~\cite{jagla2020tensorial} and a scalar elasto-plastic model in which the shear stress is represented by a scalar variable~\cite{ozawa2023elasticity} (see Appendix~\ref{sec:appendix_models} for details). In both models, thermal fluctuations are implemented through a probability rate $\tau_0^{-1} e^{-E/T}$ to trigger a plastic (mobile) event in an otherwise stable elastic (immobile) site, where $E$ is the local activation energy barrier and $\tau_0$ is a microscopic relaxation timescale \cite{ferrero2014relaxation,popovic2021thermally,ferrero2021yielding,rodriguez2023temperature}. 
The energy barrier $E$ can be related to the minimal amount of additional shear stress $x_i$ required to destabilize a site $i$. In particular, we consider $E (x)= c x^{\alpha}$, which is suggested by recent elasto-plastic models and molecular simulations~\cite{ferrero2021yielding,popovic2021thermally,fan2014thermally,lerbinger2022relevance,rodriguez2023temperature}. In this study, we set $\tau_0=1$, $c=1$, and the value $\alpha=3/2$ corresponds to a smooth microscopic potential~\cite{aguirre2018critical}. 
As stress is relaxed locally at the site of the plastic event, stress in the system changes according to the Eshelby kernel~\cite{picard2004elastic}. Specifically, in the scalar model the kernel is randomly oriented at each relaxation event. This random orientation of the Eshelby kernel is a crucial to describe for isotropic supercooled liquids in a quiescent state, in contrast to amorphous solids under shear, where Eshelby fields align along the shear direction~\cite{maloney2006amorphous}.
Note that in the tensorial model, 
after each relaxation the yielding surface is reoriented uniformly at random and 
one uses the full Eshelby propagator, hence no additional symmetrization is required~\footnote{Full isotropy is slightly broken by the choice of bi-periodic boundary conditions.}.
Besides, our elastoplastic models have zero total (macroscopic) stress, as anticipated in quiescent supercooled liquids. Detailed description of the tensorial and scalar models are found in Appendix~\ref{sec:appendix_models}. We perform numerical simulations in a two-dimensional periodic lattice ($d=2$), whereas we develop scaling arguments in general $d$ dimensions. 
Since we find that both scalar and tensor models show qualitatively and quantitatively similar results (e.g., critical exponents), we focus on the tensorial model in the main text and present the scalar one in Appendix.


\section{Dynamical heterogeneities at finite temperatures}
\label{sec:finite_T}

\begin{figure}
\centering
\includegraphics[width=\linewidth]{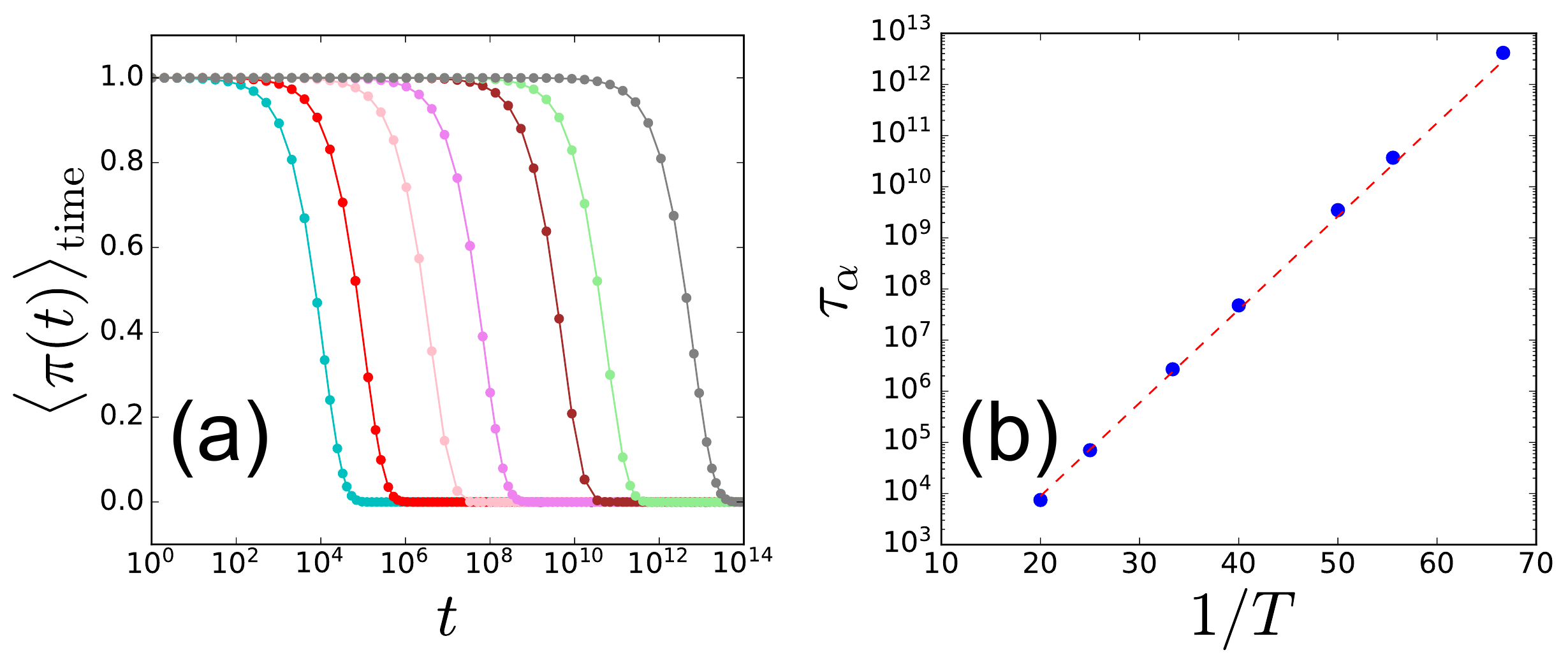}
\caption{Dynamics of the tensorial model in finite temperature simulations.
(a): Mean persistence correlation function, $\langle \pi(t) \rangle_{\rm time}$, for $L=128$ and $T=0.050, ~ 0.040, ~ 0.030, ~ 0.025, ~ 0.020, ~ 0.018$, and $0.015$ (from left to right). (b): The relaxation time $\tau_\alpha$ versus $1/T$. The red dashed-line corresponds to $\tau_\alpha \sim e^{E_c/T}$ with $E_c=x_c^{3/2}=0.42$ measured independently in Sec.~\ref{sec:zero_T}.
}
\label{fig:P_mean_and_tau}
\end{figure}

\begin{figure}
\centering
\includegraphics[width=\linewidth]{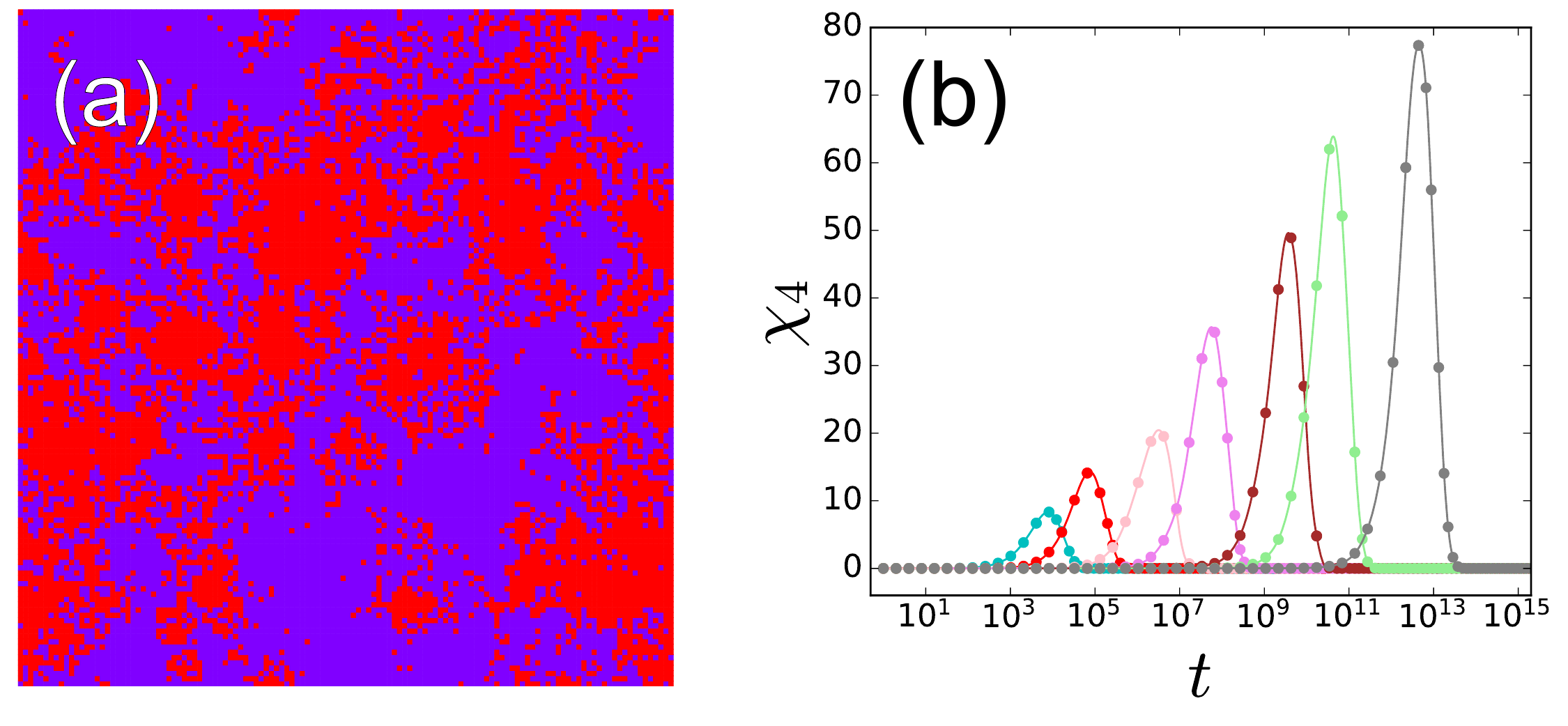}
\caption{Dynamical heterogeneity of the tensorial model in finite temperature simulations.
(a): Snapshot
characterized by local persistence, $p_i(\tau)$, when $\tau/\tau_\alpha \approx 0.53$ at $T=0.015$.
Red and purple sites correspond to mobile ($p_i(\tau)=0$) and immobile ($p_i(\tau)=1$) sites, respectively. The system size is $L=128$. (b): Four-point correlation function, $\chi_4(t)$, for $L=128$ and $T=0.050, ~ 0.040, ~ 0.030, ~ 0.025, ~ 0.020, ~ 0.018$, and $0.015$ (from left to right)..
}
\label{fig:Dynamical_heterogenity}
\end{figure}

We first perform finite temperature simulations and characterize dynamical properties.
To this end, we consider the persistence two-point time correlation function, $\langle \pi(t) \rangle_{\rm time}$, which has been used widely in the context of kinetically constrained models~\cite{ritort2003glassy}. The observable $\pi(t)$ is defined by $\pi(t) = \sum_i p_i(t)/N$,
where
$p_i(t)=1$ if the site $i$ did not exhibit a plastic event until time $t$ and remained immobile, and $p_i(t)=0$ for mobile sites that relaxed at least once. The notation 
$\langle \cdots \rangle_{\rm time}$ denotes the time average at the stationary state at temperature $T$ reached after a long enough equilibration time.
We have verified that the model does show dynamical slowing down by measuring $\langle \pi(t) \rangle_{\rm time}$ at different temperatures, as shown in Fig.~\ref{fig:P_mean_and_tau}(a).
In Fig.~\ref{fig:P_mean_and_tau}(b), we find that the associated relaxation time $\tau_\alpha$ defined by $\langle \pi(\tau_\alpha) \rangle_{\rm time}=1/2$ increases in a Arrhenius way. 
Figure~\ref{fig:Dynamical_heterogenity}(a) represents a snapshot of local persistence for $\tau/\tau_\alpha\simeq 0.53$, demonstrating spatially heterogeneous dynamics.
To quantify the magnitude 
of dynamical heterogeneity, we compute the four-point correlations function~\cite{donati2002theory}, $\chi_4(t)$, defined by
\begin{equation}
 \chi_4(t) = N \left( \langle \pi^2(t)\rangle_{\rm time} - \langle \pi(t) \rangle_{\rm time}^2 \right).   
\end{equation}
$\chi_4(t)$ is proportional to the size of the dynamically correlated region (see Appendix~\ref{sec:appendix_finite_T} for details), which is the central observable characterizing dynamical heterogeneity approaching the glass transition~\cite{berthier2011dynamical}, as it has been estimated in real experiments~\cite{berthier2005direct,dalle2007spatial,capaccioli2008dynamically,dauchot2022glass} as well as molecular simulations~\cite{donati2002theory,karmakar2009growing,coslovich2018dynamic}.

\begin{figure}
\centering
\includegraphics[width=.9\linewidth]{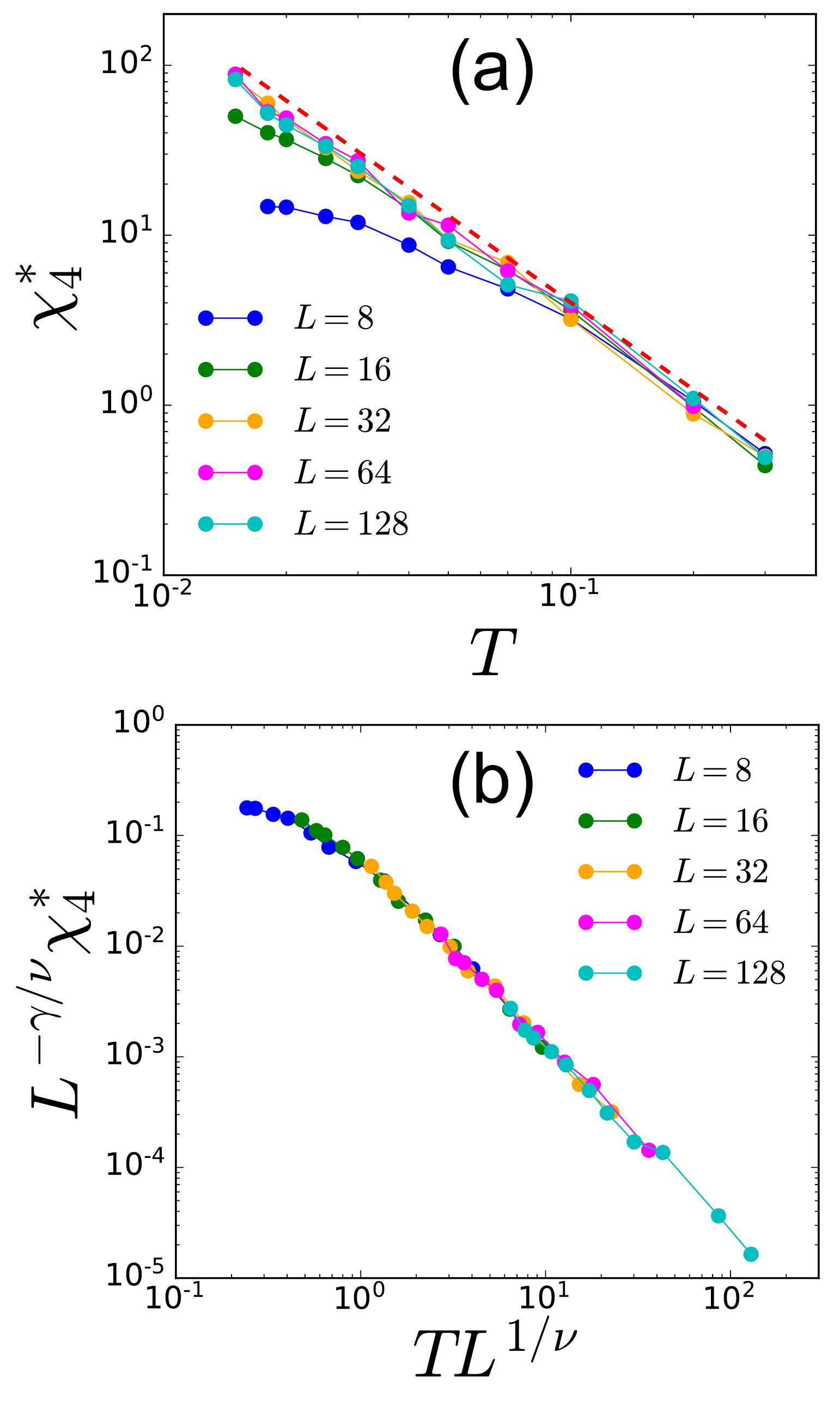}
\caption{(a): The peak of four-point correlation function, $\chi_4^{*}$, versus $T$ for several $L$ for the tensorial model. The red dashed-line follows $\chi_4^{*} \sim T^{-\gamma}$. (b): The corresponding scaling collapse.
}
\label{fig:main_finite_T}
\end{figure}

Figure~\ref{fig:Dynamical_heterogenity}(b) shows the time and temperature evolution of $\chi_4(t)$. It takes a peak near the relaxation timescale $\tau_\alpha$, and the peak grows with decreasing temperature, which is the hallmark of dynamical heterogeneity in glassy dynamics. We have then plotted the peak value of $\chi_4(t)$, denoted $\chi_4^*$, versus temperature $T$ for several system sizes in Fig.~\ref{fig:main_finite_T}(a). The system size dependence, akin to molecular simulations~\cite{karmakar2009growing,chakrabarty2017block}, allowed us to perform finite size scaling. We find that for increasing the system size $L$, $\chi_4^*$ follows a scaling form,
$\chi_4^* \sim T^{-\gamma}$ with a critical point at $T=0$ and the associated exponent $\gamma$. We obtain a scaling collapse for $\chi_4^*(L, T)$ in Fig~\ref{fig:main_finite_T}(b) (see also molecular simulation studies~\cite{karmakar2009growing,chakrabarty2017block}), indicating that the dynamics is governed by a diverging correlation length toward $T=0$, i.e., $\xi \sim T^{-\nu}$ with an exponent $\nu$.
The corresponding data for the scalar model are presented in Appendix~\ref{sec:appendix_finite_T}. Besides, we obtained a consistent result in terms of $\nu$ by directly measuring the dynamical correlation lengthscale based on the four-point structure factor~\cite{lavcevic2003spatially}.
The observed critical exponents and predictions (see below) are summarized in Table \ref{tab:critical_exponents_finite_T}.
The main goal of this paper is to provide a scaling theory for these critical behavior, connecting the dynamics at finite temperature to a zero-temperature critical point.

\begin{table}
\begin{ruledtabular}
\caption{Critical exponents ($\gamma$, $\nu$) obtained from finite $T$ simulations in the scalar and tensorial elastoplastic models in two-dimensions, compared with their predicted values proposed in Section~\ref{sec:scaling_argument}.}
\label{tab:critical_exponents_finite_T}
\begin{tabular}{c|cccc}
&Scalar model & Tensorial model & Pred. $d=2$ & Pred. MF
\\ \hline 
$\gamma$ & $1.8 \pm 0.1$  & $1.7 \pm 0.1$ & $1.60 \pm 0.05$ & $1.5$\\
$\nu$& $0.85 \pm 0.05$  & $0.80 \pm 0.05$ &$0.80 \pm 0.03$  & $0.75$ \\
\end{tabular}
\end{ruledtabular}
\end{table}

\section{Critical Point and Extremal Dynamics at $T=0^+$}
\label{sec:zero_T}

We now consider dynamics at vanishing temperature, $T=0^+$, and show that it is related to a critical point. At $T=0^+$ the site with the smallest energy barrier $E_{\rm min}= x_{\rm min}^{3/2}$, the weakest site, always yields first \footnote{We are considering either finite systems or the thermodynamic limit taken after the zero-temperature limit.}, where $x_{\rm min}$ is the corresponding stress required to destabilise the site. Therefore, one can simulate dynamics at $T=0^+$ by relaxing always the weakest site, instead of relaxing a random site weighted by the relaxation rate $e^{-E(x)/T}$. This algorithm allows us to access dynamical information even at vanishing temperature 
(see details in Appendix~\ref{sec:Appendix_extremal}).
It is an example of {\it extremal dynamics}, which is well-studied in the context of self-organized-criticality~\cite{paczuski1996avalanche} and some disordered materials under quasi-static driving~\cite{Baret2002,purrello2017creep}, including annealed glasses~\cite{kumar2022mapping}.

In the thermodynamic limit, $L \to \infty$, the extremal dynamics leads to an absorbing condition for $E \leq E_c$, where $E_c$ is a critical energy barrier as found (for $T\rightarrow 0$) in Ref.~\cite{ozawa2023elasticity}. As a result, the distribution of local (activation) energy barriers $P(E)$ vanishes for $E \leq E_c$ (see the sketch in Fig.~\ref{fig:sketch}(a)). 
This implies that $\lim_{L\rightarrow\infty} P(x)=0$ for $x\leq x_c=E_c^{2/3}$, where $P(x)$ is the distribution of $x$. Thus only a sub-extensive number of sites are found for $x \leq x_c$ at a finite $L$.
In this paper, we use $P(E)$ and $P(x)$ interchangeably since they carry essentially the same information.

\begin{figure}
\centering
\includegraphics[width=\linewidth]{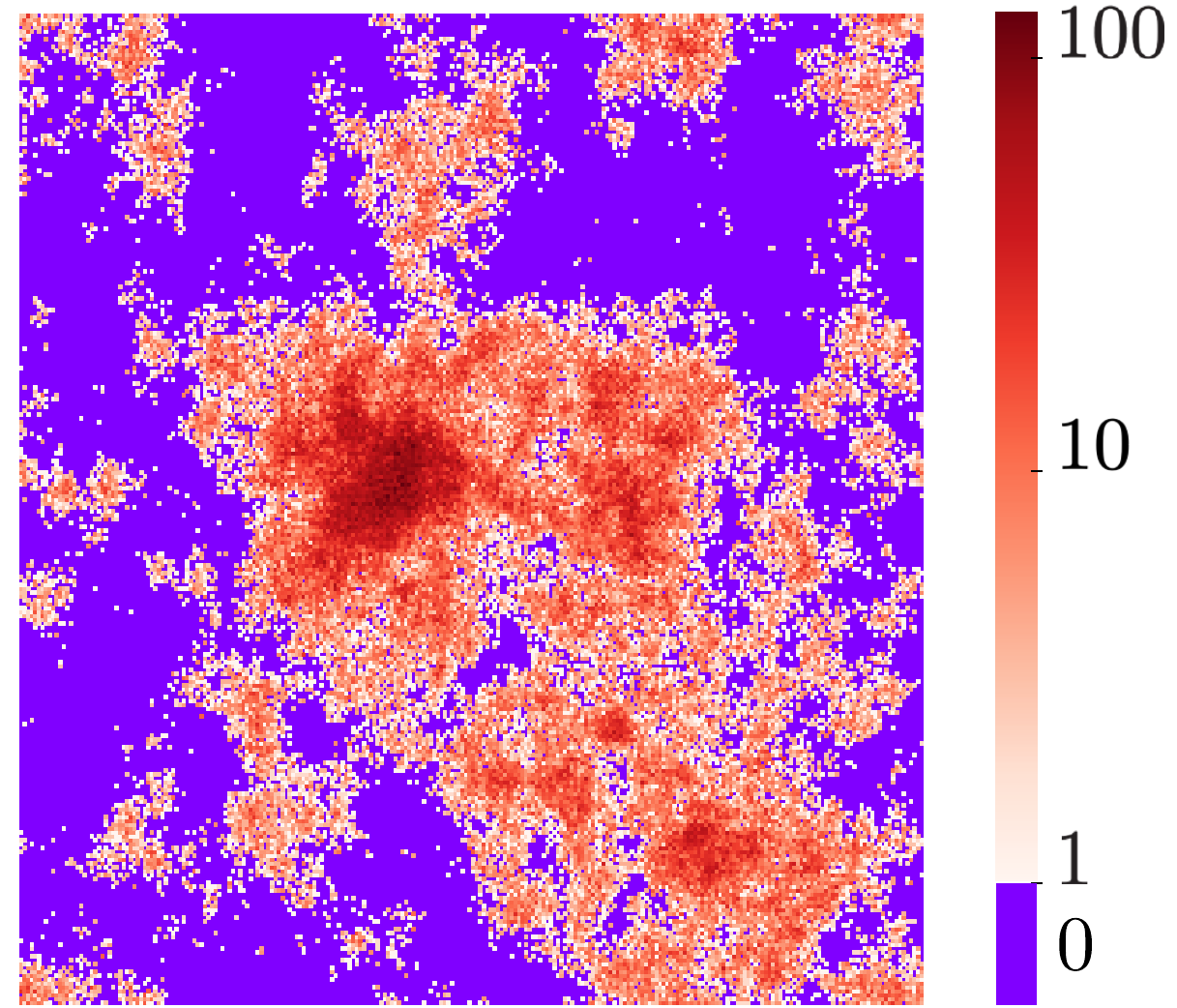}
\caption{
Snapshots of an avalanche formation for the extremal dynamics of the tensorial model with $L=256$.
Event-based avalanche size is $S \simeq 2.7 \times 10^5$, while the site-based avalanche size is $\tilde S \simeq 3 \times 10^4$.
Purple shows immobile sites (zero event), and the colorbar shows the number of relaxation events in mobile sites.
}
\label{fig:event_and_site_based_avalanches}
\end{figure}

The extremal dynamics consists of a succession of avalanches. In fact, relaxation at a site can change local energy barriers $E$ at different sites by elastic interactions. As long as those are below $E_c$ the corresponding sites belong to the same avalanche~\cite{paczuski1996avalanche}. 
This is a vivid realization of the phenomenon of dynamic facilitation.
To characterize the avalanches, we follow the method introduced in the study of extremal dynamics~\cite{paczuski1996avalanche,purrello2017creep}. 
For a finite $L$, we fix a chosen threshold stability value $x_0 \  (\leq x_c)$, and define {\it $x_0$-avalanches} as the sequences of events for which $x_{\rm min}<x_0$ (see Fig.~\ref{fig:x_min_evolution} in Appendix~\ref{sec:Appendix_extremal}). 
Two useful characterizations of avalanche size are the total number of relaxation events $S$ in a given sequence (event-based avalanche size), and the total number of sites $\tilde S$ that relaxed at least once during an avalanche (site-based avalanche size).
By construction, $\tilde S \leq S$ and $\tilde S \leq L^d$.
We show snapshots having $S$ and the corresponding $\tilde S$ in Fig.~\ref{fig:event_and_site_based_avalanches}. The two can be (and, as we shall show, are) different as a site can relax multiple times within the same avalanche.
Thus, the event-based avalanche size $S$ will allows us to quantify the accumulation of relaxation events in the mobile region (as emphasized in a recent molecular simulation study~\cite{scalliet2022thirty}), whereas the site-based avalanche size $\tilde S$ will be associated to the spatial extent of dynamically correlated regions (as it contains the essentially same information as the persistence map in Fig.~\ref{fig:Dynamical_heterogenity}(a) and hence $\chi_4$). 
Sizes of avalanches, $S$ and $\tilde S$, depend on the threshold $x_0$. They grow with increasing $x_0$ and diverge at $x_c$. 

By systematically exploring different values of $x_0$ one can determine the critical point $x_c$.
In fact, one expects that the avalanche distribution $P(S)$ during the extremal dynamics follows a power-law with a scaling form~\cite{paczuski1996avalanche,purrello2017creep,han2018critical}:
\begin{equation}
    P(S) \sim S^{-\tau} g(S/S_c),
    \label{eq:P_of_S_power_law}
\end{equation}
where $g(z)$ is a scaling function and $S_c$ is a cutoff size which takes the form: 
\begin{equation}
    S_c \sim (x_c-x_0)^{-1/\sigma}f\left( \frac{L^{d_f}}{(x_c-x_0)^{-1/\sigma}}\right),
    \label{eq:S_c_scaling}
\end{equation}
where $1/\sigma$ and $d_f$ are critical exponents and $f(z) = 1$ for $z \gg 1$ and $f(z)=z$ for $z \to 0$.
Thus, $S_c \sim L^{d_f}$ when $x_0 \to x_c$, whereas $S_c \sim (x_c-x_0)^{-1/\sigma}$ when $L \to \infty$.
The same expressions are expected to hold for the site-based avalanche size $\tilde S$, defining exponents $\tilde\tau$, $1/\tilde \sigma $, and $\tilde d_f$.
To estimate $S_c$ we use the fact that for $1 <\tau < 3$ the ratio $\langle S^3 \rangle / \langle S^2 \rangle$ is proportional to the cutoff value $S_c$, where $\langle \cdots \rangle=\int_0^{\infty} \mathrm{d} S  P(S)(\cdots)$. As we are interested in scaling of $S_c$ with system size, the numerical constant is irrelevant and we  define $S_c \equiv \langle S^3 \rangle/\langle S^2 \rangle$.
The same expression is used mutatis mutandis for the site-based avalanche size, $\tilde S$.
More details about avalanche statistics can be found in Appendix~\ref{sec:Appendix_extremal}.

\begin{figure}
\centering
\includegraphics[width=\linewidth]{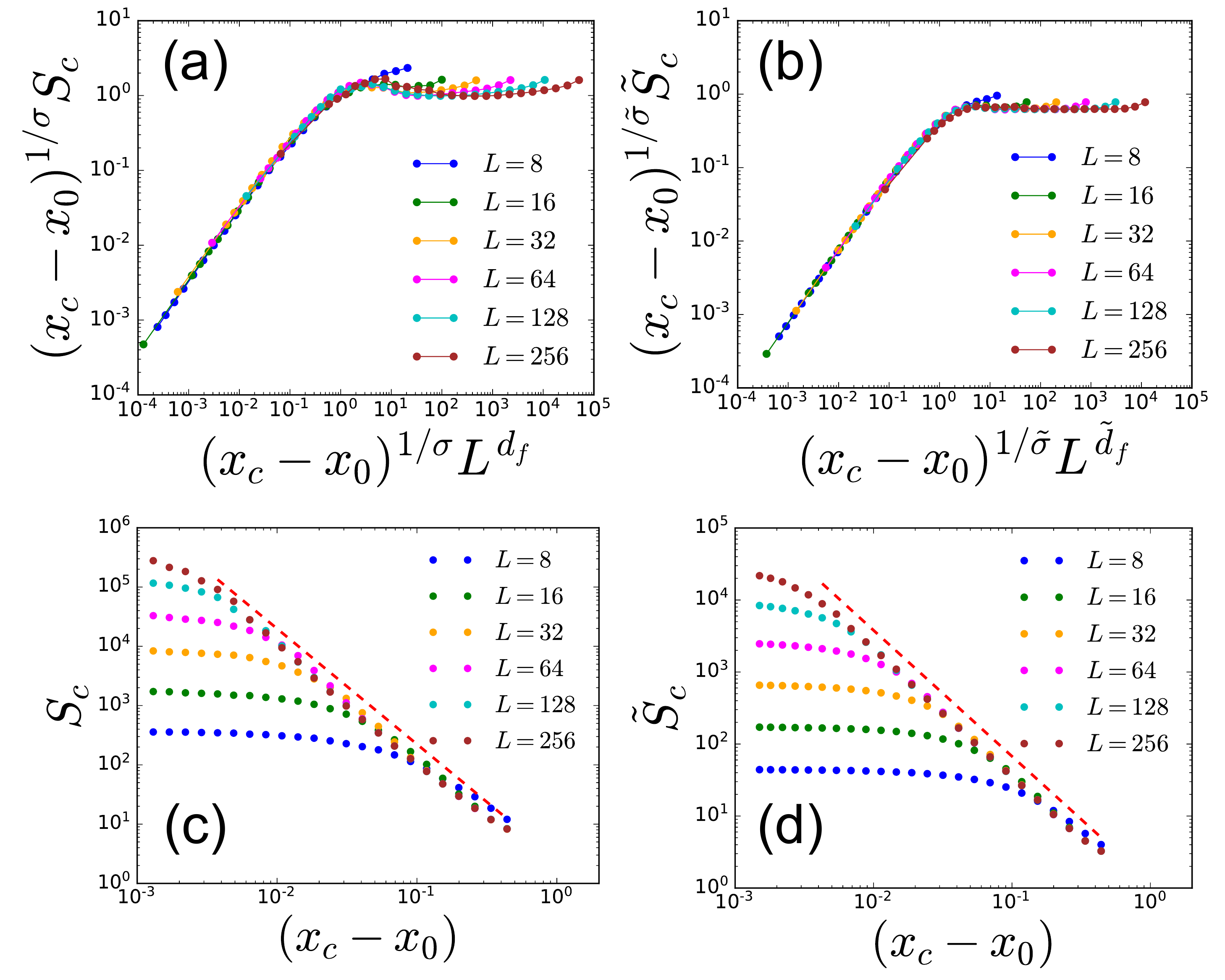}
\caption{Statistical properties of avalanches during the extremal dynamics at $T=0^+$ for the tensorial model.
(a, b): Scaling collapse for the cutoff size $S_c=\langle S^3 \rangle / \langle S^2 \rangle$ based on Eq.~(\ref{eq:S_c_scaling}), for various $L$ for the event-based (a) and site-based (b) avalanche sizes, which determines the critical threshold $x_c$ and critical exponents, $1/\sigma$, $1/\tilde\sigma$, $d_f$, and $\tilde d_f$. 
(c, d): $S_c$ versus $x_c-x_0$. The red dashed line corresponds to $S_c \sim (x_c-x_0)^{-1/\sigma}$ in (c) and $\tilde S_c \sim (x_c-x_0)^{-1/\tilde \sigma}$ in (d). 
}
\label{fig:main_zero_T_avalanches}
\end{figure}

We determine $x_c$ and the exponents $1/\sigma$, $1/\tilde{\sigma}$, $d_f$ and $\tilde{d}_f$ by measuring $S_c$ and $\tilde{S}_c$ for different values of threshold $x_0$ and system size $L$ and collapsing them using the 
scaling form in Eq. (\ref{eq:S_c_scaling}) 
for both $S_c$ and $\tilde{S_c}$, as shown in Figs.~\ref{fig:main_zero_T_avalanches} (a, b). We obtain 
$x_c= 0.560\pm 0.001$. The values of the critical exponents are presented in Table~\ref{tab:critical_exponents}.
Figures~\ref{fig:main_zero_T_avalanches} (c, d) display $S_c$ and $\tilde S_c$ as a function of $x_c-x_0$, for various system size $L$. They indeed show that $S_c$ and $\tilde S_c$ grow with $x_0$ and saturate due to finite size effects. 

\begin{figure}
\centering
\includegraphics[width=\linewidth]{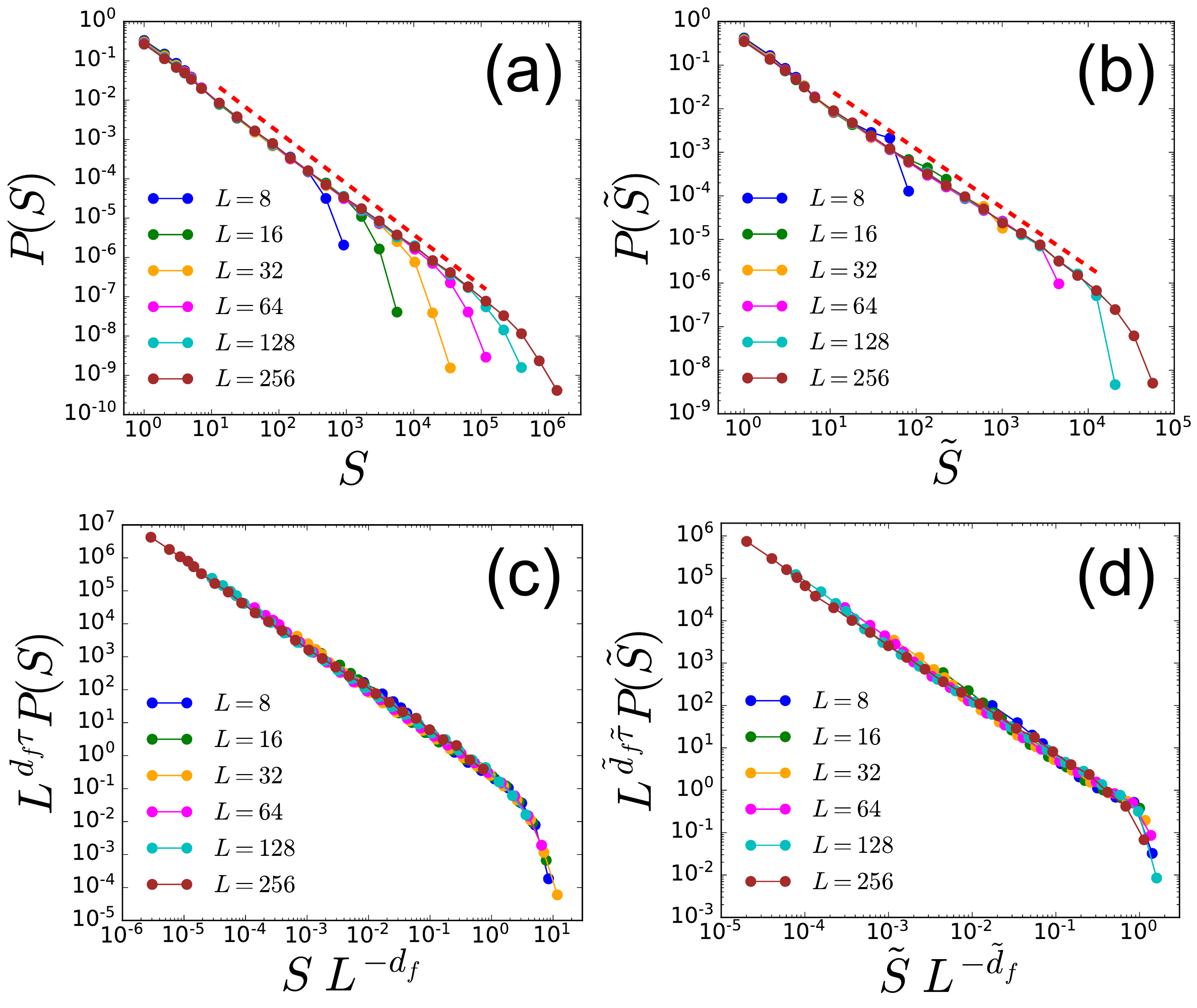}
\caption{(a, b): Distribution of avalanche size $P(S)$ (a) and $P(\tilde S)$ (b) for a stability threshold $x_0=x_c$ for the tensorial model, with varying the system size $L$. The dashed lines follow $P(S) \sim S^{-\tau}$ (a) and $P(\tilde S) \sim \tilde  S^{-\tilde  \tau}$ (b). (c, d): The corresponding scaling collapse following Eq.~(\ref{eq:P_of_S_power_law}) with $S_c \sim L^{d_f}$ (c). The same expressions are used for $P(\tilde S)$ in (d).
}
\label{fig:P_of_S_tensor}
\end{figure}

Once $x_c$ is determined, we can study the statistics of the system-spanning avalanches relevant to the thermodynamic limit. Thus, we fix $x_0=x_c$ and measure the distribution $P(S)$ and $P(\tilde S)$ of the avalanche sizes, see Figs~\ref{fig:P_of_S_tensor} (a,b). 
We find that avalanche sizes are power-law distributed with eventual cut-off,  consistent with the general assumption in Eq.~(\ref{eq:P_of_S_power_law}).
The scaling form in Eq.~(\ref{eq:P_of_S_power_law}) collapses the data using $S_c \sim L^{d_f}$ and the previously obtained values of $d_f$ and $\tilde d_f$, see Figs~\ref{fig:P_of_S_tensor} (c,d). From the data collapse we determine values of avalanche exponents $\tau$ and $\tilde \tau$ (see Table~\ref{tab:critical_exponents}).
The successful collapse of the data confirms the validity of the scaling ansatz as well as the value of critical exponents.

The above analysis reveals that the extremal dynamics of our model of glass forming-liquid displays, scale-free, avalanche-type dynamics akin to the ones of other disordered systems under external loading~\cite{sethna2001crackling,rosso2022avalanches}.

\begin{figure}
\centering
\includegraphics[width=\linewidth]{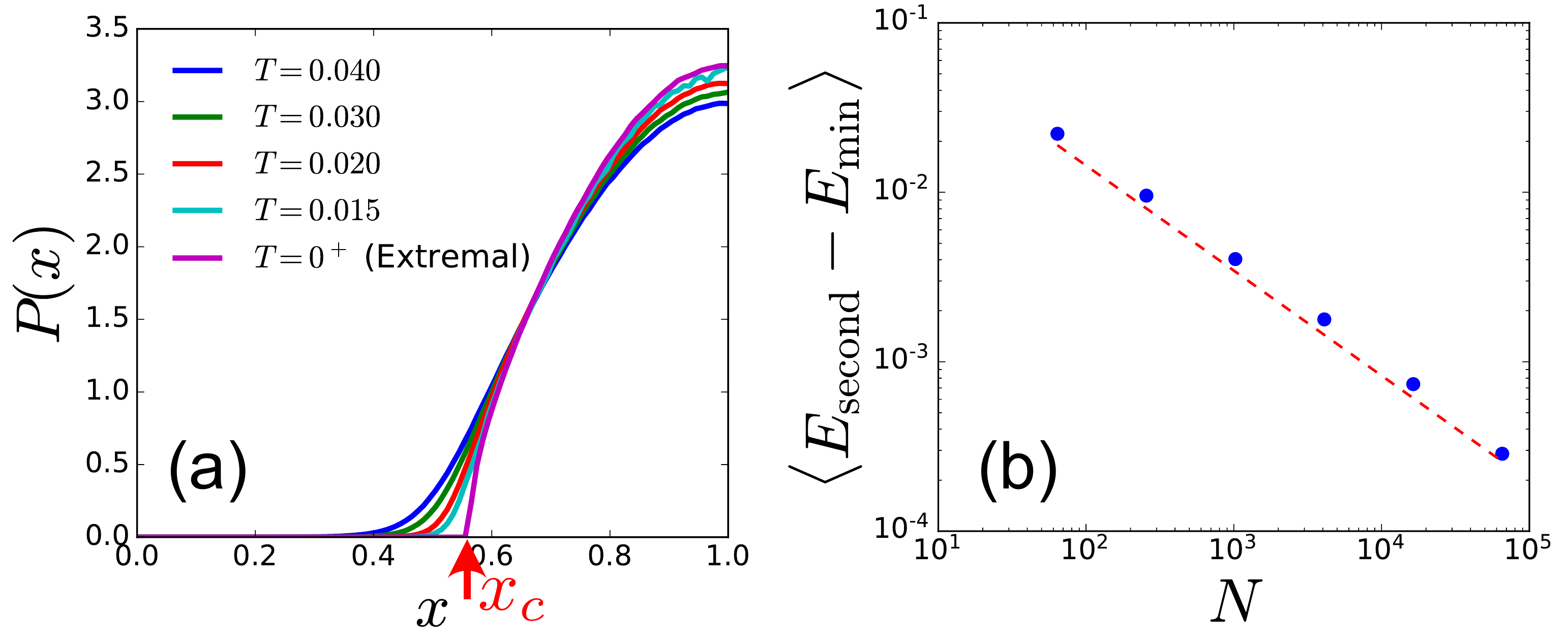}
\caption{(a): $P(x)$ from finite $T$ simulations and extremal dynamics at $T=0^+$ for the tensorial model with $L=256$. The red arrow indicates $x_c$.
(b): Average of $E_{\rm second}-E_{\rm min}$ obtained from the extremal dynamics for various $N=L^d$. The red dashed line corresponds to $\langle E_{\rm second}-E_{\rm min} \rangle \sim N^{-\delta}$.}
\label{fig:main_weak_site_distribution}
\end{figure}

Coming back to the distribution of energy barriers $P(E)$ and the corresponding distribution $P(x)$, 
one would expect that it continuously vanishes at $x_c$ as $P(x) \sim (x-x_c)^\theta$, defining an exponent $\theta$ \cite{popovic2021thermally}.
This behavior also occurs near the yielding transition of amorphous solids under shearing (with $x_c=0$)~\cite{Lin14a}, where it affects the scaling of flow properties~\cite{Lin2014b} and plasticity \cite{Karmakar2010}, and more generally in glassy systems with long-range interactions~\cite{muller2015marginal}. The underlying reason is that at each step of the extremal dynamics, each site receives a stress kick, such that $x_i$ of a site $i$ follows a random process, with an effective absorbing condition at $x_c$. If this random process was a Brownian motion, then one would obtain $\theta=1$. Yet the kicks are much more broadly distributed, and in higher dimensions, this random process is akin to a Levy flight~\cite{Lemaitre07}, which can be shown to imply $\theta_{\rm MF}=1/2$ ~\cite{lin2016mean}. 
In Fig.~\ref{fig:main_weak_site_distribution}(a), we plot $P(x)$ for the extremal dynamics, measured from configurations just before (or after) each avalanche defined as $x_{\rm min}>x_0=x_c$, together with $P(x)$ obtained from finite temperature simulations studied in Sec.~\ref{sec:finite_T}. 
At higher $T$, $P(x)$ shows a broader distribution with a smooth decay. As $T$ is decreased, $P(x)$ converges to the one obtained by the extremal dynamics at $T=0^+$, whose shape is consistent with $P(x)\sim (x-x_c)^\theta$.

The following two features of $P(x)$ connect the extremal and finite temperature dynamics. First, the point $x_c$ or the corresponding energy scale $E_c=x_c^{3/2}$ controls the effective energy barrier associated to $\tau_\alpha$. Indeed, at small temperature the dynamics proceeds by relaxing the sites with smallest barriers, i.e., sites having barriers close to $E_c$ ~\cite{ozawa2023elasticity,popovic2021thermally}. This is in agreement with the relaxation time $\tau_\alpha$ observed in Fig.~\ref{fig:P_mean_and_tau}(b), which scales as $ \tau_\alpha \sim e^{E_c/T}$.

Second, more importantly for what follows, in a finite system of size $N=L^d$, the typical scale of $x_{\rm min}$ and the second lowest $x$, denoted as $x_{\rm second}$, is expected to follow
a power law, $\langle x_{\rm second}-x_{\rm min} \rangle \sim N^{-\delta}$. As we shall show, this plays a key role in the characterization of thermal avalanches.
Since the energy difference between the lowest and second lowest activation energies is given by $E_{\rm second}-E_{\rm min} \sim x_{\rm second}-x_{\rm min}$, we conclude
\begin{equation}
\langle E_{\rm second}-E_{\rm min} \rangle \sim N^{-\delta}.
\label{eq:delta_scaling}
\end{equation}
Thus the exponent $\delta$ characterizes an important feature of the energy barrier relevant for low temperature dynamics.
We numerically confirmed this scaling in Fig.~\ref{fig:main_weak_site_distribution}(b), and the obtained value of $\delta$ is reported in Table~\ref{tab:critical_exponents}.
Extreme value statistics argument would suggest $\delta=1/(1+\theta)$~\cite{Lin2014b} (although near the yielding point deviations from this relation have been reported in finite dimension \cite{ferrero2021properties}, and explained in \cite{korchinski2021signatures}).   
Within the mean-field theory~\cite{lin2016mean},  one finds $\delta_{\rm MF}=1/(1+\theta_{\rm MF})=2/3$, a value close to what we observe in Fig.~\ref{fig:main_weak_site_distribution}(b).


The results presented in this section show that the extremal dynamics of the tensorial elastoplastic model of super-cooled liquids is governed by system-spanning avalanches. We have fully characterized the associated zero temperature critical point by obtaining all relevant exponents, summarized in Table~\ref{tab:critical_exponents}.
We also obtained qualitatively and quantitatively (e.g., critical exponents) similar results in the scalar model, which are presented in Appendix~\ref{sec:Appendix_extremal}.
Remarkably, we find that the scalar and tensorial models display very close values of the critical exponents and likely correspond to the same universality class.

\begin{table}
\begin{ruledtabular}
\caption{Critical exponents obtained from extremal dynamics simulations at $T=0^+$ for the scalar and tensorial elastoplastic models in two-dimensions. The reported error for the measured exponents corresponds to the range of parameters over which the power law behaviour successfully collapses the data.}
\label{tab:critical_exponents}
\begin{tabular}{cc}
Scalar model & Tensorial model
\\ \hline 
$\delta = 0.64 \pm 0.01$ & $\delta = 0.62 \pm 0.01$ \\

$\tau = 1.25 \pm 0.05$  & $\tau = 1.30 \pm 0.05$  \\
$d_f = 2.3 \pm 0.1$  & $d_f = 2.3 \pm 0.1$  \\
$1/\sigma = 2.2 \pm 0.1$  & $1/\sigma = 1.95 \pm 0.03$  \\

$\tilde \tau = 1.25 \pm 0.05$  & $\tilde \tau = 1.35 \pm 0.03$  \\
$\tilde d_f = 2.00 \pm 0.02$  & $\tilde d_f = 1.95 \pm 0.05$  \\
$1/\tilde \sigma = 1.9 \pm 0.1$  & $1/\tilde \sigma = 1.75 \pm 0.03$  \\


\end{tabular}
\end{ruledtabular}
\end{table}

\section{Scaling theoretical arguments}
\label{sec:scaling_argument}

In the following, we develop a scaling theory that connects dynamical heterogeneities observed in finite temperature simulations (Sec.~\ref{sec:finite_T}) and the zero-temperature critical point, and associated avalanches, of the extremal dynamics (Sec.~\ref{sec:zero_T}). We will also discuss other important physical consequences, such as the time evolution of avalanche sizes and the  Stokes-Einstein violation.

\subsection{Length scale of dynamical heterogeneity}

We consider the effect of finite temperature $T$ on the extremal dynamics. As we will discuss below, the breakdown of the condition for the extremal dynamics naturally leads to the lengthscale $\xi$ of dynamical heterogeneity at finite $T$. The proposed picture is schematically depicted in Fig.~\ref{fig:sketch}.

At finite but low $T$, the dynamics is expected to be extremal on large but finite lengthscales. To understand the underlying mechanism, let us consider a finite size system. For a fixed system size, if $T$ is small enough, the site with the lowest activation energy $E_{\rm min}$ typically relaxes first like for $T=0^+$ (the probability to relax the site with the second lowest energy is negligible). 
This relaxation corresponds to a slow "nucleation" event when $E_{\rm min} \approx E_c$ that occurs on a timescale $\tau_\alpha \sim e^{E_c/T}$, as schematically shown in Fig.~\ref{fig:sketch}(a). The nucleation event lowers energy barriers of some other sites of the system, which we call "facilitated" sites, due to elastic interactions, and causing a sequence of faster events when $E_{\rm min}<E_c$, as shown in Fig.~\ref{fig:sketch}(b). This cascade process forms an avalanche, which eventually stops. A new avalanche starts again once another nucleation event occurs. The dynamics is thus intermittent with the power-law avalanches discussed in Sec.~\ref{sec:zero_T}. Clearly, there is a typical size $N_T$ above which this picture breaks down.   
Indeed, the above description holds when the thermal energy $T$ is much smaller than the typical energy difference between the lowest and second lowest activation energies, $\langle E_{\rm second} - E_{\rm min} \rangle$, otherwise sites with higher energy barriers (such as $E_{\rm second}$) might relax and dynamics cease to be extremal. 
According to Eq.~(\ref{eq:delta_scaling}), $\langle E_{\rm second} - E_{\rm min} \rangle$ depends on the system size $N$.
Thus, the condition for the extremal dynamics which interplays $T$ and $N$ is given by $T \ll \langle E_{\rm second} - E_{\rm min} \rangle \sim N^{-\delta}$, i.e., $N \ll N_T\sim T^{-1/\delta}$.

In consequence, when the system size $N$ is too large at fixed $T$, the above description cannot hold. In this case, in particular in the thermodynamic limit, multiple nucleation events followed by avalanches will take place (independently) in parallel in the system, as sketched in Fig.~\ref{fig:sketch}(c). The cut-off length scale $\xi$ encompassing a single avalanche is not limited by system size but by other avalanches in the system, which can be estimated assuming that the locations of these avalanches are independent. Such an approximation will be accurate if structural spatial correlations are not prepunderant in this system, as discussed below. Assuming finite size scaling, $\xi^{d}$ must then corresponds to the largest system size $N_T$ for which extremal dynamics at finite $T$ holds, i.e., $\xi^{d} \sim N_T$. This provides the link between finite size zero-temperature avalanches and thermal ones. Moreover, it also directly predicts that the size of dynamically correlated regions characterized by $\chi_4^*$, which is given by the cross-over size above which the condition for the extremal dynamics breaks down: $\chi_4^* \sim \xi^{\tilde d_f}$ leading to $\chi_4^* \sim T^{-\tilde{d_f}/d\delta}$.
To conclude, we derive two scaling relations for thermal avalanches:
\begin{eqnarray}
    \gamma &=& \frac{\tilde{d_f}}{d \delta} ,\\
    \nu &=& \frac{1}{d \delta} .
    \label{eq:3}
\end{eqnarray}
These scaling relations connect dynamical heterogeneities at finite $T$ (characterized by $\gamma$ and $\nu$) and the distribution of local energy barriers at $T=0^+$ (characterized by $\delta$) together with the morphology of avalanches  during the extremal dynamics (by $\tilde d_f$).
We thus predict $\gamma$ and $\nu$ 
by using $\delta$ and $\tilde d_f$ measured in the extremal dynamics in Sec.~\ref{sec:zero_T}. 
These predictions are in good agreement with finite temperature simulations in Sec.~\ref{sec:finite_T}, as summarized in Table~\ref{tab:critical_exponents_finite_T}. 
We also find that the mean-field predictions 
using $\delta_{\rm MF}=2/3$ and $\tilde d_f=d=2$ lead to similar values.



This argument is expected to break down if large spatial correlations characterize the structure of the system, causing the locations where avalanches start to be correlated. Indeed, as is always the case in disordered materials at zero temperature, any critical threshold such as $x_c$ or $E_c$ must display finite-size fluctuations $\Delta x_c\sim \Delta E_c$. These fluctuations are described by some exponent $\nu'$, such that in a  system of finite size $L$, one has $\Delta E_c\sim L^{-1/\nu'}$. 
The value of $\nu'$ is affected by  spatial correlations. In general one must have $\nu'\geq \nu$, since the fluctuations of $\Delta x_c$ or $\Delta E_c$ must be at least as large as the typical distance between most unstable sites in  a quiescent system $ E_{\rm second} - E_{\rm min}\sim L^{-1/\nu}$: indeed, $E_c$ cannot be more precisely defined than this difference. We expect our argument above to hold when $\nu=\nu'$, corresponding to $\Delta E_c\sim E_{\rm second} - E_{\rm min}$.
$\nu'$ can be related to previously introduced exponents, $1/\tilde \sigma$ and $\tilde d_f$, as follows. If a finite system displays a system spanning avalanche $\tilde S \sim \tilde S_c \sim L^{\tilde d_f}$ and entirely rearranges, then the change of $x_c$, or equivalently that of $E_c$, must be of order $\Delta x_c$. According to Eq.~(\ref{eq:S_c_scaling}), an avalanche of size $\tilde S \sim \tilde S_c \sim L^{\tilde d_f}$ is associated with a characteristic change of the critical threshold $\Delta x_c = x_c - x_0 \sim L^{-\tilde \sigma \tilde d_f}$ (when $z \approx 1$ in $f(z)$), implying that $\xi \sim \Delta x_c^{-1/(\tilde \sigma \tilde d_f)}$. Since $\Delta x_c \sim \Delta E_c$ and $\Delta E_c \sim \xi^{-1/\nu'}$, we obtain $\nu'=1/(\tilde \sigma \tilde d_f)$. 
Using numerical values for $\tilde \sigma$ and  $\tilde d_f$, this expression leads to $\nu' \approx 0.95$ and $\nu'\approx 0.9$, respectively, for the scalar and tensorial models. We thus have in the present system $\nu\approx \nu'$, supporting our assumption of independent avalanches. Note however that  such an equality does not need to hold in general, especially in large $d$. 



\subsection{Time evolution of thermal avalanches}

We now focus on the time evolution of the size of thermal avalanches, based on the scaling results for the extremal dynamics. In Sec.~\ref{sec:zero_T}, we introduced {\it $x_0$-avalanches} to more generally probe the critical behavior at $x_0 = x_c$. This turns out to be useful also to work out the relationship between time and length scales for thermal avalanches. In fact, $x_0$ is associated to a typical energy scale $E(x_0)=x_0^{3/2}$. Hence, it carries information both about the typical time-scale  
$\tau(x_0) \sim e^{E(x_0)/T}$ and the avalanche (cut-off) size $\tilde S_c(x_0)$.

According to the scaling argument discussed before, even at a finite temperature $T$, the extremal dynamics can be applied to a finite system, in particular, smaller than $N_T$.
The cutoff size of $x_0$-avalanches, $\tilde S_c$, follows $\tilde S_c \sim (x_c-x_0)^{-1/\tilde \sigma}$, see Eq.~(\ref{eq:S_c_scaling}) and Fig.~\ref{fig:main_zero_T_avalanches}(b).
During an extremal dynamics, the duration of an avalanche will be dominated by the relaxation of the most stable site it involves, of characteristic timescale $\tau(x_0) \sim e^{E(x_0)/T}$. We thus obtain a relation connecting the duration of avalanches, $\tau(x_0)$ (measured in the unit of the relaxation time $\tau_\alpha \sim e^{E_c/T}$), with the cutoff size $\tilde S_c$ as
$\tilde S_c^{-\tilde \sigma} \sim (x_c-x_0) \sim E_c-E(x_0) \sim T \ln(\tau_\alpha/\tau(x_0))$.
Therefore, avalanches in such a finite system and over a finite time interval $t$ (smaller than the relaxation time scale $\tau_\alpha$) are intermittent. Namely, the system is quiescent most of the time, yet when it is not, its avalanche size $\tilde S(t)$ is power-law distributed with a cutoff size $\tilde S_c(t)$, which is given by
\begin{equation}
    \tilde S_c(t) \sim \left[T\ln(\tau_\alpha/t)\right]^{-1/\tilde \sigma}.
    \label{eq:prediction_S_c}
\end{equation}
Thus, we find that the size of avalanches grows very slowly -- only logarithmically -- with time.

\begin{figure}
\centering
\includegraphics[width=\linewidth]{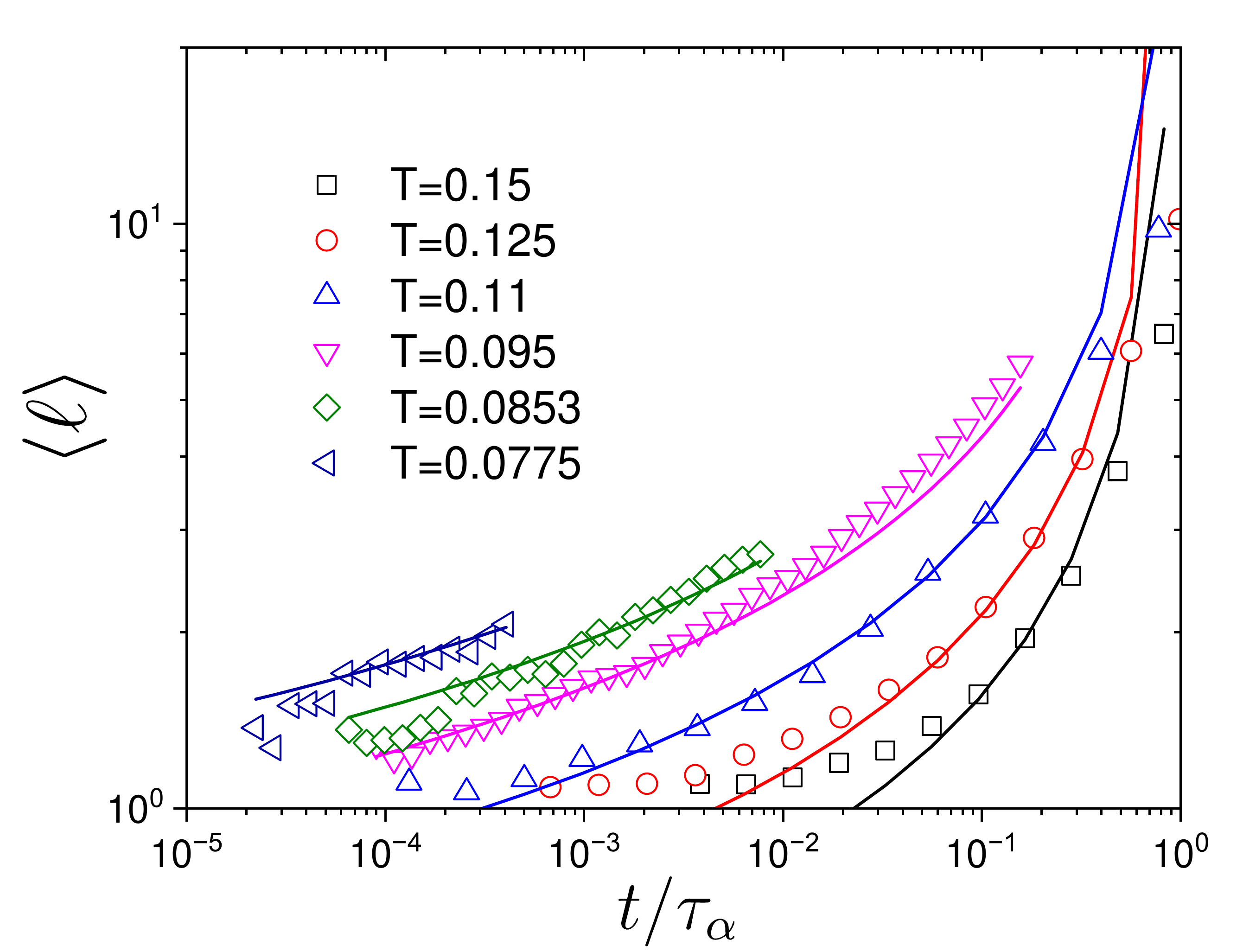}
\caption{Comparison between the molecular simulation data (empty points) in Ref.~\cite{scalliet2022thirty} and our theoretical prediction (solid curves) in Eq.~(\ref{eq:prediction_for_ell}) for the time evolution of a linear avalanche size $\langle \ell \rangle$. We set $1/\tilde \sigma=1.8$ and $\tilde d_f=2$, based on Table~\ref{tab:critical_exponents}.
We use $A=0.6, 0.7, 0.9, 1.1, 1.2$ and $1.3$ for $T=0.15, 0.125, 0.11, 0.095, 0.0853$ and $0.0775$, respectively.}
\label{fig:MD_vs_prediction}
\end{figure}

\begin{figure}[bht!]
\centering
\includegraphics[width=1\linewidth]{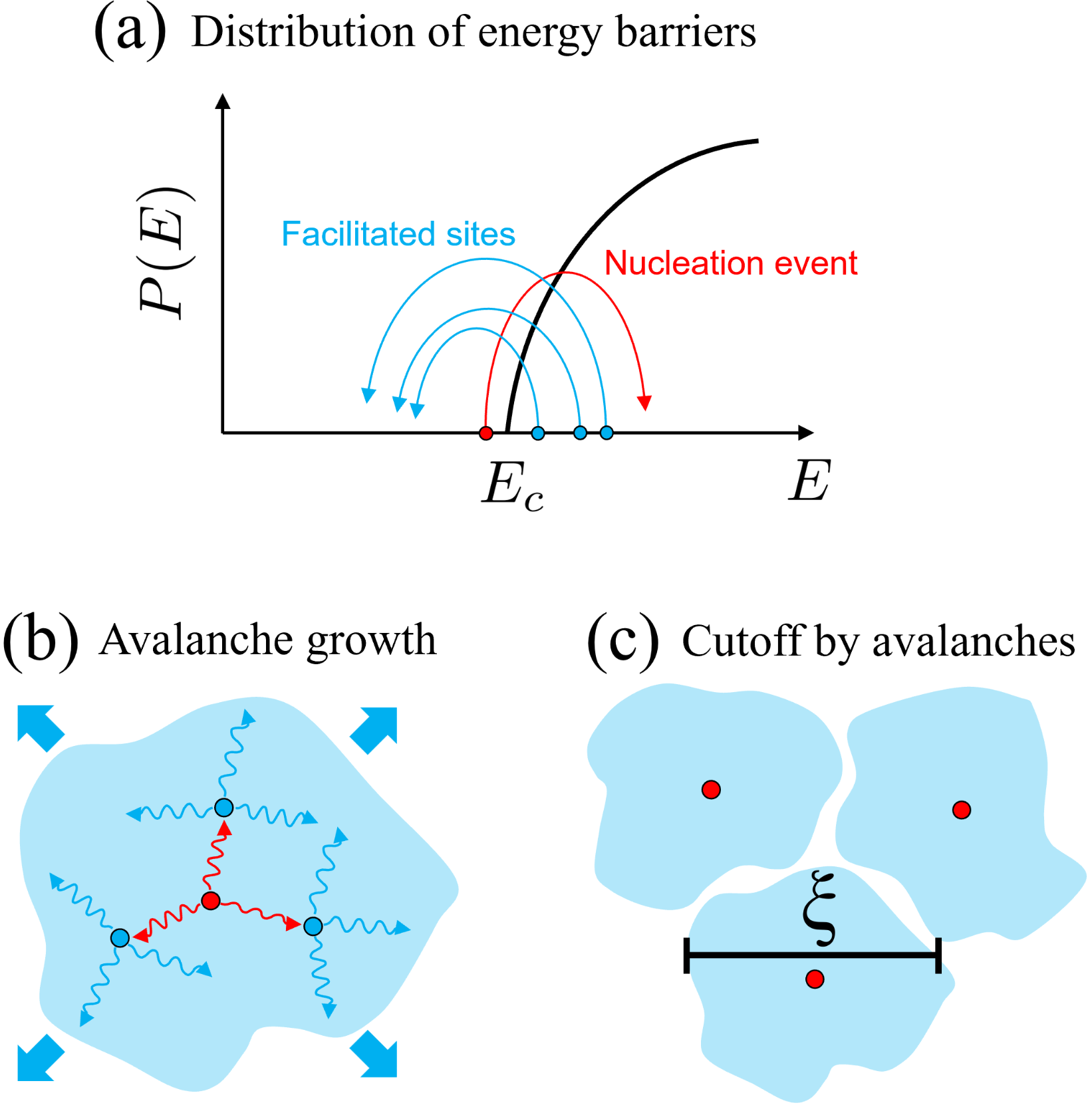}
\caption{
Proposed picture for dynamical heterogeneities in glass-forming liquids. (a): At low temperatures, the distribution of activation energy barriers $P(E)$ presents a gap below some energy $E_c$. On a time scale $\tau_\alpha \sim e^{E_c/T}$, a site with a barrier near $E_c$ relaxes (red arrow), which we call a "nucleation event". As a result, due to elastic interactions, other sites may display lower barriers $E<E_c$ (blue arrows), which we call "facilitated sites". They relax on a time scale much faster than $\tau_\alpha$, leading to a rapid sequence of events, forming a thermal avalanche. The corresponding real space picture is shown in (b, c). 
(b): A nucleation event (red circle) triggers facilitated sites (blue circles) by elastic interactions (wavy arrows). These induced events again induce other sites, forming an avalanche growth. 
(c): Avalanche growth is cut off due to other avalanches originating from different nucleation events taking place simultaneously in the system.
The cutoff length $\xi$ corresponds to the maximum size for which extremal dynamics applies, which defines the length of dynamical heterogeneity at finite temperature. }
\label{fig:sketch}
\end{figure}

The above prediction can be tested experimentally or in molecular dynamics simulations~\cite{candelier2010spatiotemporal,scalliet2022thirty}.
In Ref.~\cite{scalliet2022thirty}, the time evolution of a chord length $\langle \ell \rangle$ characterizing the linear size of mobile domains has been measured. Since we expect $\tilde S_c \sim \langle \ell \rangle^{\tilde d_f}$, our prediction for $\langle \ell \rangle$ is given by
\begin{equation}
    \langle \ell \rangle^{\rm Pred} = A(T) \left[T\ln(\tau_\alpha/t)\right]^{-1/(\tilde \sigma \tilde d_f)},
    \label{eq:prediction_for_ell}
\end{equation}
where $A(T)$ is a function with a finite limiting value as $T\rightarrow 0$.
In Fig.~\ref{fig:MD_vs_prediction}, we compare the molecular simulation data for a two-dimensional polydisperse mixture~\cite{scalliet2022thirty} and our prediction.
The prediction is very good, in particular, at lower temperatures where the elastoplastic description is supposed to work well.

In Appendix~\ref{sec:appendix_finite_T}, we also connect $\tilde S_c(t)$ and $\chi_4(t)$ explicitly. This allows us to predict the time evolution of $\chi_4(t)$ for times $1\ll t \ll \tau_\alpha$ -- a prediction that agrees with the numerics.


\subsection{Decoupling between diffusion and relaxation}

We now show that the scaling theory developed above directly implies  decoupling between diffusion and relaxation as found in super-cooled liquids, the so-called Stokes-Einstein violation~\cite{tarjus1995breakdown,ediger2000spatially,sengupta2013breakdown,charbonneau2014hopping,kawasaki2017identifying}.

One of the remarkable aspects found in the numerical simulations in Sec.~\ref{sec:zero_T} is that sites relax {\it multiple times} within the same avalanche. In terms of a dynamical trajectory, a site waits a long time (remains immobile) before relaxing, but once it relaxes, it redoes multiple times within a short period of time. This is characterized by a difference between the so-called persistence time and exchange time~\cite{jung2004excitation,berthier2004length,hedges2007decoupling} (or the caging time in molecular simulations~\cite{ciamarra2016particle,pastore2021breakdown}), which has been argued to be the key ingredient of decoupling between diffusion and relaxation in super-cooled liquids \cite{jung2004excitation,berthier2004length,hedges2007decoupling,chaudhuri2007universal,pastore2021breakdown}. 

In our case, this effect originates from the difference between the event-based and site-based avalanche sizes characterized by $d_f>\tilde{d_f}$. To connect it to the zero temperature critical point, let us focus on the formation of a thermal avalanche whose time-scale is the order of $\tau_\alpha$ and liner size is $\xi$. 
During the formation, 
a single site relaxes order of $\xi^{d_f}/\xi^{\tilde d_f}$ times. Assuming that each relaxation gives a random displacement to particles in the neighborhood of this site, their diffusion constant $D$ must be proportional to the rate for the relaxation event, $\xi^{(d_f-\tilde{d_f})}/\tau_\alpha$, leading to a Stoke-Einstein breakdown $D \tau_\alpha$ of order $D \tau_\alpha\sim \xi^{(d_f-\tilde{d_f})}\sim T^{-h}$ that diverges at vanishing temperature with $h=\nu(d_f-\tilde{d_f})$. 
Conventionally, the Stokes-Einstein violation is considered a consequence of spatially heterogeneous dynamics~\cite{ediger2000spatially}. We directly connect the former and the length scale of dynamical heterogeneity $\xi$. On top of that, our scaling theory emphasizes accumulations of multiple events inside a mobile region as the microscopic mechanism leading to the Stokes-Einstein violation. 

Note that our scaling argument does not predict a fractional Stokes-Einstein violation in which $D \sim \tau_\alpha^{-\zeta}$, or $D \tau_\alpha \sim \tau_\alpha^{1-\zeta}$,  but instead $D\tau_\alpha \sim (\log \tau_\alpha)^{1-\zeta'}$ where $1-\zeta'=\nu(d_f-\tilde{d_f})$. The former is the fit that is usually conjectured from experimental data \cite{swallen2003self,mallamace2010transport} with $\zeta \approx 0.8$. However, given the small value of $1-\zeta$ the latter fit is also a viable option 
\footnote{In fact, in some kinetically constrained models~\cite{garrahan2011kinetically}, the fractional Stokes-Einstein violation (with a similar value of $\zeta$ with experiments) was conjectured based on numerical data and physical arguments. Recent rigorous results~\cite{blondel2014there} showed that this is not what happens, but there is likely a violation as the one we find. This suggests that the logarithmic Stokes-Einstein violation can be easily misinterpreted as a fractional one with a small exponent $1-\zeta$.}. 

\begin{figure}
\centering
\includegraphics[width=\linewidth]{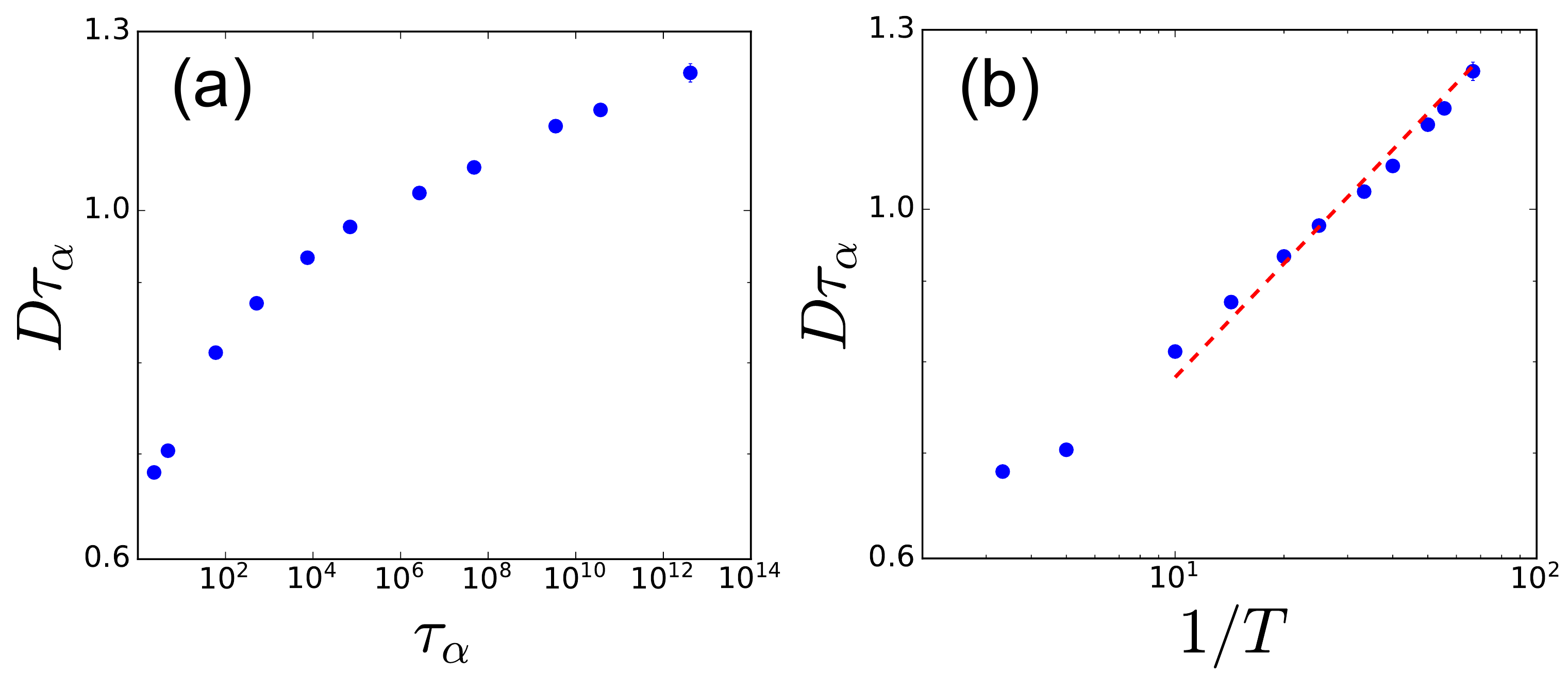}
\caption{The Stoke-Einstein decoupling $D \tau_\alpha$, where $D$ is the diffusion coefficient of tracers, for the tensorial elasto-plastic model with $L=128$. (a): $D \tau_\alpha$ versus $\tau_\alpha$  could suggest an effective power-law behaviour (the fractional Stokes-Einstein violation), with two apparent exponents in the range $\tau_\alpha<10^4$ and $\tau_\alpha>10^4$. (b): Our prediction suggests instead a power-law behaviour in terms of $T$. The red dashed line represents $D \tau_\alpha \sim T^{-\nu(d_f-\tilde d_f)}$.}
\label{fig:Stokes_Einstein_violation}
\end{figure}

We have numerically tested our prediction for the Stokes-Einstein violation in Fig.~\ref{fig:Stokes_Einstein_violation}, showing the product $D\tau_\alpha$ in the tensorial model. In this plot,  $D$ is estimated numerically using tracer particles that jump randomly by one lattice spacing each time relaxation occurs in their current site, similarly to what was originally done for kinetic constrained models~\cite{jung2004excitation,berthier2004length} (see Appendix~\ref{sec:appendix_finite_T} for details). 
$D\tau_\alpha$ increases with decreasing $T$, following the scaling prediction $D\tau_\alpha \sim T^{-h}$ with $h=\nu(d_f - \tilde d_f)$ measured independently in Secs.~\ref{sec:finite_T} and \ref{sec:zero_T}.
The observed amount of the violation we find is not large and hence more representative of strong glass-forming liquids than fragile ones since it has been reported that the magnitude of the violation and fragility are correlated~\cite{ediger2000spatially,ozawa2016tuning}. We will come back to this point in Sec.~\ref{sec:conclusion}. 



\section{Conclusion and discussion}
\label{sec:conclusion}

We provided a theoretical description of dynamical heterogeneities in super-cooled liquids based on the assumption that local rearrangements are elastically coupled. The elasto-plastic models we studied offer a quantitative solution for how dynamical correlations can emerge 
even in cases in which local barriers control the dynamics~\cite{glarum1960dielectric,dyre2006colloquium,lerbinger2022relevance,hasyim2021theory,massimo}. 
Our main result is the theoretical explanation of dynamical correlations in terms of a zero-temperature critical point with the associated scaling relations. This leads to quantitative predictions on the power-law statistics of thermal avalanches testable in more realistic systems. 
Our study suggests that dynamical heterogeneities in super-cooled liquids should be investigated in terms of the temperature $T$ to seek power-law relations, rather than in terms of the relaxation time scale. 

One important aspect of the models we studied is that they encode in a very simple and natural way the coupling of local relaxation and elastic interaction. Their simplicity, combined with the richness of the dynamical behavior -- in particular, the emergence of facilitation and dynamical correlations -- is a remarkable aspect, as it shows which salient facts one can obtain with minimal physical ingredients. 
Kinetically constrained models have instead abstract kinetic rules and show a large variety of behaviors depending on the kinetic constraints \cite{ritort2003glassy}. Although local excitations identified in the dynamic facilitation scenario~\cite{isobe2016applicability,keys2011excitations} would have connections with local activations in our framework, the physical interpretation of kinetic rules in kinetic constrained models is still an open and crucial challenge. Along this line of thought, it would be interesting to devise a kinetic constraint rule effectively incorporating elastoplasticity.
Nevertheless, various concepts and theoretical tools developed in the study of kinetically constrained models provide important guidelines and, in fact, played an essential role for our analysis of thermal elasto-plastic models.
It has been demonstrated in the dynamic facilitation theory that a critical point in an extended non-equilibrium phase diagram influences glassy dynamics~\cite{elmatad2010finite,turci2017nonequilibrium}. It would also be interesting to study whether such a critical point exists in elastoplastic models.

Quantitatively, the magnitude of dynamical heterogeneities we simulated is comparable to most super-cooled liquids (as the estimated $\chi_4^*$ increases by about two decades as the glass transition is reached~\cite{dalle2007spatial,dauchot2022glass}), whereas the magnitude of the Stoke-Einstein breakdown is comparable to that of rather strong glass-forming liquids \cite{ediger2000spatially}.
According to our scaling argument, such a breakdown will increase if the system shows more intensive accumulations of multiple relaxations characterized by larger $\nu(d_f-\tilde d_f)$.
This effect could be achieved by imposing that some of the model parameters (such as the local values of the yield stress, or how sites are coupled to the elastic field) are randomly distributed~\cite{agoritsas2015relevance}, instead of being single-valued as assumed here for simplicity. These effects are expected in glass-forming liquids due to the presence of structural heterogeneity of local orders~\cite{tanaka2010critical,royall2015role,tanaka2019revealing,paret2020assessing}. Such generalization would allow us to study the structure-dynamics relationship~\cite{widmer2004reproducible,widmer2008irreversible,hocky2014correlation} in elastoplastic models.
Further improvements may be achieved by adding fluctuations and non-linearities to the propagator, which are present at short-range \cite{lemaitre2021anomalous,Lerner14}. An interesting line of research to develop quantitative models is a
mapping from a molecular simulation to an elastoplastic
model~\cite{liu2021elastoplastic,castellanos2021insights,castellanos2022history}
or the one pursued in Refs.~\cite{tah2022fragility,xiao2023machine}, which uses machine learning methods and the so-called softness field to obtain quantitative effective models.



Our results also underline important themes to study in the future, including the nature of local rearrangements in glass-forming liquids, and their connection to fragility~\cite{ediger1996supercooled}. Concerning the latter, the current elastoplastic models correspond to strong glass-formers with Arrhenius behavior, with activation energy given by the magnitude of the gap $E_c$ entering the distribution of local barriers~\cite{popovic2021thermally,ozawa2023elasticity}. This point results from the simplifying assumption that the energy scale of local rearrangements does not vary with temperature. In an improved (still simplified) model where all elastic energies follow the high-frequency elastic modulus $G_{\infty}(T)$ of the material, the activation energy $E_c$ will be proportional to $G_{\infty}(T)$, a correlation known to exist in some glass-forming liquids~\cite{dyre2006colloquium,hecksher2015review}. More recent works relate this energy scale of local rearrangements to local (instead of global) elasticity~\cite{JeppeEdan}, plasticity~\cite{lerbinger2022relevance}, or alternatively to the varying geometry of elementary rearrangements under cooling \cite{Wencheng22}. 
Thus, it is worthwhile to reveal the relationship between the activation energy barriers in liquids and low-energy excitations in glasses~\cite{scalliet2019nature,mizuno2020intermittent,richard2023detecting}.
Local energy barriers would also be related to locally-favored structures in microscopic perspectives~\cite{coslovich2007understanding,royall2015role}.
More progress along those lines will be instrumental to understand what controls fragility in glass-forming liquids.

Note that although we mainly focused our attention on the theoretical scenario based on local barriers driving the dynamics \cite{JeppeEdan,lerbinger2022relevance,massimo} together with facilitation and avalanches, the physical phenomena discussed in this work are more general. For example, they can also apply to cases in which cooperative rearrangements take place.  In these perspectives, in particular within Random First Order Transition theory, the local relaxation event would correspond to a cooperative rearrangement \cite{scalliet2022thirty,biroli2022rfot}. Obviously, our current elastoplastic models do not take into account growing cooperativity as a static correlation, and they have a singularity only at $T=0$ in contrast to RFOT with a finite temperature singularity at $T_{\rm K}>0$. One could incorporate a growing static correlation in the models by increasing the number of sites involving a thermal activation process. It would be interesting to work out precise predictions in this case.  

Finally, we expect our scaling theory which connects spatial correlations at finite temperature to extremal dynamics at zero temperature, to be relevant for a broader class of problems beyond the context of the glass transition studied here. Phenomena in which the implications of these arguments could be studied include the creep flow of disordered materials  \cite{castellanos2018avalanche,Bauer2006, Caton2008, Divoux2011, Siebenbuerger2012, Grenard2014, Leocmach2014} or that of pinned elastic interfaces  \cite{bustingorry2007thermal, purrello2017creep,kolton2005creep} below the threshold force where they spontaneously flow.


\section*{Acknowledgement}

We thank the Simons collaboration for discussions, in particular J. Baron, L. Berthier, C. Brito, C. Gavazzoni, E. Lerner, C. Liu, D. R. Reichman, and C. Scalliet. We thank S. Patinet for insightful discussions. 
We appreciate C. Scalliet for sharing the data in Ref.~\cite{scalliet2022thirty}.
GB thanks J.P. Bouchaud for discussions, in particular on the Bak-Sneppen model and models of dynamic facilitation.  MW thanks A. Rosso and M. Muller for discussions on avalanche-type response over the years, and T. de Geus, W. Ji and M. Pica Ciamarra for discussions. MW acknowledges
support from the Simons Foundation Grant (No. 454953 Matthieu Wyart) and from the SNSF under Grant No. 200021-165509. GB 
acknowledges
support from the Simons Foundation Grant (No. 454935 Giulio Biroli).
\appendix

\section{Thermal elastoplastic models}
\label{sec:appendix_models}

\subsection{Scalar model}

We study a scalar elastoplastic model~\cite{ozawa2023elasticity} in a two-dimensional lattice whose linear box length is $L$ using the lattice constant as the unit of length. For each site, we assign local shear stress $\sigma_i$ (scalar variable) at a position ${\bf r}_i$.

The dynamical rule for the simulation model is akin to Monte-Carlo dynamics~\cite{berthier2007monte}. 
We pick a site, say $i$, up randomly among $L^2$ sites. 
If $\sigma_i$ is greater (or lower) than or equal to a threshold $\sigma_Y>0$ (or $-\sigma_Y<0$), namely,
$|\sigma_i| \geq \sigma_Y$, this site shows a plastic event: $\sigma_i \to \sigma_i - \delta \sigma_i$, where $\delta \sigma_i$ is the local stress drop.
We use an uniform threshold, $\sigma_Y=1$.
Instead, if $|\sigma_i| < \sigma_Y$, with probability $e^{-E(\sigma_i)/T}$, where $ E(\sigma_i)$ is a local energy barrier and $T$ is the temperature, this site shows a plastic event: $\sigma_i \to \sigma_i - \delta \sigma_i$. This corresponds to a plastic rearragement induced by a local thermal activation. We employ $E(\sigma_i)=(\sigma_Y-|\sigma_i|)^\alpha$ with $\alpha=3/2$~\cite{maloney2006energy}.
By introducing the local stress distance to threshold, $x_i=\sigma_Y-|\sigma_i|$, we can rewrite $E(x)=x^{3/2}$.
This specific form of the local energy barrier is suggested by molecular simulation studies~\cite{fan2014thermally,lerbinger2022relevance} and previous elastoplastic models under shear~\cite{popovic2021thermally,ferrero2021yielding}.
The stress drop $\delta \sigma_i$ associated with a plastic event is a stochastic variable. In this paper, we use $\delta \sigma_i=(z+|\sigma_i|-\sigma_Y){\rm sgn}(\sigma_i)$, where ${\rm sgn}(x)$ is the sign function and $z >0$ is a random number drawn by an exponential distribution, $p(z)=\frac{1}{z_0} e^{-z/z_0}$. $z_0$ is the mean value and we set $z_0=1$. This exponential distribution would be realistic according to molecular simulations in Ref.~\cite{barbot2018local}.

A local plastic event at site $i$ influences all other sites ($\forall j \neq i$) as 
\begin{equation}
    \sigma_j \to \sigma_j + {G}^{\psi_i}_{{\bf r}_{ji}} \ \delta \sigma_i,
\end{equation}
where ${\bf r}_{ji}={\bf r}_{j}-{\bf r}_{i}$ and $\psi_i \in [0, \pi/2)$ is a random orientation of the Eshelby kernel ${G}^{\psi}_{\bf r}$. Numerical implementation of ${G}^{\psi}_{\bf r}$ is described in Ref.~\cite{ozawa2023elasticity}.

Similar to the Monte-Carlo dynamics, we repeat the above attempt $L^2$ times, which corresponds to unit time.

For the initial condition, we draw the local stress $\sigma_i$ $(\forall i)$ randomly while keeping the force balance, i.e., the sum of stresses along
each row and column of lattice sites is strictly zero~\cite{popovic2018elastoplastic,pollard2022yielding}.
To study dynamical properties at the steady-state, we monitor the waiting time dependence of observables, and we report them only at the steady-state, discarding the initial transient part.

\subsection{Tensorial model}

We have implemented a two-dimensional elastoplastic model in which we account for the tensorial nature of the shear stress tensor. In this tensorial version of the elastoplastic model, the state of each site $i$ is described by its local shear stress tensor $\tilde{\sigma}_i$. Note that symbols $\sigma$ and $\tilde \sigma$ are also used as critical exponents in the main text, not to be confused with local shear stress defined here. The shear stress tensor is traceless and symmetric and hence in two dimensions it is defined by two independent components: $\tilde{\sigma}_{xx, i}$ and $\tilde{\sigma}_{xy, i}$.

\begin{figure}
    \centering
    \includegraphics[width=0.6\linewidth]{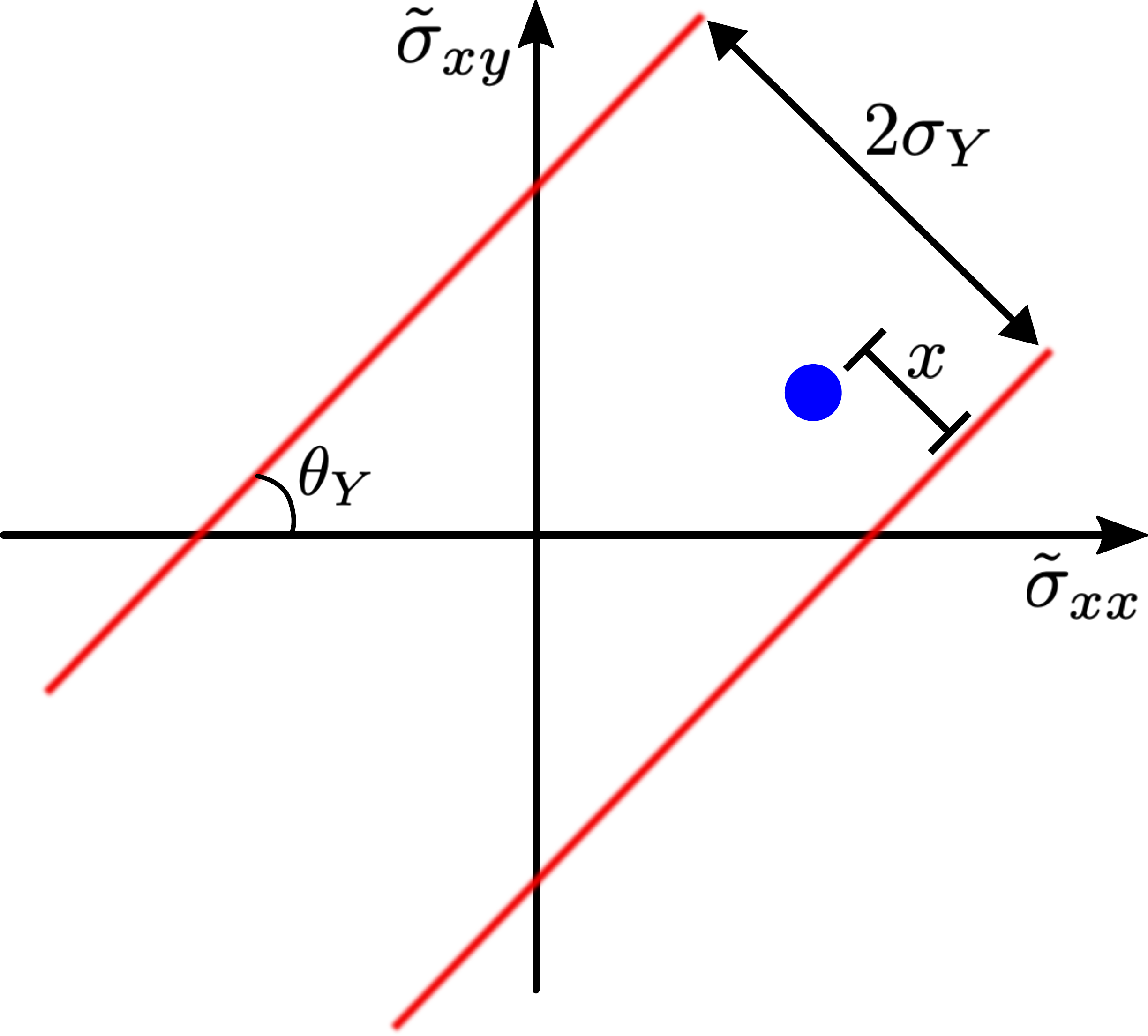}
    \caption{A schematic plot of the yield surface for the tensorial model. The blue dot shows the state of a site in the shear stress space. The yield surface is described by two parallel lines separated by $2\sigma_Y$ and each line makes an angle of $\theta_Y$ with $\tilde \sigma_{xx}$ axis. The distance from the yield surface is shown by $x$.}.
    \label{fig:yield_surface_schemati}
\end{figure}

The local yield stress is defined by a surface in the shear stress space, with the region inside and outside the surface corresponding to mechanically stable (elastic, immobile) and unstable (plastic, mobile) states, respectively.
The minimum amount of shear stress required to make a site unstable is the distance to the yield surface, and we denote its magnitude by $x$, as schematically shown in Fig~\ref{fig:yield_surface_schemati}.
We choose the local yield surface to consist of two parallel lines at an angle $\theta_{Y}$ with respect to the $\tilde\sigma_{xx}$ axis in shear stress space, centered at zero shear stress and separated by $2 \sigma_Y$ (see Fig \ref{fig:yield_surface_schemati}).
The local yield surface is assigned for each site $i$.
We initiate $\theta_{Y,i}$ with a uniformly distributed random number in $[0, 2\pi)$.

When a site $i$ becomes unstable it undergoes a plastic event over a timescale $\tau_0$: $\tilde \sigma_{xx,i} \to \tilde \sigma_{xx,i} - \delta \tilde \sigma_{xx,i}$ and $\tilde \sigma_{xy,i} \to \tilde \sigma_{xy,i} - \delta \tilde \sigma_{xy,i}$, where the amount of stress drops $\delta \tilde \sigma_{xx,i}$ and $\delta \tilde \sigma_{xy,i}$ are given by
\begin{align}
\delta \tilde \sigma_{xx,i} &= -(z - x) \sin(\theta_{Y,i}) \sgn({\tilde\sigma_{xx,i}}), \\
\delta \tilde \sigma_{xy,i} &= -(z - x) \cos(\theta_{Y,i}) \sgn({\tilde\sigma_{xy,i}}),
\end{align}
respectively.
$\sgn(x)$ is the sign function and $z$ is a random number drawn from an exponential distribution $p(z)=e^{-z/z_0}/z_0$ with $z_0=1$. The duration of a plastic event $\tau_0$ is accounted for by triggering the relaxation with a probability per unit time $1/\tau_0$ whenever the site is unstable.
In an athermal system sites can only relax by first becoming unstable. At finite temperature $T$ stable sites  undergo relaxation at the rate  $e^{-E(x)/T}/\tau_0$, where $E(x) = x^{3/2}$ is the local energy barrier~\cite{popovic2021thermally}.
After each plastic event, we redraw the angle of the yield surface $\theta_{Y, i}$ from a uniform random distribution.

Note that to simulate such a dynamics at low temperatures we implement a Gillespie type of algorithm~\cite{popovic2021thermally}, which operates as follows.  Consider an event occurring at some time $t$. Following it, a relaxation time $\tau_i$ for each site $i$ is chosen with an exponential distribution of mean  $e^{-E(x_i)/T}/\tau_0$. The next event corresponds to the smallest $\tau_i$, leading to a plastic event on the corresponding site, which occurs at  a time $t+\tau_i$. At that point, stresses are computed once again,  the variables $\tau_i$ are sampled from their new distributions. this algorithm is then repeated iteratively.

To maintain the force balance, a plastic event at site $i$ redistributes the shear stress field in the other sites following the force dipole propagator.
In Fourier space the elements of elastic kernel are given by,
\begin{align}
\widehat G_{xx,xx}({\bf q}) &= -\frac{(q_x^2-q_y^2)^2}{(q_x^2+q_y^2)^2} , \\
\widehat G_{xy,xy}({\bf q}) &= -\frac{4 q_x^2 q_y^2}{(q_x^2+q_y^2)^2} , \\
\widehat G_{xx,xy}({\bf q}) = \widehat  G_{xy,xx}({\bf q}) &= - \frac{2q_xq_y (q_x^2-q_y^2)}{(q_x^2+q_y^2)^2} ,
\end{align}
where ${\bf q}=(q_x,q_y)$ is the Fourier vector.
For a discrete system with periodic boundary conditions, we introduce a correction term to the Fourier modes, given by $q_x^2 = 2 - 2\cos{(2\pi n_x / L)}$ , $q_y^2 = 2 - 2\cos{(2\pi n_y / L)}$ and $q_x q_y = 2\sin{(2 \pi n_x /L)} \sin{(2 \pi n_y / L)}$ where $n_\alpha = -L/2 + \lbrace 1 , \cdots , L \rbrace$ with $\alpha=x, y$.



\section{Characterizations of dynamical heterogeneities for finite temperature simulations}
\label{sec:appendix_finite_T}

\subsection{Four-point correlation function}

We explain how to analyze correlation functions for finite temperature simulations.
We first consider the persistence two-point time correlation function, $\langle \pi(t) \rangle_{\rm time}$, which is defined by $\pi(t) = \frac{1}{L^d} \sum_i p_i(t)$,
where
$p_i(t)=1$ (immobile) if the site $i$ did not show a plastic event until time $t$ from $t=0$, and $p_i(t)=0$ (mobile) otherwise.
$\langle \cdots \rangle_{\rm time}$ denotes the time average at the stationary state.
$\langle \pi(t) \rangle_{\rm time}$ for the scalar and tensorial models are presented in Fig. 1 of Ref.~\cite{ozawa2023elasticity} and Fig.~\ref{fig:P_mean_and_tau}(a) in the main text, respectively.
We then measure a four-point correlations function, $\chi_4(t)$, defined by
\begin{equation}
 \chi_4(t) = L^d \left( \langle \pi^2(t)\rangle_{\rm time} - \langle \pi(t) \rangle_{\rm time}^2 \right).   
\end{equation}
$\chi_4(t)$ for the scalar and tensorial models are presented in Fig. 3 of Ref.~\cite{ozawa2023elasticity} and Fig.~\ref{fig:Dynamical_heterogenity}(b) in the main text, respectively.
$\chi_4(t)$ quantifies the size of the dynamically correlated region, because one can rewrite it as 
\begin{eqnarray}
    \chi_4(t) &=& \frac{1}{L^d} \sum_{i,j} \left( \langle p_i(t)p_j(t) \rangle_{\rm time} - \langle \pi(t)\rangle_{\rm time}^2 \right) \label{eq:chi_4_equation} \\
    &=& \frac{1}{L^d} \sum_i \sum_k \langle \phi_i(t) \phi_{i+k}(t) \rangle_{\rm time},
\end{eqnarray}
where $\phi_i(t)=p_i(t)-\langle \pi(t)\rangle_{\rm time}$. For example, $\phi_i(\tau_\alpha)= \pm 1/2$.
Therefore, $\chi_4(t)$ is proportional to the average number of sites correlated dynamically.
Therefore, its peak value, $\chi_4^*$, contains essentially the same information as $\tilde S$, in particular,  $\chi_4^* \sim \tilde S_c$ at $T=0^+$.

\begin{figure}
\centering
\includegraphics[width=0.8\linewidth]{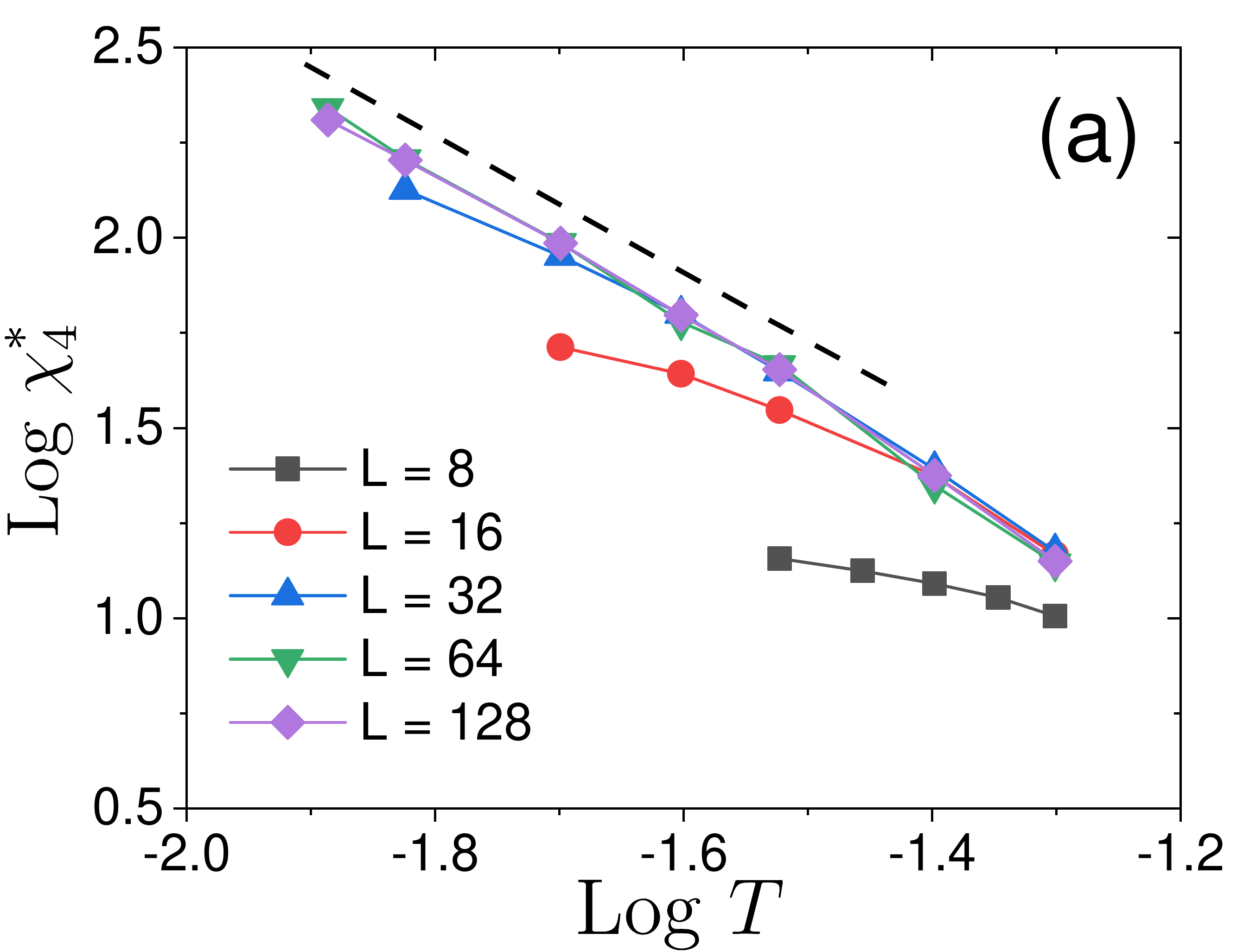}
\includegraphics[width=0.8\linewidth]{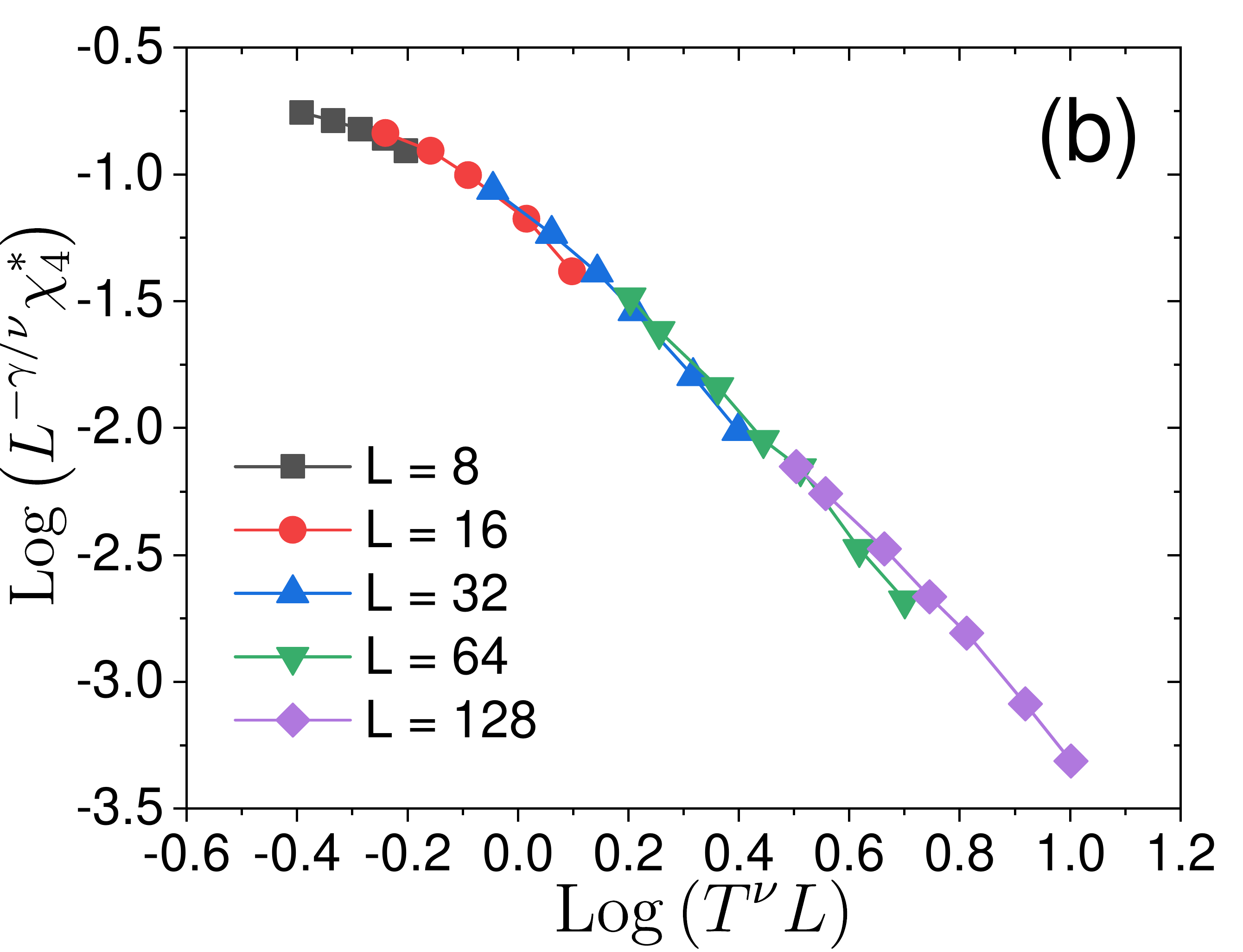}
\caption{The four-point correlation function $\chi_4$ for the scalar model. (a): The peak value $\chi_4^*$ versus $T$ for several $L$ in a Log-Log plot. The dashed line follows $\chi_4^* \sim T^{-\gamma}$. (d): Scaling collapse of $\chi_4^*(L,T)$, which determines $\nu$. 
}
\label{fig:scalar_finite_T}
\end{figure}

Figure~\ref{fig:scalar_finite_T}(a) shows $\chi_4^*$ versus $T$ for several system sizes $L$ for the scalar model.
One can see a scaling regime, $\chi_4^* \sim T^{-\gamma}$ at lower $T$ and larger $L$.
A scaling collapse is obtained in Fig.~\ref{fig:scalar_finite_T}(b), which determines another critical exponent $\nu$ associated with a lengthscale of dynamical heterogeneity.
The obtained values for the critical exponents are reported in Table~\ref{tab:critical_exponents}.

\subsection{Dynamical correlation lengthscale}

\begin{figure}
\centering
\includegraphics[width=0.49\linewidth]{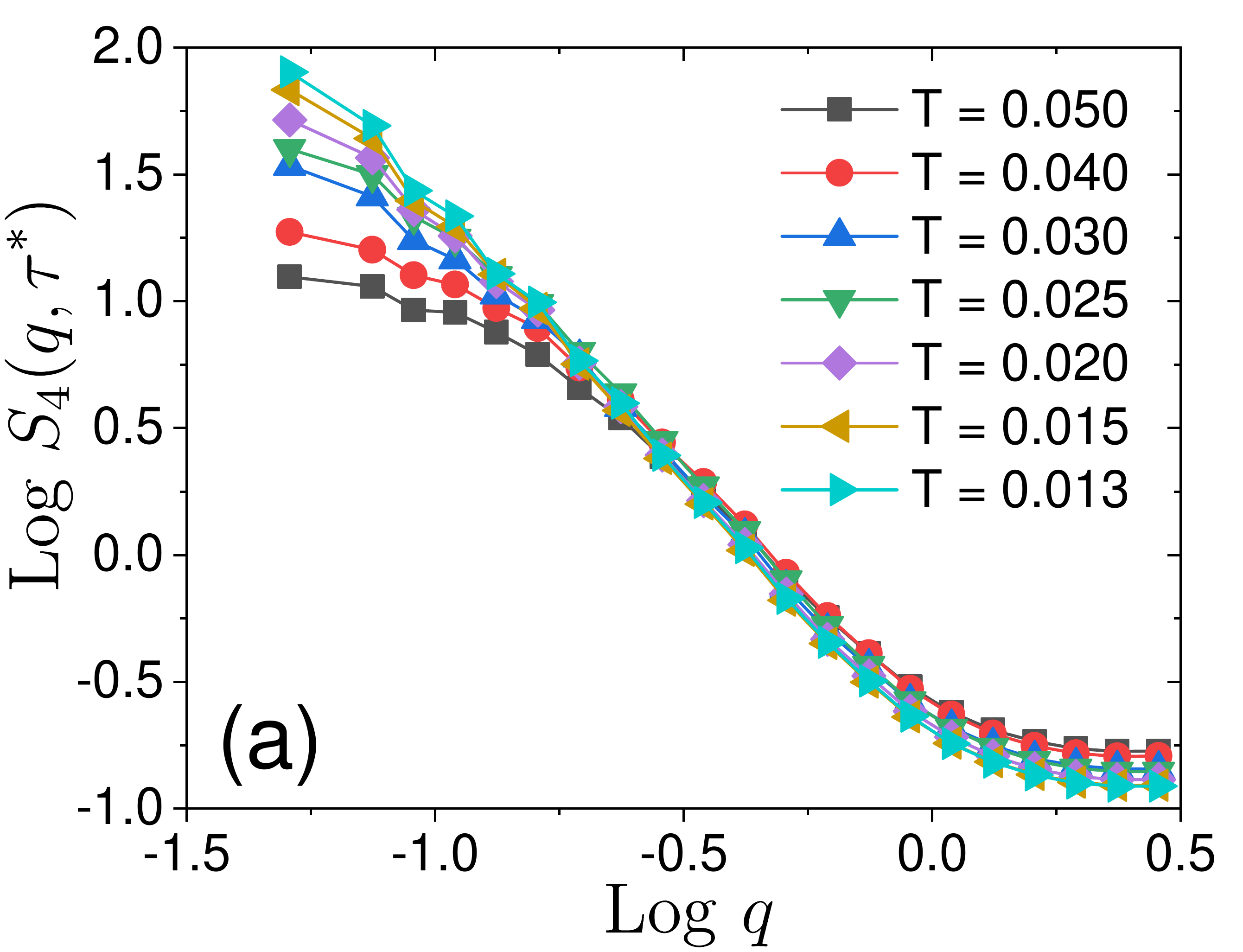}
\includegraphics[width=0.49\linewidth]{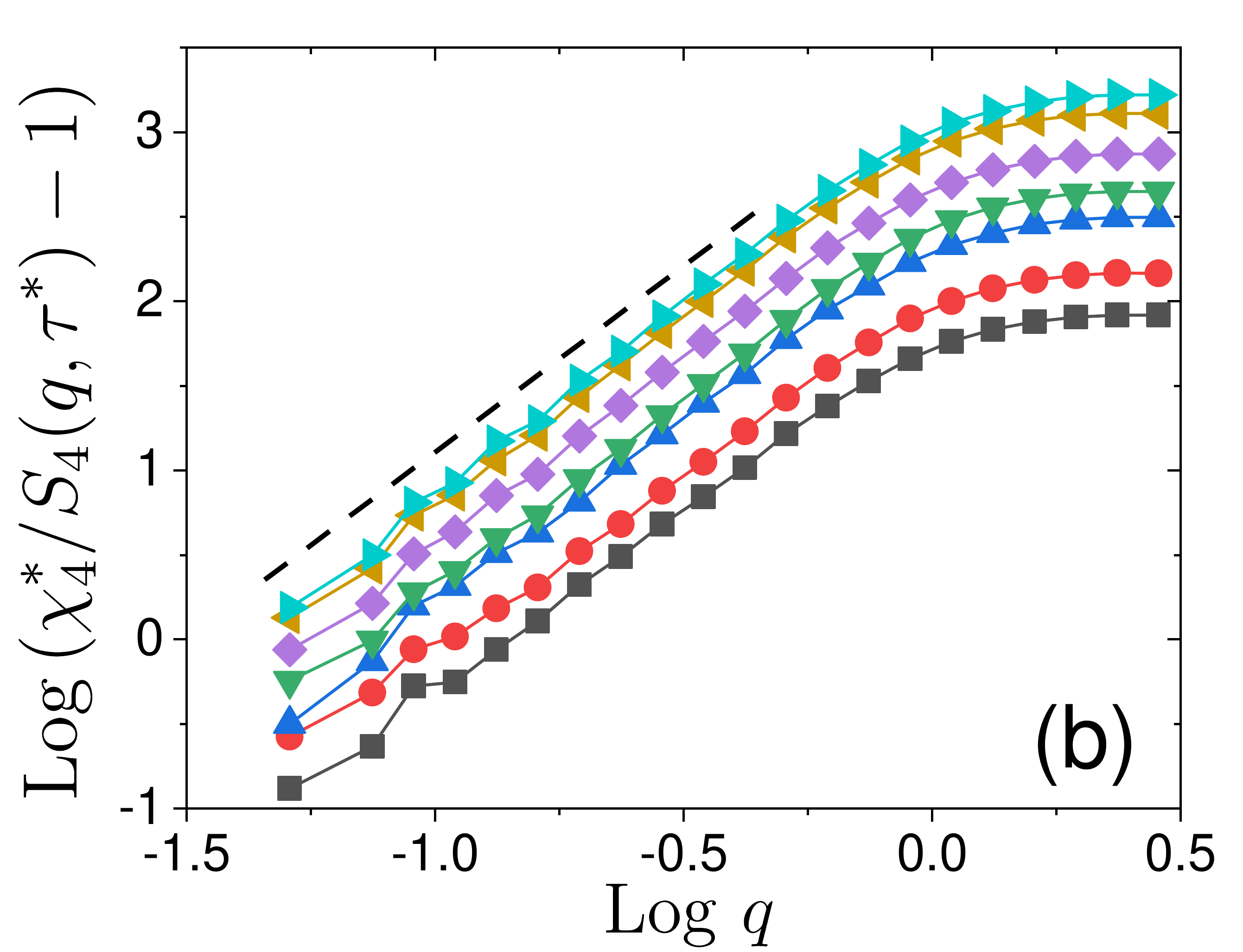}
\includegraphics[width=0.49\linewidth]{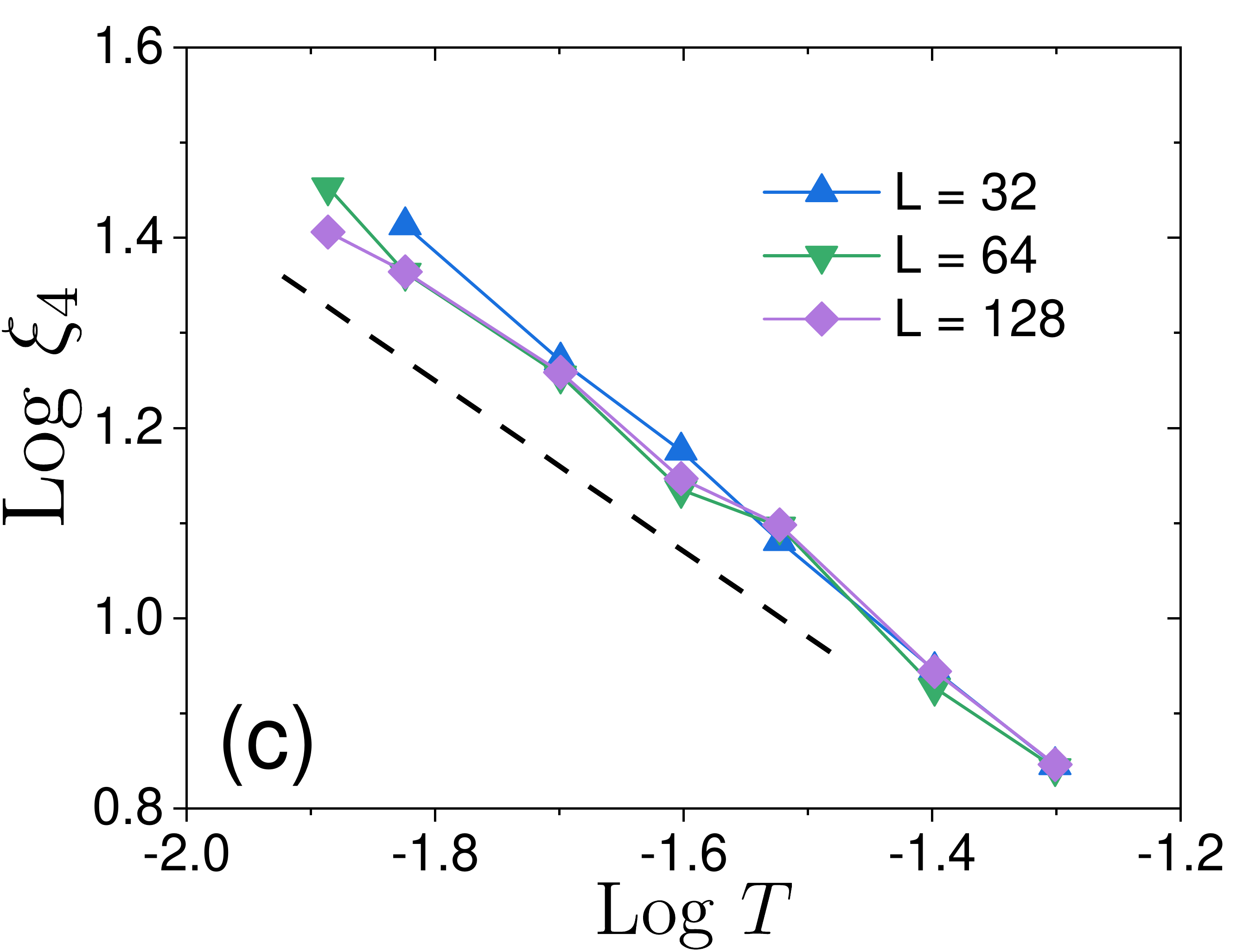}
\includegraphics[width=0.49\linewidth]{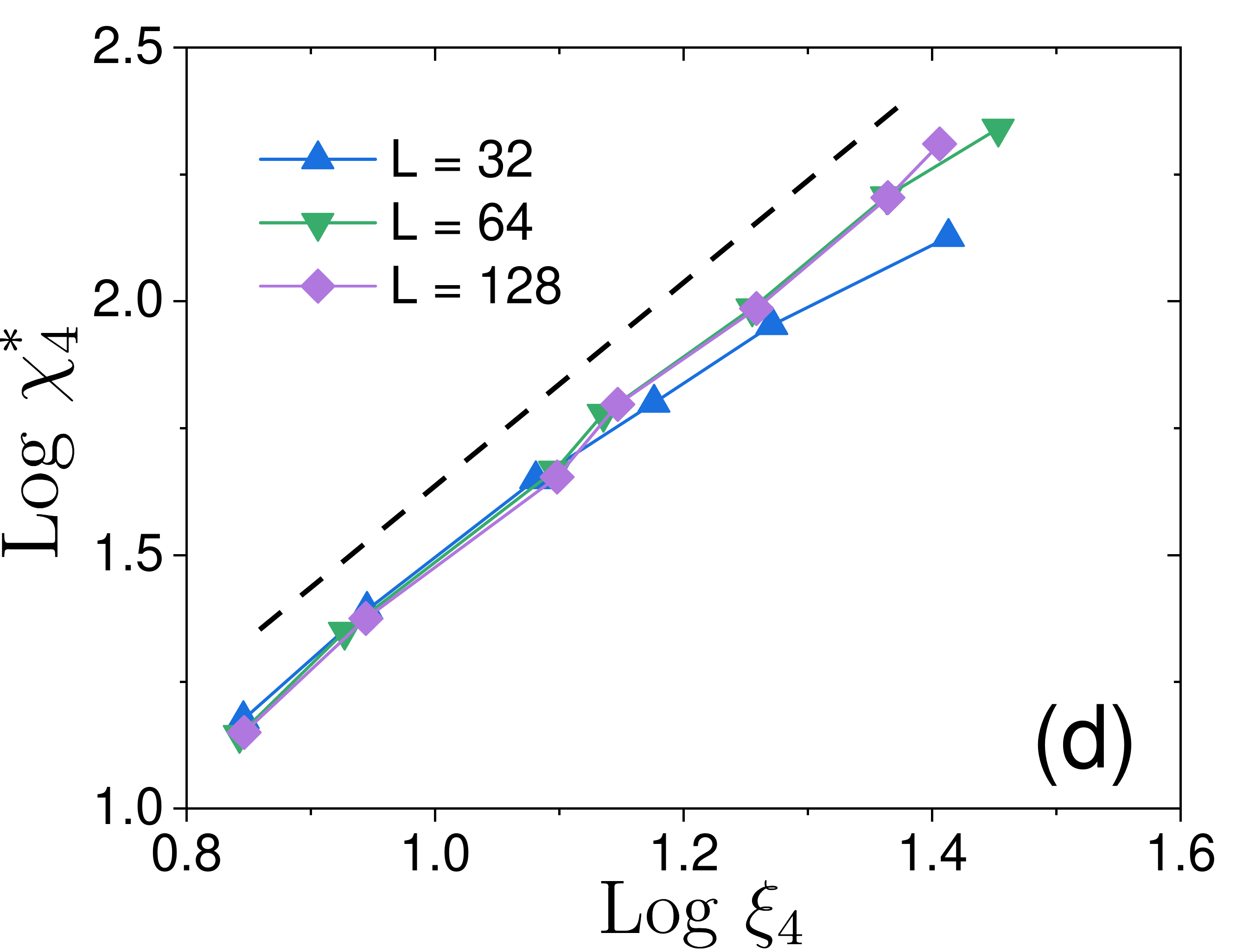}
\caption{The four-point structure factor $S_4(q, t)$ and the associated correlation length $\xi_4$ for the scalar model.
(a): $S_4(q, \tau^*)$ for several temperatures for $L=128$. (b): The corresponding plot for the Ornstein-Zernike
form in Eq.~(\ref{eq:OZ}). The dashed-line defines a slope corresponding to the exponent $a$.
(c): The extracted $\xi_4$ versus $T$. The dashed line follows $\xi_4 \sim T^{-\nu}$.
(b): $\chi_4^*$ versus $\xi_4$. The dashed line follows $\chi_4^* \sim \xi_4^{\tilde d_f}$.}
\label{fig:S4}
\end{figure}

We consider extracting a correlation length directly instead of performing finite-size scaling. 
To this end, we measure the spatial dependence of the four-point structure factor, $S_4(q, t)$~\cite{lavcevic2003spatially}, defined by
\begin{equation}
    S_4(q,t) = \frac{1}{L^d} \sum_{ij} \left( \langle p_i(t)p_j(t)\rangle_{\rm time} - \langle \pi(t) \rangle_{\rm time}^2 \right)e^{i{\bf q} \cdot ({\bf r}_i-{\bf r}_j)},
\end{equation}
where $q=|{\bf q}|$.
In Fig.~\ref{fig:S4}(a), we show $S_4(q,t)$ for the scalar model at $t=\tau^*$ when $\chi_4(t)$ takes the peak value, $\chi_4^*=\chi_4(\tau^*)$. We find that $\tau^*$ is close to $\tau_\alpha$, and thus $S_4(q,\tau^*)$ encodes heterogeneity associated with structural relaxation. 
We note that at the long wave-length limit, $S_4(q,t)$ converges to $\chi_4(t)$, namely, $\lim_{q \to 0} S_4(q,t) = \chi_4(t)$.
We then assume the Ornstein-Zernike
form at lower $q$, 
\begin{equation}
    S_4(q,\tau^*) = \frac{\chi_4^*}{1+(q\xi_4)^a},
    \label{eq:OZ}
\end{equation}
where $\xi_4$ is the dynamical correlation length extracted and $a$ is an exponet.
As shown in Fig.~\ref{fig:S4}(b), we find this scaling with $a=2.2$.
From this plot, we extract $\xi_4$ and present its temperature dependence in Fig.~\ref{fig:S4}(c).
This analysis can be done only for larger systems, $L=32$, $64$, and  $128$, since the scaling regime cannot be reached in smaller systems within our simulations.
We find $\xi_4 \sim T^{-\nu}$ with $\nu=0.9$, which is consistent with the one estimated from the finite size scaling in Fig.~\ref{fig:scalar_finite_T}. Moreover, the $\chi_4^*$ versus $\xi_4$ plot~\cite{flenner2014universal,kim2013multiple} in Fig.~\ref{fig:S4}(d) provides us with the fractal dimensions, $\tilde d_f=2$, which is also consistent with the one measured in the extremal dynamics in Fig.~\ref{fig:S_c_scalar}.

\subsection{Prediction for the four-point correlation function}

We connect the cutoff size for the site-based avalanche size, $\tilde S_c(t)$, in a given time interval $t$, and the time evolution of the four-point correlation function, $\chi_4(t)$.

We first compute the site-based avalanche size by $\tilde S(t) = \sum_{i=1}^{N} n_i(t)$, where
$n_i(t)=0$ if the site $i$ did not exhibit a plastic event until time $t$, and $n_i(t)=1$ for mobile sites that relaxed at least once. Thus, $n_i(t)$ can be written by $n_i(t)=1-p_i(t)$.
The first and second moments of $\tilde S(t)$ measured by the time average are given by
\begin{eqnarray}
    \langle \tilde S(t) \rangle_{\rm time} &=& N(1-\langle \pi(t) \rangle_{\rm time}), \nonumber \\
    \langle \tilde S^2(t) \rangle_{\rm time} &=& \sum_{i,j} \left\{ \langle p_i(t) p_j(t) \rangle_{\rm time} -(2\langle \pi(t) \rangle_{\rm time}-1) \right\}, \nonumber
\end{eqnarray}
respectively.
Consider a correlation volume of linear extension $\xi$. On this  length scale,  $\chi_4(t)$ crosses-over toward its value for an infinite system. It is also the largest length  for which extremal dynamics applies, implying that $\tilde S(t)$ is distributed in a power-law fashion. Thus, $\tilde S_c(t)$ can be estimated by $\langle \tilde S^2(t) \rangle_{\rm time}/\langle \tilde S(t) \rangle_{\rm time}$, as given by
\begin{equation}
    \tilde S_c(t) = \frac{\sum_{i,j} \left\{ \langle p_i(t) p_j(t) \rangle_{\rm time} -(2\langle \pi(t) \rangle_{\rm time}-1) \right\}}{N(1-\langle \pi(t) \rangle_{\rm time})}
.
\label{eq:S_c_explicit}
\end{equation}
In general, one can expect that
$\langle \pi(t) \rangle_{\rm time}$ follows the (stretched) exponential decay,
$\langle \pi(t) \rangle_{\rm time} \simeq e^{-(t/\tau_\alpha)^\beta}$, where $\beta$ is an exponent. Typically, $0<\beta \leq 1$ for equilibrium supercooled liquids.
Our elastoplastic models show nearly exponential relaxation with $\beta \simeq 1$. 
We now consider the early time stage, where $\langle \pi(t) \rangle_{\rm time}$ can be approximated by
$\langle \pi(t) \rangle_{\rm time} \simeq 1 -(t/\tau_\alpha)^\beta$.
Under such a circumstance, Eq.~(\ref{eq:S_c_explicit}) and $\chi_4(t)$ defined in
Eq.~(\ref{eq:chi_4_equation})  suggest that
$\tilde S_c(t) \simeq (t/\tau_\alpha)^{-\beta} \chi_4(t)$.
Together with Eq.~(\ref{eq:prediction_S_c}) in the main text, we predict the time evolution of $\chi_4(t)$ as
\begin{equation}
    \chi_4(t) \simeq A(t/\tau_\alpha)^{\beta} \left[ T \ln(\tau_\alpha/t) \right]^{-1/\tilde \sigma},
    \label{eq:chi4_prediction}
\end{equation}
where $A$ is a constant which does not depend on $t$ and $T$.

\begin{figure}
\centering
\includegraphics[width=0.49\linewidth]{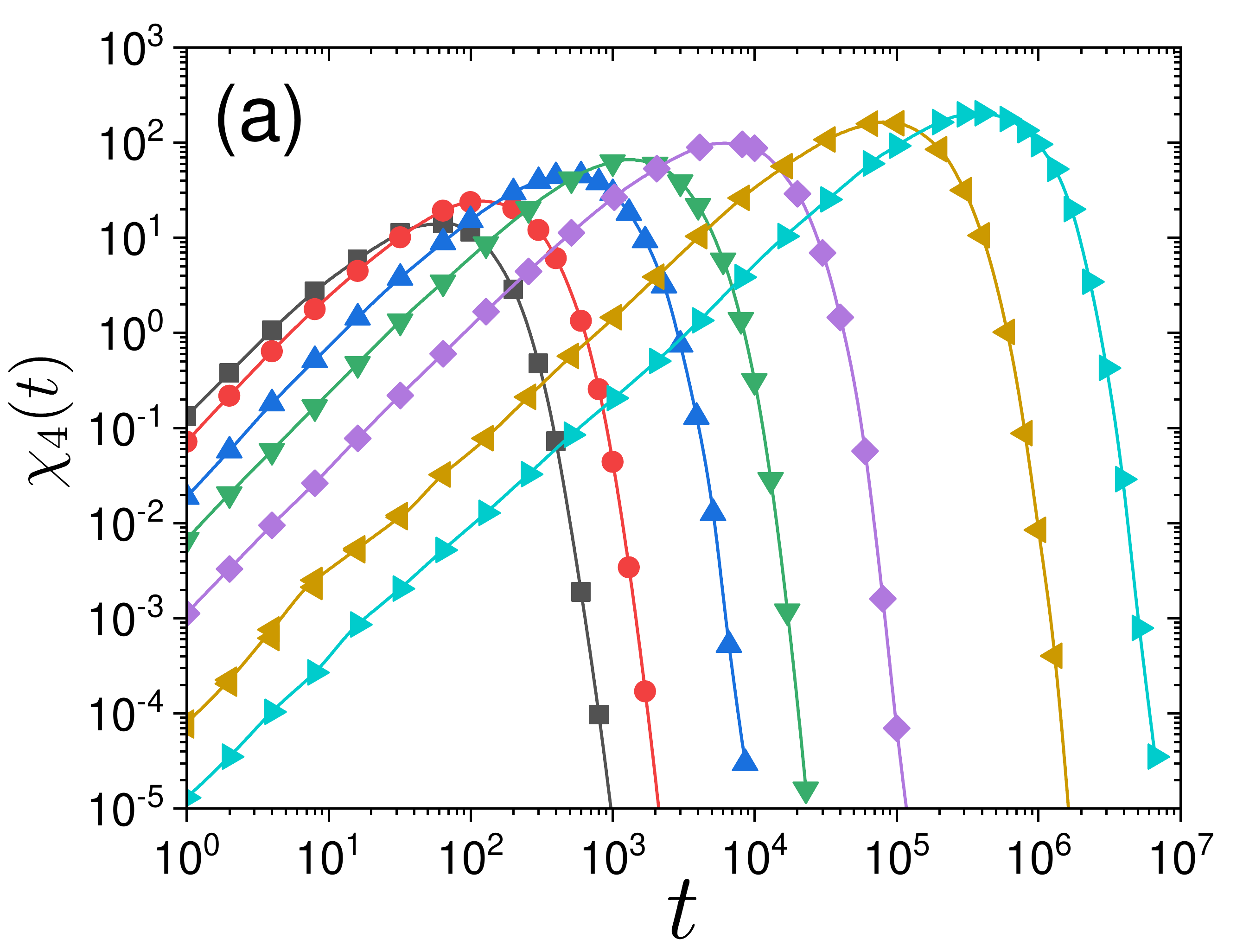}
\includegraphics[width=0.49\linewidth]{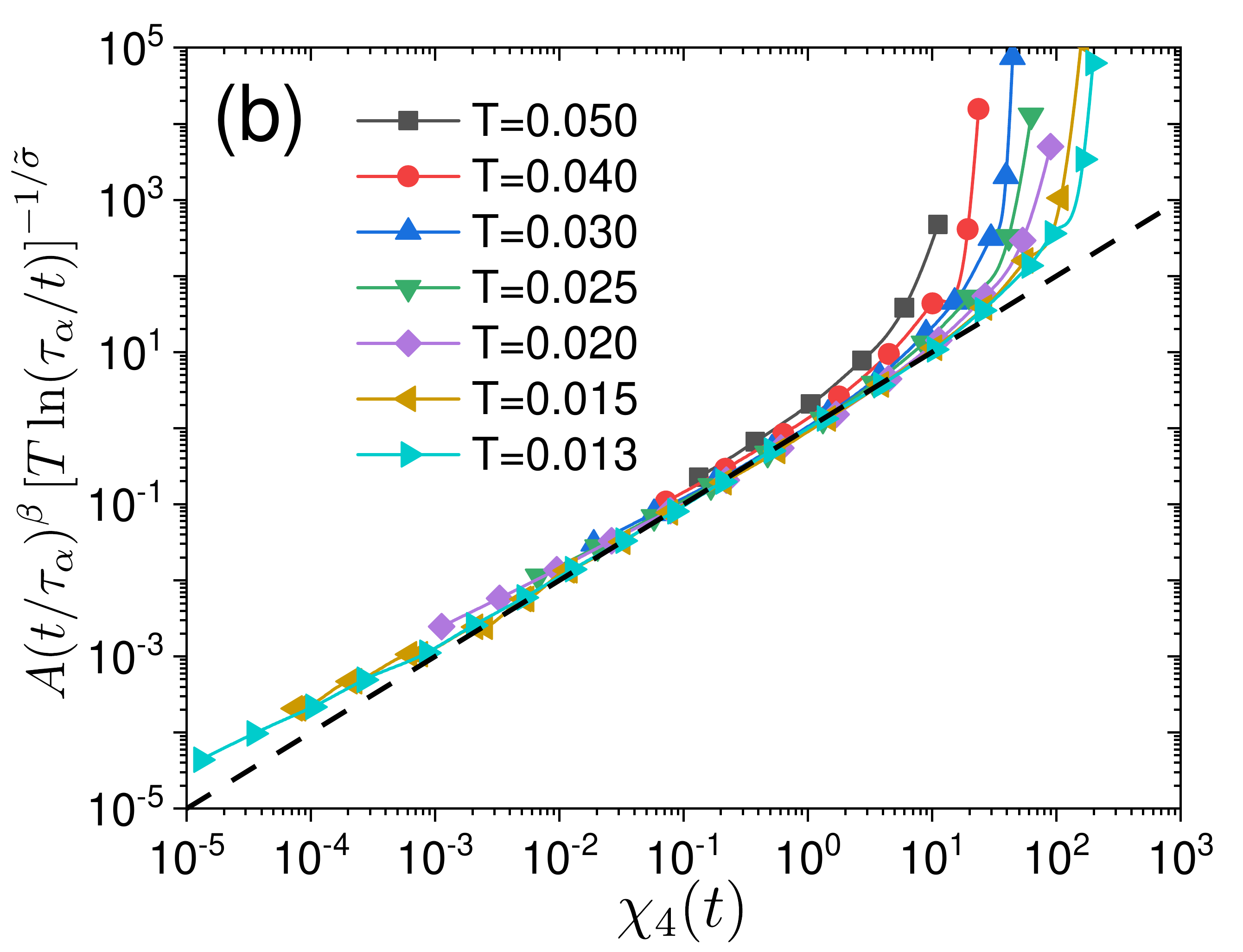}
\caption{(a): Log-Log plot for $\chi_4(t)$ for the scalar model with $L=128$ for $T=0.050, 0.040, 0.030, 0.025, 0.020, 0.015$, and $0.013$ (from left to right). (b): Parametric plot to test the prediction in Eq.~(\ref{eq:chi4_prediction}). The dashed line defines the linear relation.}
\label{fig:chi4_prediction}
\end{figure}

Figure~\ref{fig:chi4_prediction}(a) shows a Log-Log plot for $\chi_4(t)$ measured at finite temperature simulations for the scalar model.
The initial growth can be fitted effectively by a power-law, $\chi_4(t) \sim t^b$~\cite{biroli2022dynamical,flenner2016dynamic}, with $b \simeq 1.4$.
Instead, our argument in Eq.~(\ref{eq:chi4_prediction}) predicts a linear growth with a logarithmic correction.
In Fig.~\ref{fig:chi4_prediction}(b), we show a parametric plot to numerically test Eq.~(\ref{eq:chi4_prediction}) with  $\beta=1$, $A=0.5$, and $1/\tilde \sigma =1.9$ (see Table~\ref{tab:critical_exponents}). 
We find that the simulated $\chi_4(t)$ for different temperatures follows our prediction at  early times. 
Deviations from the prediction can be observed on very short timescales. This is presumably due to the fact that on such time scales, the corresponding energy scale is too small compared with $E_c$, which violates the assumption underlying the asymptotic argument of Sec.~\ref{sec:scaling_argument}.

\subsection{Tracer particles}

\begin{figure}
\centering
\includegraphics[width=\linewidth]{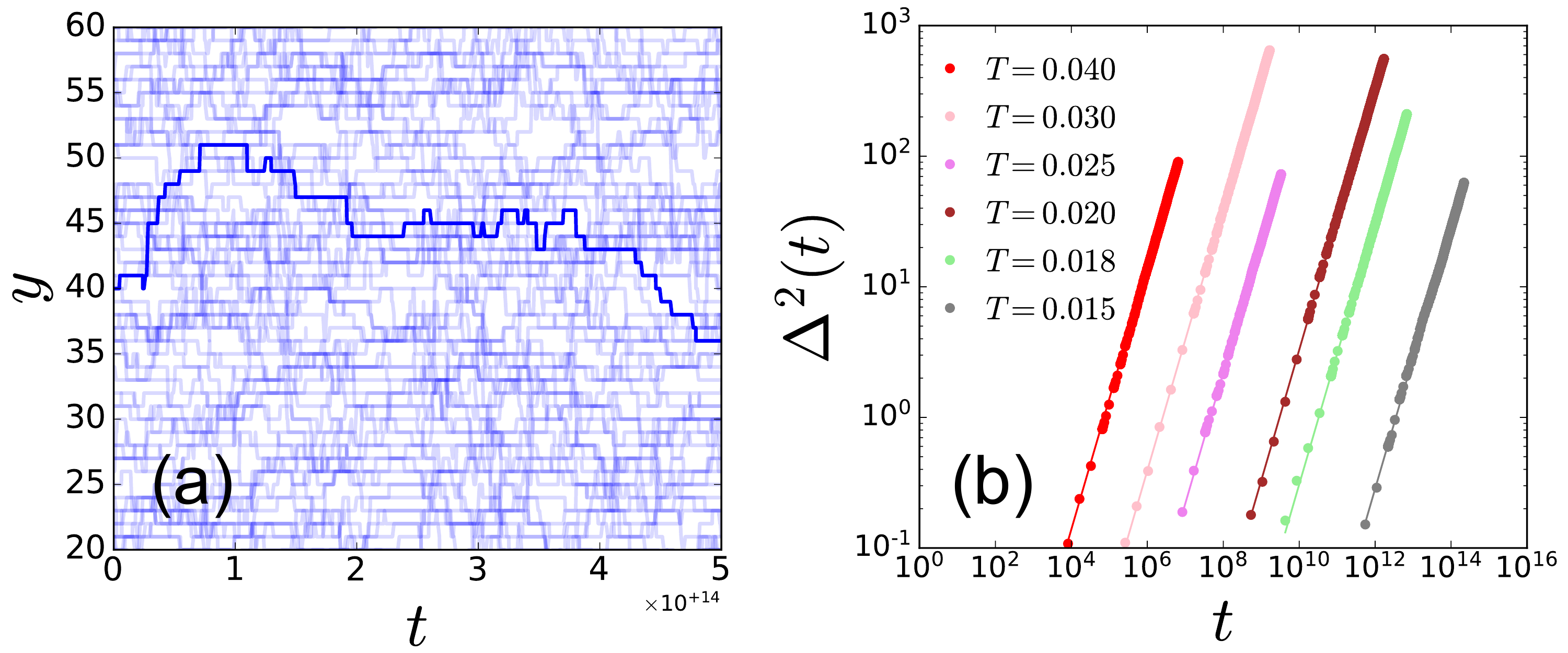}
\caption{Diffusion of tracer particles for the tensorial model with $L=128$. (a): The $y$ component of typical trajectory of a tracer particle with $T=0.015$ ($\tau_\alpha = 4.14 \times 10^{12}$).
(b): Mean-squared displacement $\Delta^2(t)$ of tracer particles are shown with points. The solid lines follow $\Delta^2(t) = D t$, from which we extract the diffusion coefficient $D$. 
}
\label{fig:tracer_particles}
\end{figure}

We monitor the diffusion of tracer particles \cite{jung2004excitation,berthier2004length} due to the local relaxations.
We consider one tracer particle in each site of the elastoplastic model; each tracer particle moves randomly to one of the four nearest neighbors (in $d=2$) after a plastic event in that site.
The trajectory of the $k$-th tracer particle is specified by  $(x_k(t),y_k(t))$.
Typical trajectories of the tracer particles are shown in Fig.~\ref{fig:tracer_particles}(a) as a function of time.
In a time scale comparable to $\tau_\alpha$, the tracer travels over multiple sites in a very short time and spends most of the time without any activity.
We then define the mean-squared displacement $\Delta^2(t)$ of the tracers by
\begin{equation}
    \Delta^2(t) = \frac{1}{N_{\rm t}} \sum_{k=1}^{N_{\rm t}} \langle \Delta r_k^2(t) \rangle_{\rm time} ,
\end{equation}
where $\Delta r_k^2(t)=(x_k(t)-x_k(0))^2 + (y_k(t)-y_k(0))^2$ and $N_{\rm t}$ is the number of tracer particles.
In Fig.~\ref{fig:tracer_particles}(b) we show $\Delta^2(t)$ for different temperatures.
We find a diffusive behavior at a larger time, $\Delta^2(t) = Dt$, from which we extract the diffusion coefficient $D$ for each temperature.

\section{Avalanches at $T=0^+$}
\label{sec:Appendix_extremal}

\subsection{Extremal dynamics}

We explain the extremal dynamics at $T=0^+$.
In the finite temperature simulations described in Sec.~\ref{sec:appendix_models}, we take into account local thermal activation for a plastic event based on the probability $e^{-E(x)/T}$, where $E(x)=x^{3/2}$ at $0 \leq x \leq1$.
At vanishing temperature, $T=0^+$, this probability is extremely small. Therefore, the site with the smallest $x$, denoted as $x_{\rm min}$, associated with the lowest energy barrier $E_{\rm min}=x_{\rm min}^{3/2}$, shows the next plastic event. 
Thus, in practice, one can choose the weakest site having $x_{\rm min}$ sequentially, instead of asking $e^{- E(x)/T}$ each time and waiting until it shows an event. 
This algorithm enormously accelerates dynamics and allows us to access information about plastic activities even at $T=0^+$. 
This is the so-called {\it extremal dynamics}~\cite{paczuski1996avalanche,Baret2002,purrello2017creep}.

Finding one $x_{\rm min}$ corresponds to one simulation step. This is not directly related to physical time (that is why one can simulate it even at $T=0^+$), yet one can associate the simulation step with the size of an avalanche (see below).  
Simulations start with the same initial condition used in the finite temperature simulations. 
The system enters the stationary state after passing the initial transient regime. We carefully checked the stationarity by monitoring the waiting time dependence of $P(x)$. We report data taken only from the stationary state. 

\begin{figure}
\centering
\includegraphics[width=0.95\linewidth]{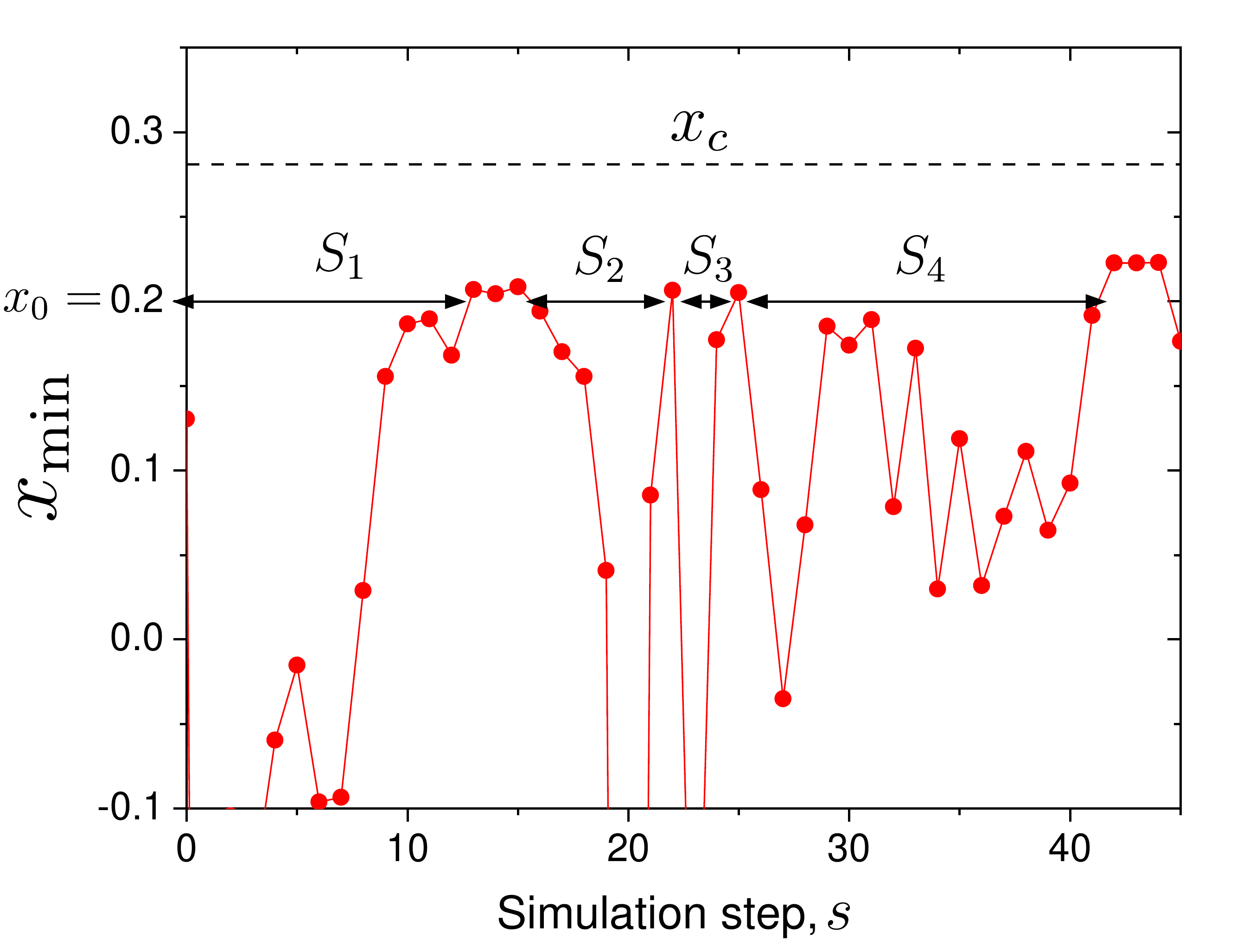}
\caption{
An example of the evolution of $x_{\rm min}$ during the extremal dynamics at a steady state for the scalar model. The system size is $L=256$.
A series of event-based avalanche sizes, $S_1, S_2, ...$, are presented based on the threshold value $x_0=0.2$.
$x_c=0.281$ is indicated by the horizontal dashed line.
}
\label{fig:x_min_evolution}
\end{figure}

In Fig.~\ref{fig:x_min_evolution}, we show a representative trajectory of $x_{\rm min}$ during an extremal dynamics simulation at the stationary state, which is an analog of FIG. 5 in Ref.~\cite{paczuski1996avalanche} for a model for self-organized criticality.
Typically, the weakest site with $x_{\rm min}(s)$ at a simulation step $s$ induces the next weakest site at step $s+1$ with $x_{\rm min}(s+1)$ at a neighbor region because of elastic interactions. In particular, $x_{\rm min}(s) > x_{\rm min}(s+1)$, when the previous weakest site at step $s$ destabilizes the next weakest site at step $s+1$. Therefore, a sequence of the weakest sites is dynamically correlated, forming an avalanche until the last weakest site is found at an uncorrelated place with a higher value of $x_{\rm min}$. The determination of uncorrelation and hence the termination of an avalanche has some ambiguity. Thus, following previous works, we introduce the threshold $x_0$ below which a sequence of $x_{\rm min}$ is correlated.
In particular, we define the size of {\it event-based} avalanche, $S$, by the number of chosen $x_{\rm min}$ forming a sequence with $x_{\rm min}<x_0$ (in other words, the duration of simulation steps with $x_{\rm min}<x_0$). As shown in Fig.~\ref{fig:x_min_evolution}, one can extract a series of event-based avalanche sizes, $S_1, S_2,...$, from the trajectory of the extremal dynamics simulation.
A given site may be chosen as the weakest site several times during one avalanche formation, which all contribute to $S$. 
Instead, one can define the size of {\it site-based} avalanche, $\tilde S$, by the number of sites participating in a single avalanche. By construction, $\tilde S \leq S$. The distinction between $S$ and $\tilde S$ provides us with important physical information about the accumulation of multiple relaxation activities, which leads to an argument about the Stokes-Einstein violation, as discussed in Sec.~\ref{sec:scaling_argument}.

By construction, $S$ and $\tilde S$ depend on the threshold value $x_0$. As $x_0$ is increased, the size of avalanches, $S$ and $\tilde S$, increases.
As we will discuss further below, avalanches become a system spanning when $x_0 \to x_c$, where $x_c$ is the critical value associated with the critical energy gap $E_c=x_c^{3/2}$. We will vary $x_0$ systematically to probe the critical behavior associated with $x_c$ (see below).

\subsection{Avalanche statistics}

\begin{figure}
\centering
\includegraphics[width=0.48\linewidth]{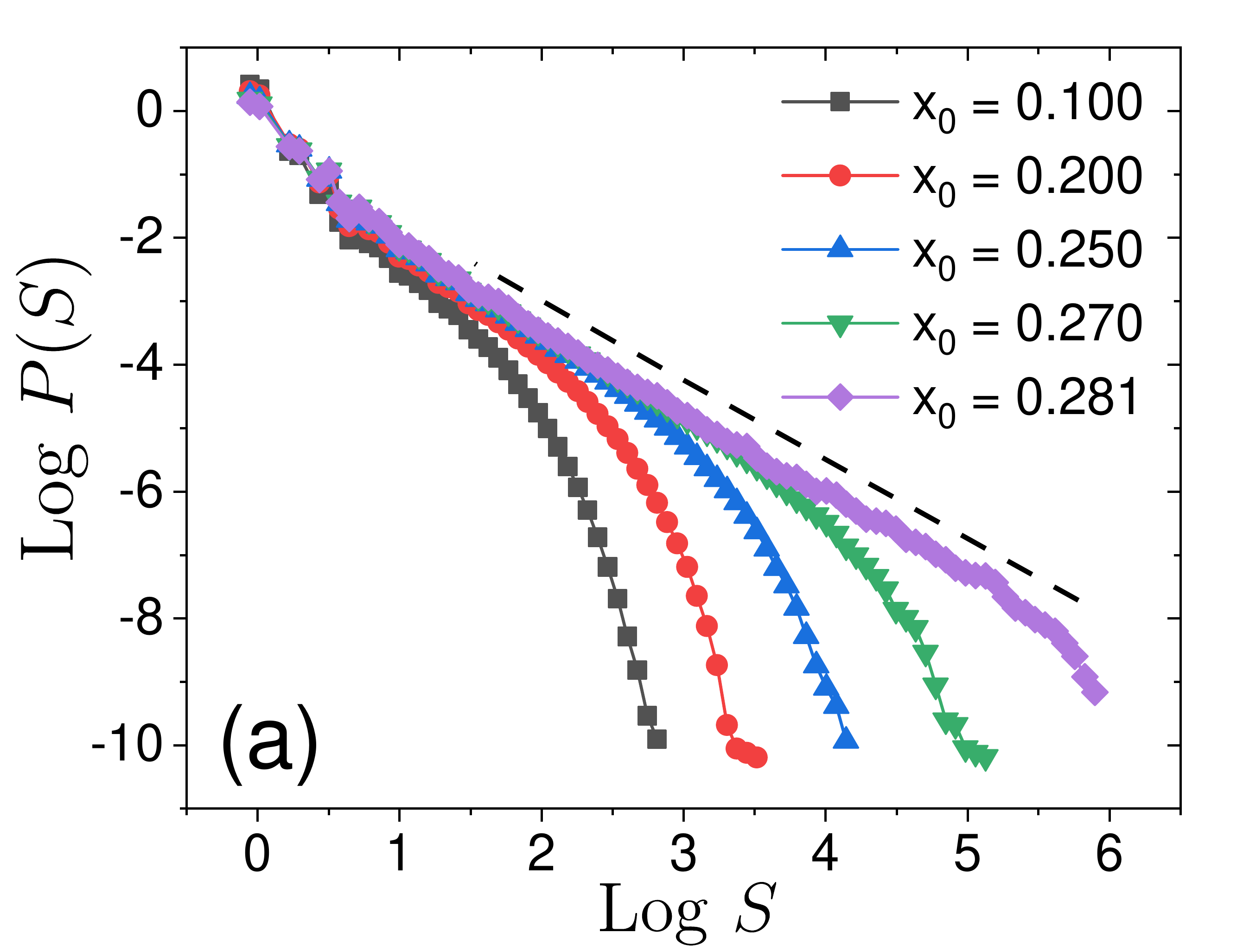}
\includegraphics[width=0.48\linewidth]{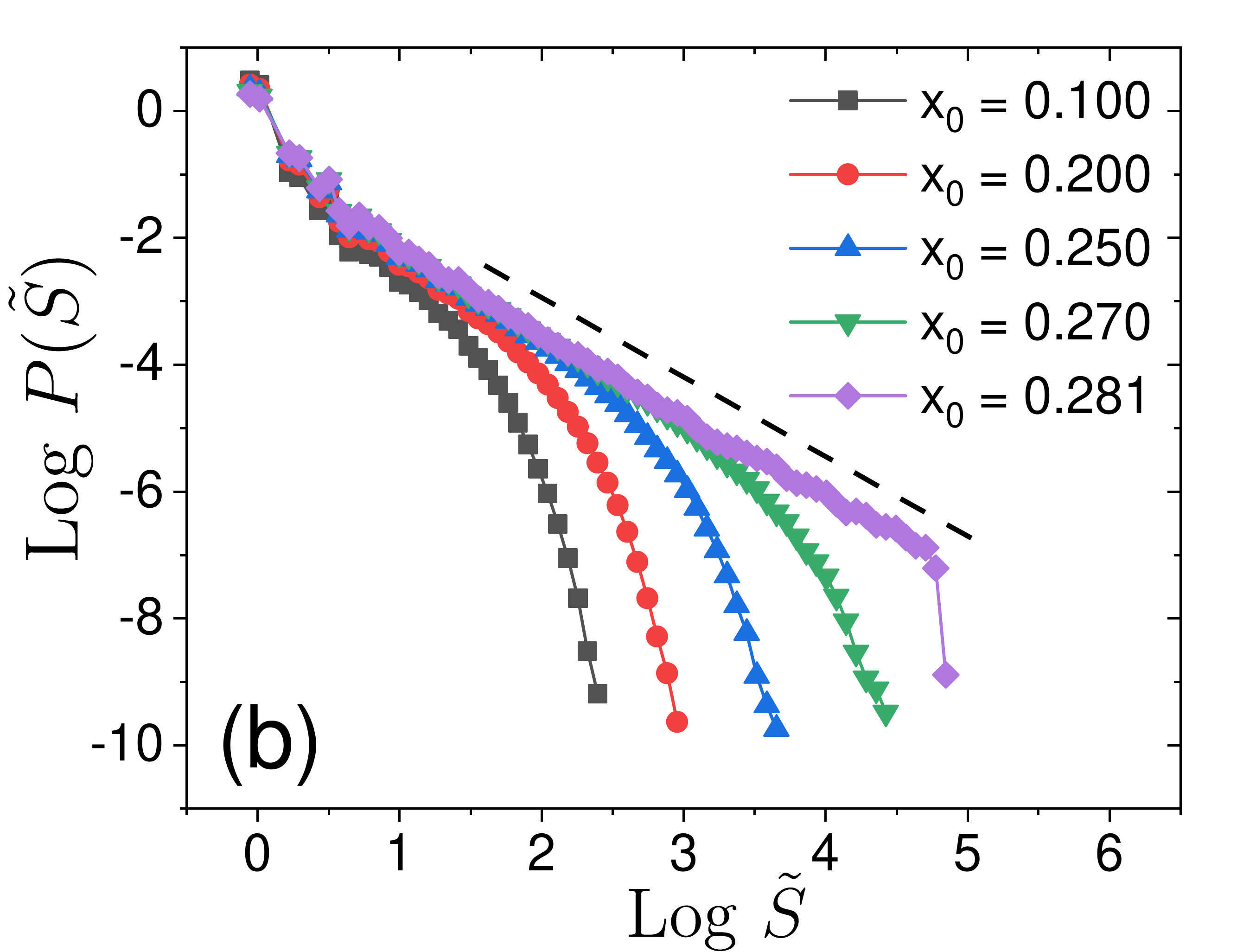}
\caption{
Avalanche distributions $P(S)$ (a) and $P(\tilde S)$ (b) for several $x_0$ approaching the critical point $x_c=x_0=0.281$ for the scalar model with $L=256$. The dashed lines in (a) and (b) follow $P(S) \sim S^{-\tau}$ and  $P(\tilde S) \sim S^{-\tilde \tau}$, respectively.
}
\label{fig:P_of_S_x0}
\end{figure}

We describe how to analyze the avalanche data obtained during $T=0^+$ extremal dynamics simulations.
During simulations, we record the series of the event and site-based avalanche sizes, given by $\{ S_1, S_2, ..., S_M\}$ and $\{ \tilde S_1, \tilde S_2, ..., \tilde S_M\}$, respectively, where $M$ is the number of data points.
In this paper, we analyze $S$ and $\tilde S$ in parallel. Below we will explain how to analyze the data using $S$, but the same procedures are applied for $\tilde S$.
We first define the $m$-th moments of avalanche distribution ($m=1, 2, ...$) by
\begin{equation}
    \langle S^m \rangle = \int_0^{\infty} \mathrm{d} S P(S) S^m = \frac{1}{M} \sum_{k=1}^M S_k^m,
\end{equation}
where $P(S)$ is the distribution of avalanches.
One expects that $P(S)$ follows a power law distribution,
\begin{equation}
    P(S) \sim S^{-\tau} g(S/S_c),
    \label{eq:power_law}
\end{equation}
where $\tau$ is a critical exponent, $S_c$ is a cutoff size, and $g(z)$ is a scaling function.
Figure~\ref{fig:P_of_S_x0} shows $P(S)$ and $P(\tilde S)$ for several $x_0$ for the scalar model. 
These plots demonstrate that the size of avalanches (both $S$ and $\tilde S$) grow with increasing $x_0$, as expected. In particular, a scale-free, power-law behavior (with the eventual cut-off) is being developed by approaching the critical point $x_c$, which proves that $x_0$ is the relevant parameter that dictates the critical behavior of the system.

\begin{figure}
\centering
\includegraphics[width=0.48\linewidth]{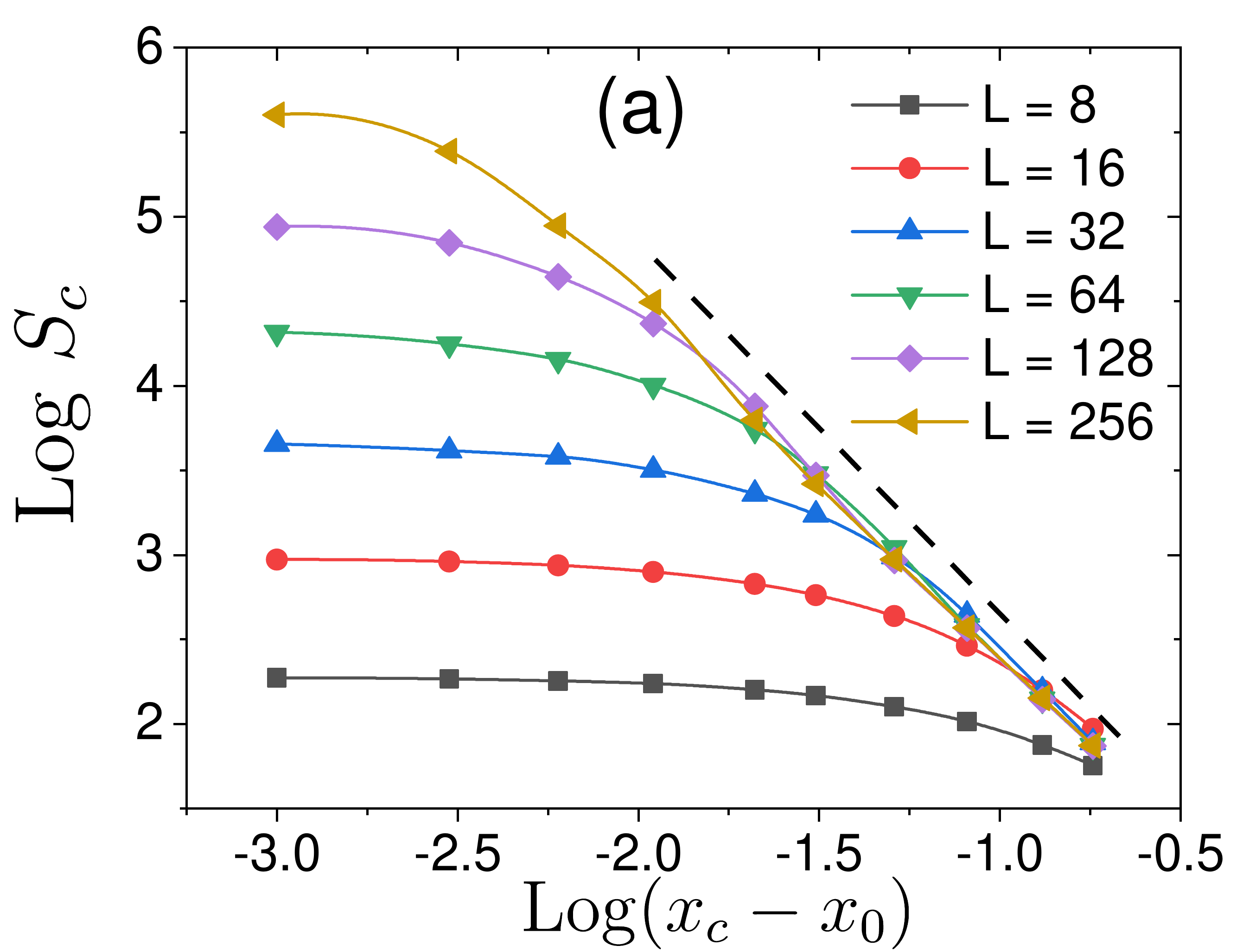}
\includegraphics[width=0.48\linewidth]{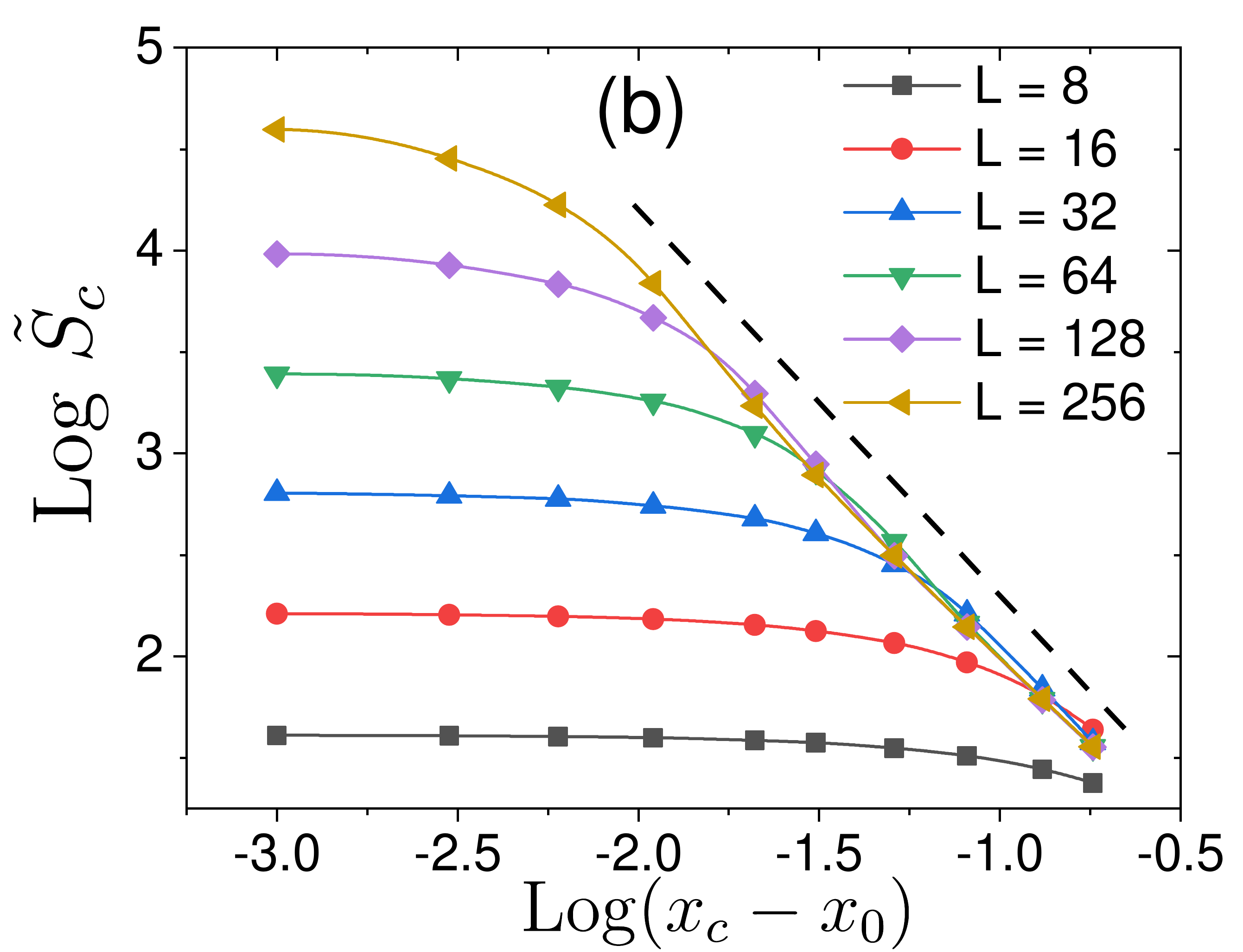}
\includegraphics[width=0.48\linewidth]{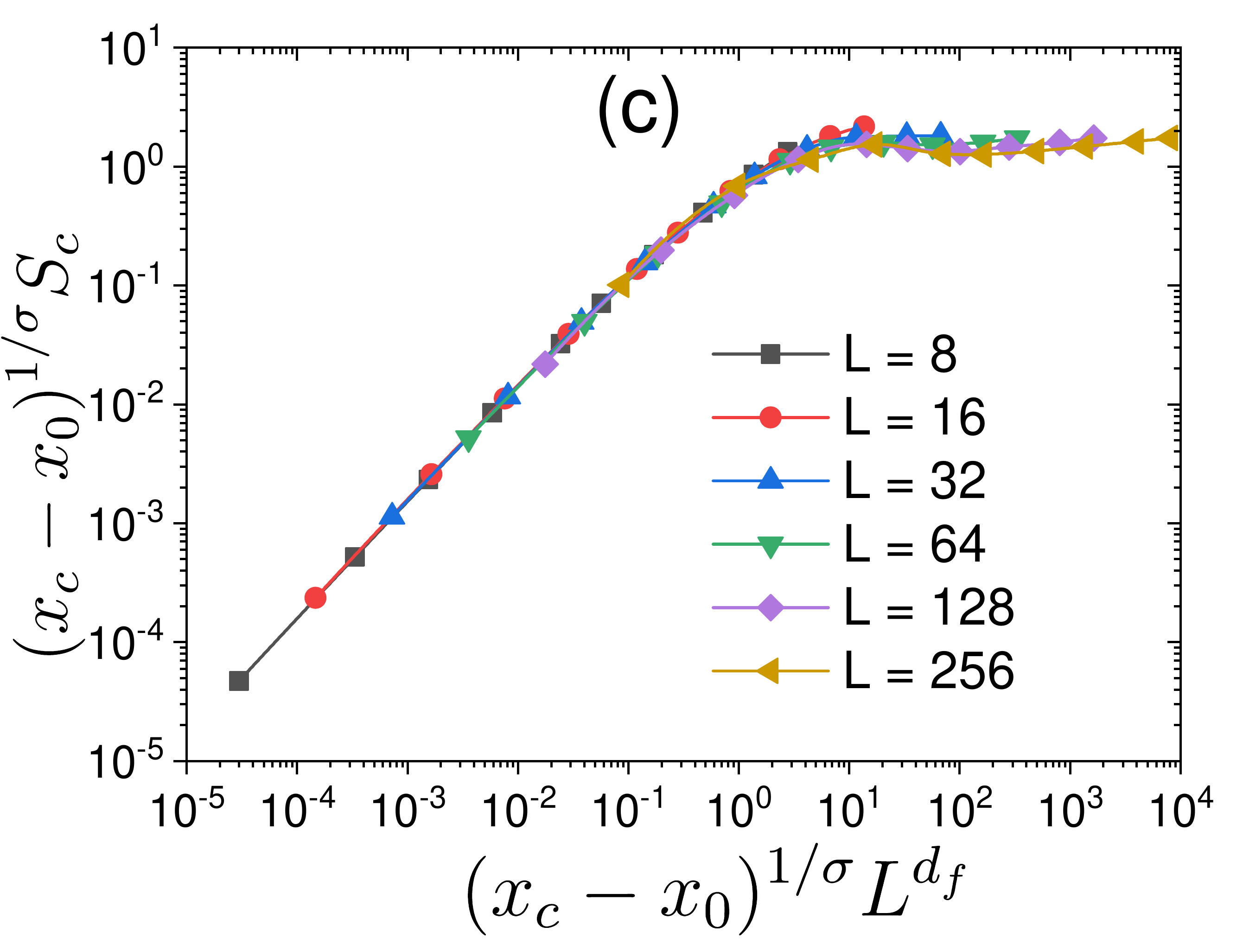}
\includegraphics[width=0.48\linewidth]{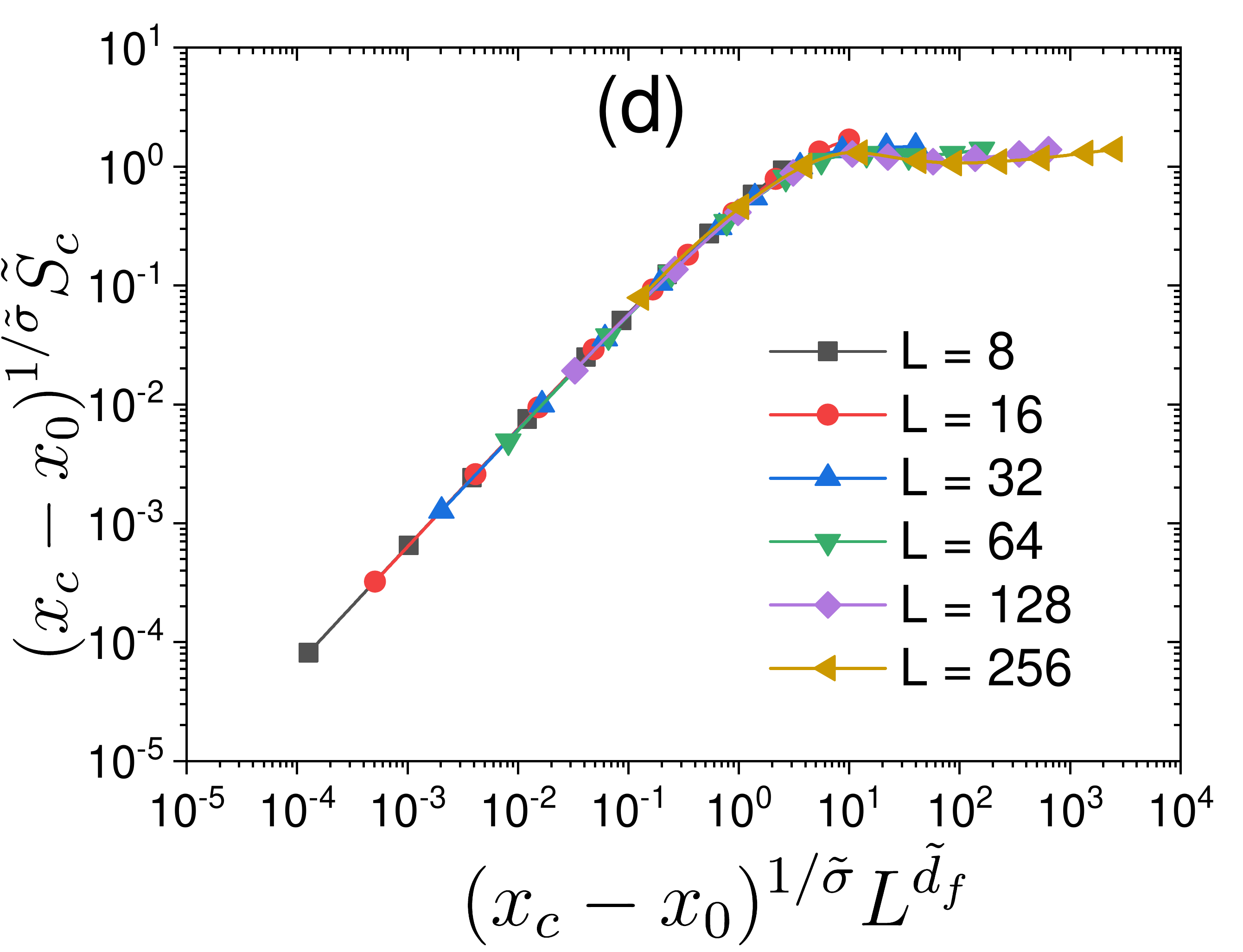}
\caption{Cut-off in avalanche distributions for the scalar model. (a, b): $S_c=\langle S^3 \rangle/\langle S^2 \rangle$ (a) and $\tilde S_c = \langle \tilde S^3 \rangle/\langle \tilde S^2 \rangle$ (b) versus $x_c-x_0$ for various system sizes. The dashed lines in (a) and (b) follow $S_c \sim (x_c-x_0)^{-1/\sigma}$ and $\tilde S_c \sim (x_c-x_0)^{-1/\tilde  \sigma}$, respectively.
(c,d): Scaling collapse assuming Eq.~(\ref{eq:S_cut_x0}).
The critical point $x_c$, exponents $\sigma$, $\tilde \sigma$, $d_f$, and $\tilde d_f$ are determined by these scaling plots.}
\label{fig:S_c_scalar}
\end{figure}

Assuming Eq.~(\ref{eq:power_law}) and $1<\tau<2$, one obtains
\begin{equation}
    \langle S^m \rangle \sim \int_0^{\infty} \mathrm{d}S S^{m-\tau} g(S/S_c) \sim S_c^{m+1-\tau},
\end{equation}
which implies $S_c \sim \langle S^{m+1} \rangle/\langle S^m \rangle$. Thus, in practice, we {\it define} $S_c$ by $S_c = \langle S^3 \rangle/\langle S^2 \rangle$.
Following Ref.~\cite{han2018critical}, we assume
\begin{equation}
    S_c \sim (x_c-x_0)^{-1/\sigma}f\left( \frac{L^{d_f}}{(x_c-x_0)^{-1/\sigma}}\right),
    \label{eq:S_cut_x0}
\end{equation}
where $f(z) = 1$ for $z \gg 1$ and $f(z)=z$ for $z \to 0$.
Thus, $S_c \sim L^{d_f}$ when $x_0 \to x_c$ and $S_c \sim (x_c-x_0)^{-1/\sigma}$ when $L \to \infty$.
Figures~\ref{fig:S_c_scalar}(a,b) show $S_c$ and $\tilde S_c$ approaching $x_0 \to x_c$ for the scalar model.
Both $S_c$ and $\tilde S_c$ increase with increasing $x_0$ with an eventual saturation due to a finite-size effect. One can see the expectated behavior, $S_c \sim (x_c-x_0)^{-1/\sigma}$ at larger $L$ ($\tilde S_c$ as well). 
We then perform the scaling collapse in Figs.~\ref{fig:S_c_scalar}(c,d) following the scaling form in Eq.~(\ref{eq:S_cut_x0}).
These scaling plots determine the critical point $x_c=0.281$ and the critical exponents, $1/\sigma$, $1/\tilde \sigma$, $d_f$, and $\tilde d_f$, for the scalar model.
The obtained values are reported in Table~\ref{tab:critical_exponents}.



\begin{figure}
\centering
\includegraphics[width=0.49\linewidth]{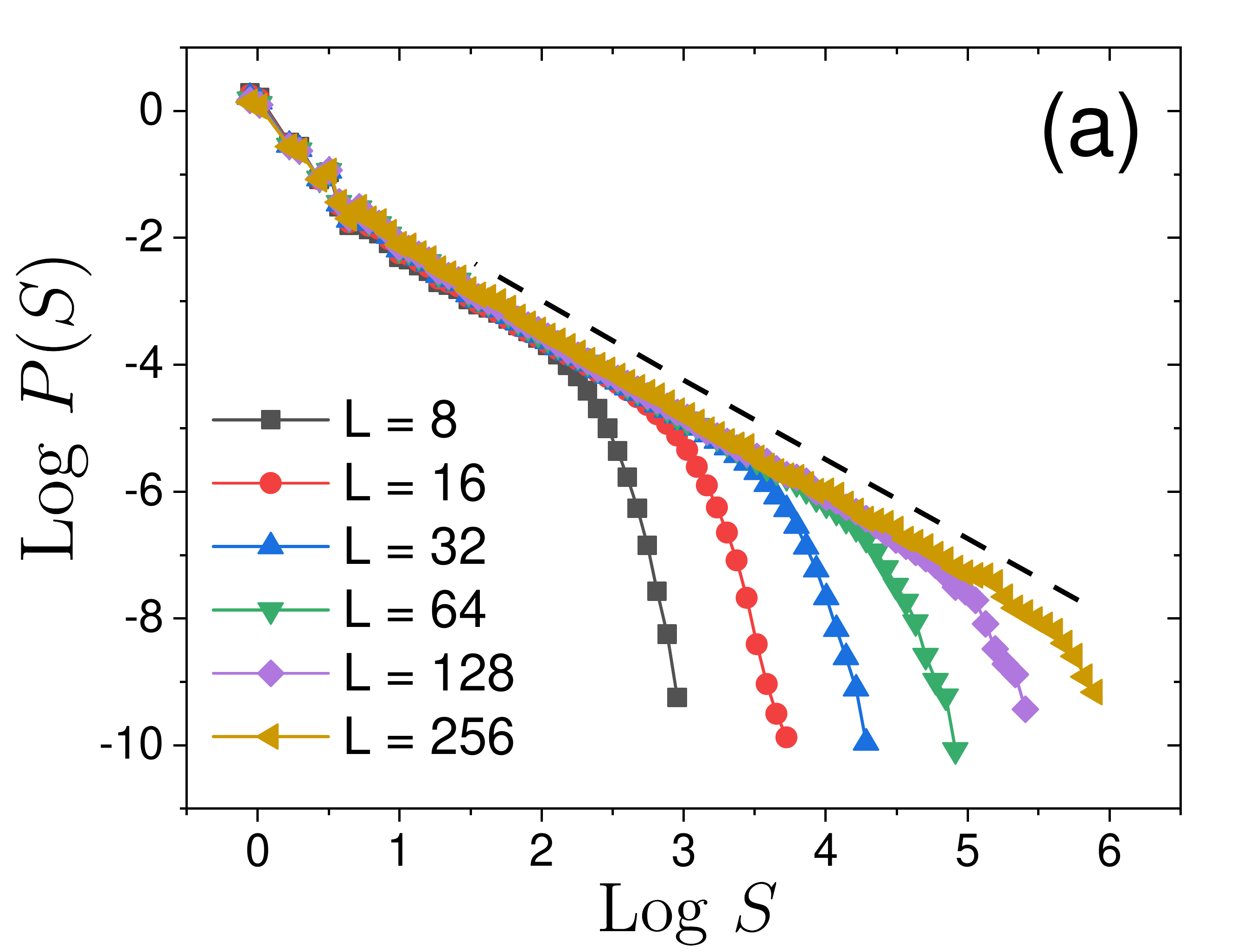}
\includegraphics[width=0.49\linewidth]{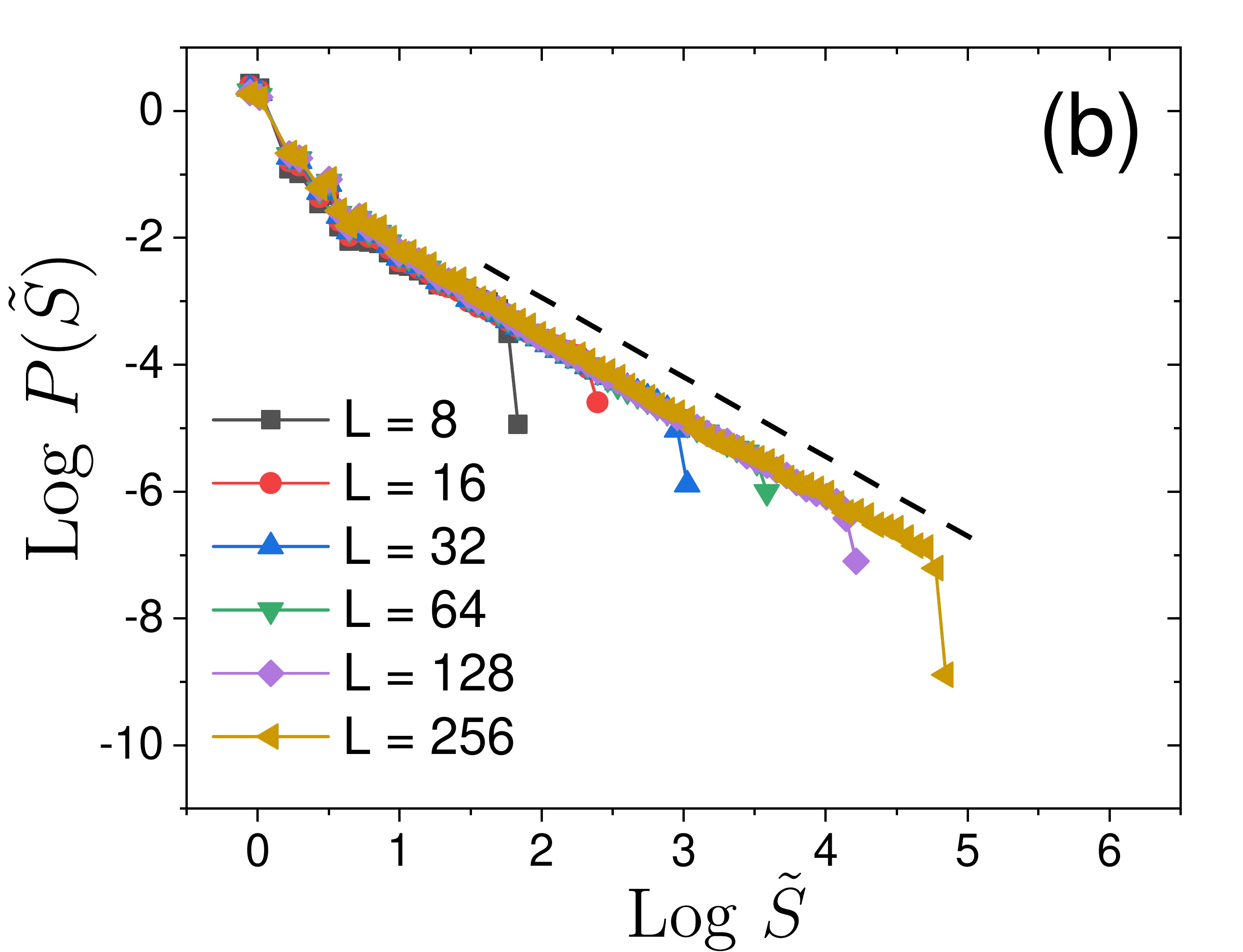}
\includegraphics[width=0.49\linewidth]{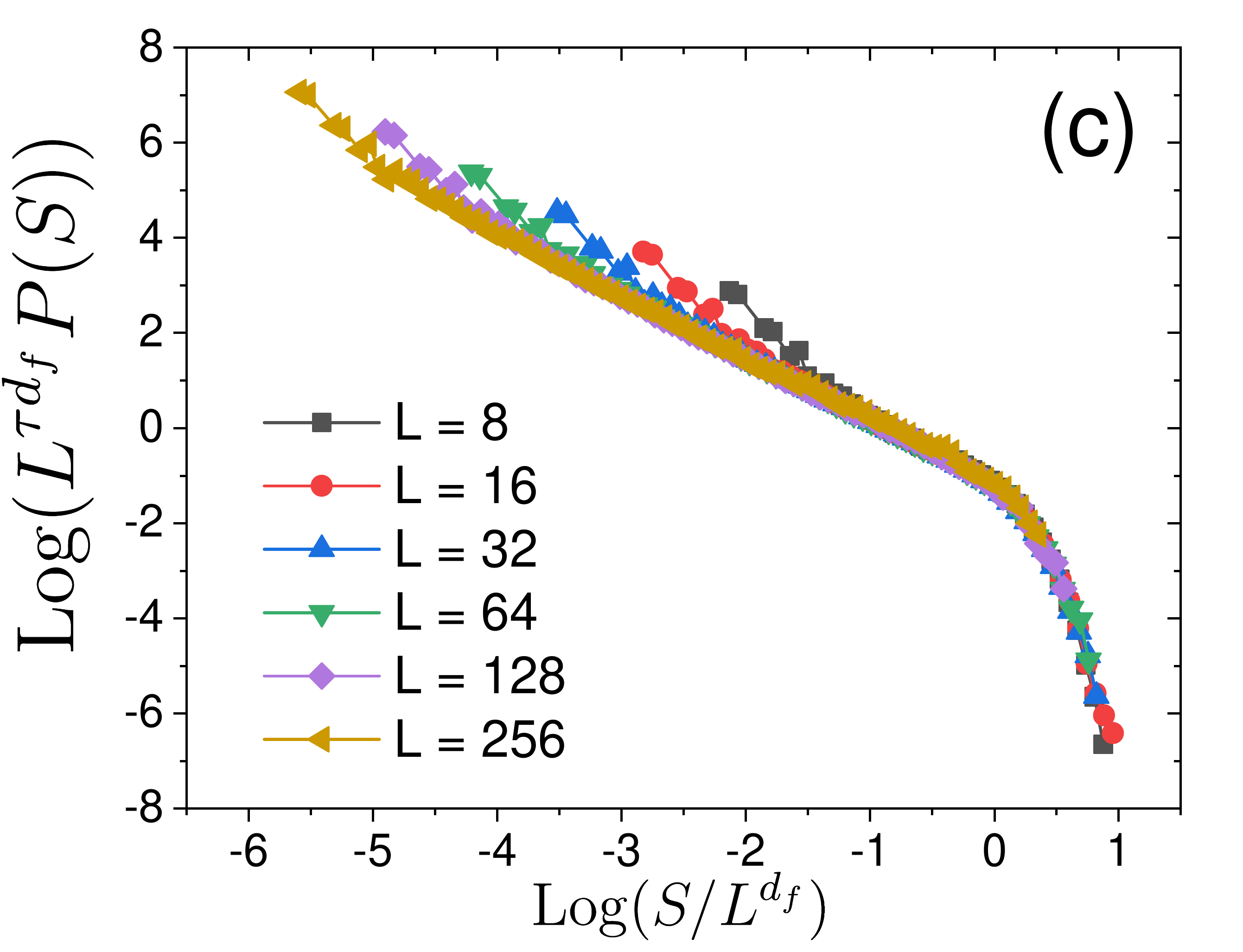}
\includegraphics[width=0.49\linewidth]{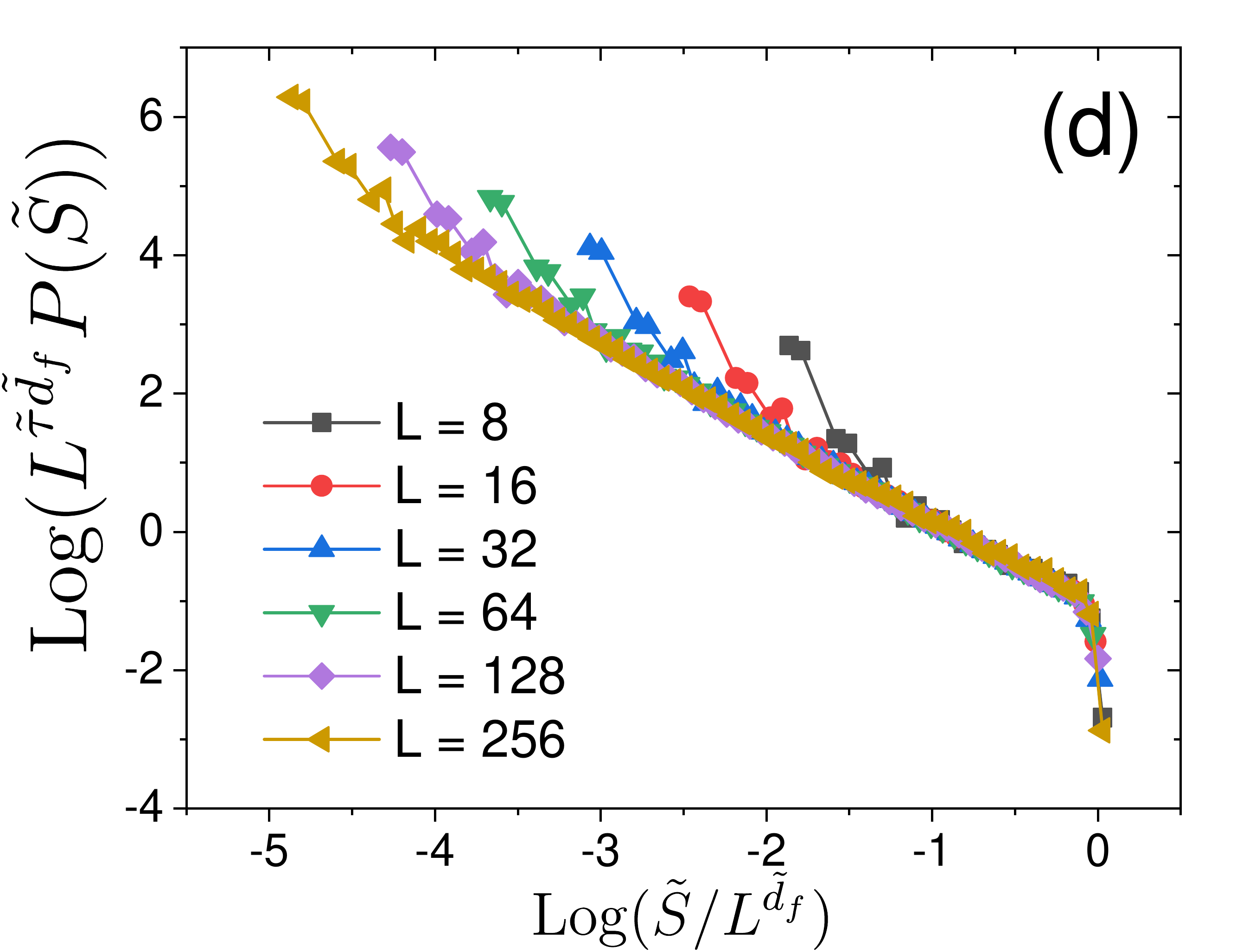}
\caption{(a,b): Distribution of avalanche size $P(S)$ (a) and $P(\tilde S)$ (b) for a stability threshold $x_0=x_c$ for the scalar model, with varying the system size $L$. The dashed lines follow $P(S) \sim S^{-\tau}$ (a) and $P(\tilde S) \sim \tilde  S^{-\tilde  \tau}$ (b). (c): The corresponding scaling collapse following Eq.~(\ref{eq:power_law}) with $S_c \sim L^{d_f}$. (d): Same for $P(\tilde S)$.
}
\label{fig:scalar_zero_T}
\end{figure}

Once the critical threshold $x_c$ is determined, we measure avalanche distributions $P(S)$ and $P(\tilde S)$ at $x_0=x_c$, as shown in Figs.~\ref{fig:scalar_zero_T}(a,b) for the scalar model. They show characteristic power-law behavior with cut-off $S_c$ and $\tilde S_c$ due to a finite size effect, which scales as $S_c \sim L^{d_f}$ and $\tilde S_c \sim L^{\tilde d_f}$, respectively.
The power-law behavior determines $\tau$ and  $\tilde \tau$ whose values are reported in Table~\ref{tab:critical_exponents}.
We then perform scaling collapses in Figs.~\ref{fig:scalar_zero_T}(c,d), following Eq.~(\ref{eq:power_law}), which validates the scaling ansatz and measured critical exponents.

\begin{figure}
\centering
\includegraphics[width=0.49\linewidth]{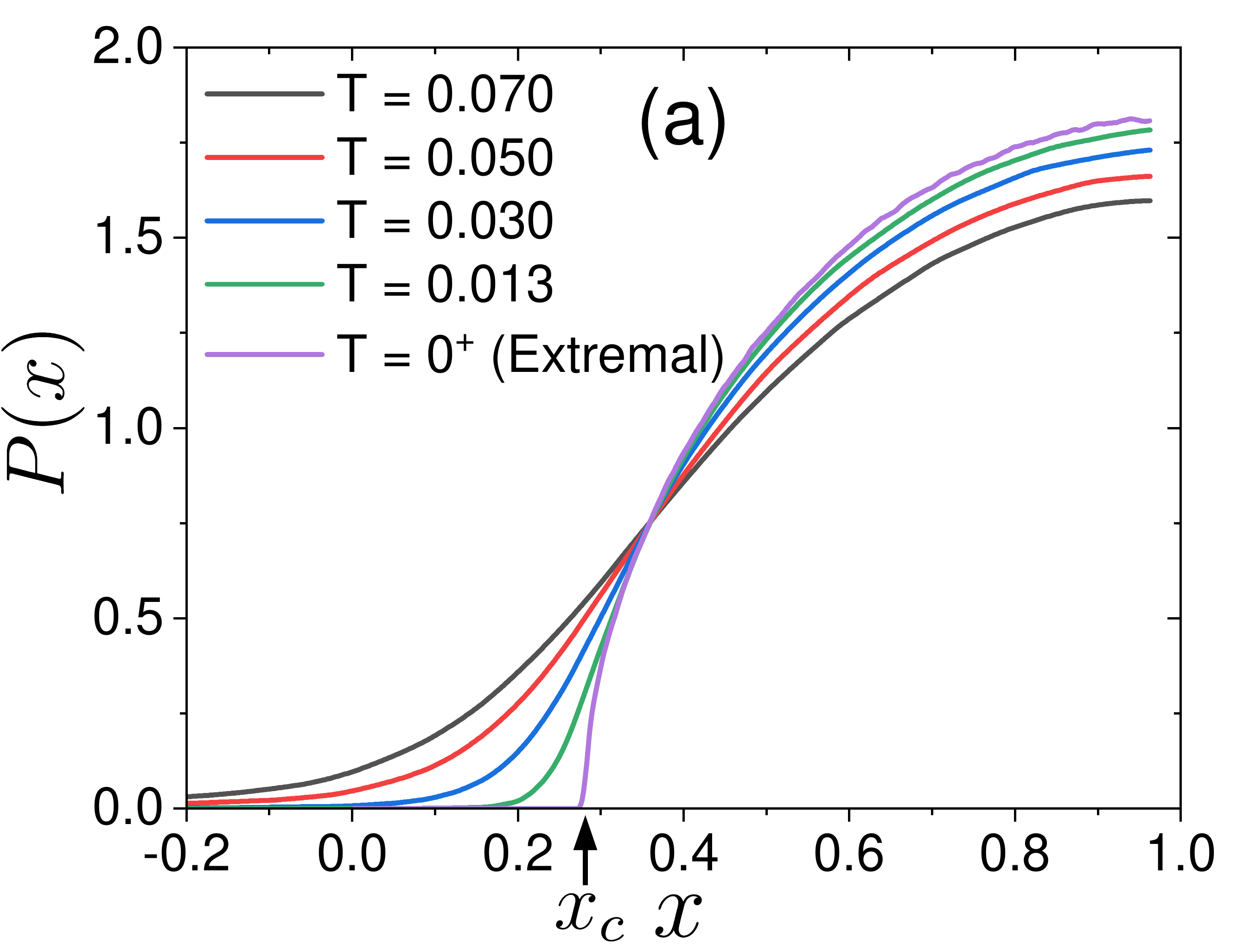}
\includegraphics[width=0.49\linewidth]{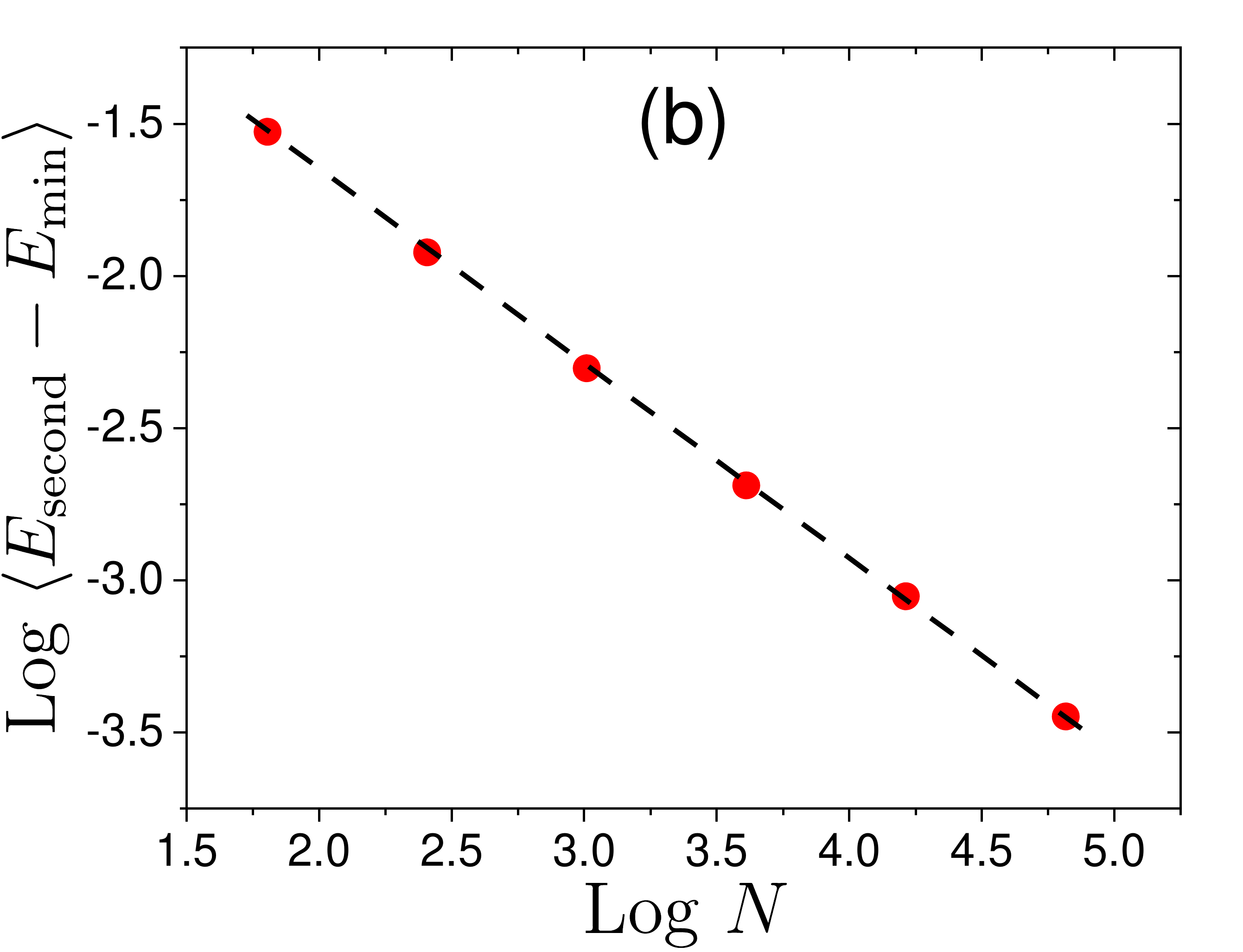}
\caption{(a): Distribution $P(x)$ for the scalar model with $L=256$, obtained from the finite temperature simulations and the extremal dynamics.
The vertical arrow locates the critical threshold $x_c$.
(b): $\langle E_{\rm second}-E_{\rm min}\rangle$, where the average is taken over configurations right before (or after) each avalanche formation, i.e., for which $x_{\rm min} > x_0=x_c$. The dashed line follows $\langle E_{\rm second}-E_{\rm min}\rangle \sim N^{-\delta}$.}
\label{fig:scalar_P_of_x_zero}
\end{figure}

Finally, we compute the distribution $P(x)$ from the configuration right before (or after) each avalanche defined by $x_0=x_c$ starts (or ends). Thus we exclude configurations during each avalanche and focus only on stable configurations expected to hold at strictly $T=0$, where dynamics is not allowed.
In Fig.~\ref{fig:scalar_P_of_x_zero}(a), we show the measured $P(x)$ for the scalar model together with $P(x)$ obtained from the finite temperature simulations. 
As $T$ is lowered, $P(x)$ for finite $T$ converges to $P(x)$ obtained from the extremal dynamics, whose functional form is consistent with $P(x) \sim (x-x_c)^{\theta}$, expected from other disordered systems.
We then measure 
$\langle E_{\rm second}-E_{\rm min}\rangle \sim N^{-\delta}$ in Fig.~\ref{fig:scalar_P_of_x_zero}(b), which encodes an important feature of the distribution of activation energy barriers at $T=0$ (see Sec.~\ref{sec:zero_T} for detail discussions).
The obtained exponent $\delta$ is reported in Table~\ref{tab:critical_exponents}.

\bibliography{new.bib}

\begin{thebibliography}{141}%
\makeatletter
\providecommand \@ifxundefined [1]{%
 \@ifx{#1\undefined}
}%
\providecommand \@ifnum [1]{%
 \ifnum #1\expandafter \@firstoftwo
 \else \expandafter \@secondoftwo
 \fi
}%
\providecommand \@ifx [1]{%
 \ifx #1\expandafter \@firstoftwo
 \else \expandafter \@secondoftwo
 \fi
}%
\providecommand \natexlab [1]{#1}%
\providecommand \enquote  [1]{``#1''}%
\providecommand \bibnamefont  [1]{#1}%
\providecommand \bibfnamefont [1]{#1}%
\providecommand \citenamefont [1]{#1}%
\providecommand \href@noop [0]{\@secondoftwo}%
\providecommand \href [0]{\begingroup \@sanitize@url \@href}%
\providecommand \@href[1]{\@@startlink{#1}\@@href}%
\providecommand \@@href[1]{\endgroup#1\@@endlink}%
\providecommand \@sanitize@url [0]{\catcode `\\12\catcode `\$12\catcode
  `\&12\catcode `\#12\catcode `\^12\catcode `\_12\catcode `\%12\relax}%
\providecommand \@@startlink[1]{}%
\providecommand \@@endlink[0]{}%
\providecommand \url  [0]{\begingroup\@sanitize@url \@url }%
\providecommand \@url [1]{\endgroup\@href {#1}{\urlprefix }}%
\providecommand \urlprefix  [0]{URL }%
\providecommand \Eprint [0]{\href }%
\providecommand \doibase [0]{http://dx.doi.org/}%
\providecommand \selectlanguage [0]{\@gobble}%
\providecommand \bibinfo  [0]{\@secondoftwo}%
\providecommand \bibfield  [0]{\@secondoftwo}%
\providecommand \translation [1]{[#1]}%
\providecommand \BibitemOpen [0]{}%
\providecommand \bibitemStop [0]{}%
\providecommand \bibitemNoStop [0]{.\EOS\space}%
\providecommand \EOS [0]{\spacefactor3000\relax}%
\providecommand \BibitemShut  [1]{\csname bibitem#1\endcsname}%
\let\auto@bib@innerbib\@empty
\bibitem [{\citenamefont {Anderson}(1995)}]{Anderson95}%
  \BibitemOpen
  \bibfield  {author} {\bibinfo {author} {\bibfnamefont {D.L.}\ \bibnamefont
  {Anderson}},\ }\bibfield  {title} {\enquote {\bibinfo {title} {Through the
  glass lightly},}\ }\href@noop {} {\bibfield  {journal} {\bibinfo  {journal}
  {Science}\ }\textbf {\bibinfo {volume} {267}},\ \bibinfo {pages} {1618--1618}
  (\bibinfo {year} {1995})}\BibitemShut {NoStop}%
\bibitem [{\citenamefont {Ediger}\ \emph {et~al.}(1996)\citenamefont {Ediger},
  \citenamefont {Angell},\ and\ \citenamefont {Nagel}}]{ediger1996supercooled}%
  \BibitemOpen
  \bibfield  {author} {\bibinfo {author} {\bibfnamefont {Mark~D}\ \bibnamefont
  {Ediger}}, \bibinfo {author} {\bibfnamefont {C~Austen}\ \bibnamefont
  {Angell}}, \ and\ \bibinfo {author} {\bibfnamefont {Sidney~R}\ \bibnamefont
  {Nagel}},\ }\bibfield  {title} {\enquote {\bibinfo {title} {Supercooled
  liquids and glasses},}\ }\href@noop {} {\bibfield  {journal} {\bibinfo
  {journal} {The journal of physical chemistry}\ }\textbf {\bibinfo {volume}
  {100}},\ \bibinfo {pages} {13200--13212} (\bibinfo {year}
  {1996})}\BibitemShut {NoStop}%
\bibitem [{\citenamefont {Debenedetti}\ and\ \citenamefont
  {Stillinger}(2001)}]{Debenedetti01}%
  \BibitemOpen
  \bibfield  {author} {\bibinfo {author} {\bibfnamefont {P.G.}\ \bibnamefont
  {Debenedetti}}\ and\ \bibinfo {author} {\bibfnamefont {F.H.}\ \bibnamefont
  {Stillinger}},\ }\bibfield  {title} {\enquote {\bibinfo {title} {Supercooled
  liquids and the glass transition},}\ }\href@noop {} {\bibfield  {journal}
  {\bibinfo  {journal} {Nature}\ }\textbf {\bibinfo {volume} {410}},\ \bibinfo
  {pages} {259--267} (\bibinfo {year} {2001})}\BibitemShut {NoStop}%
\bibitem [{\citenamefont {Berthier}\ and\ \citenamefont
  {Biroli}(2011)}]{berthier2011theoretical}%
  \BibitemOpen
  \bibfield  {author} {\bibinfo {author} {\bibfnamefont {Ludovic}\ \bibnamefont
  {Berthier}}\ and\ \bibinfo {author} {\bibfnamefont {Giulio}\ \bibnamefont
  {Biroli}},\ }\bibfield  {title} {\enquote {\bibinfo {title} {Theoretical
  perspective on the glass transition and amorphous materials},}\ }\href@noop
  {} {\bibfield  {journal} {\bibinfo  {journal} {Reviews of modern physics}\
  }\textbf {\bibinfo {volume} {83}},\ \bibinfo {pages} {587} (\bibinfo {year}
  {2011})}\BibitemShut {NoStop}%
\bibitem [{\citenamefont {Kob}\ \emph {et~al.}(1997)\citenamefont {Kob},
  \citenamefont {Donati}, \citenamefont {Plimpton}, \citenamefont {Poole},\
  and\ \citenamefont {Glotzer}}]{kob1997dynamical}%
  \BibitemOpen
  \bibfield  {author} {\bibinfo {author} {\bibfnamefont {Walter}\ \bibnamefont
  {Kob}}, \bibinfo {author} {\bibfnamefont {Claudio}\ \bibnamefont {Donati}},
  \bibinfo {author} {\bibfnamefont {Steven~J}\ \bibnamefont {Plimpton}},
  \bibinfo {author} {\bibfnamefont {Peter~H}\ \bibnamefont {Poole}}, \ and\
  \bibinfo {author} {\bibfnamefont {Sharon~C}\ \bibnamefont {Glotzer}},\
  }\bibfield  {title} {\enquote {\bibinfo {title} {Dynamical heterogeneities in
  a supercooled lennard-jones liquid},}\ }\href@noop {} {\bibfield  {journal}
  {\bibinfo  {journal} {Physical review letters}\ }\textbf {\bibinfo {volume}
  {79}},\ \bibinfo {pages} {2827} (\bibinfo {year} {1997})}\BibitemShut
  {NoStop}%
\bibitem [{\citenamefont {Yamamoto}\ and\ \citenamefont
  {Onuki}(1998)}]{yamamoto1998dynamics}%
  \BibitemOpen
  \bibfield  {author} {\bibinfo {author} {\bibfnamefont {Ryoichi}\ \bibnamefont
  {Yamamoto}}\ and\ \bibinfo {author} {\bibfnamefont {Akira}\ \bibnamefont
  {Onuki}},\ }\bibfield  {title} {\enquote {\bibinfo {title} {Dynamics of
  highly supercooled liquids: Heterogeneity, rheology, and diffusion},}\
  }\href@noop {} {\bibfield  {journal} {\bibinfo  {journal} {Physical Review
  E}\ }\textbf {\bibinfo {volume} {58}},\ \bibinfo {pages} {3515} (\bibinfo
  {year} {1998})}\BibitemShut {NoStop}%
\bibitem [{\citenamefont {Dalle-Ferrier}\ \emph {et~al.}(2007)\citenamefont
  {Dalle-Ferrier}, \citenamefont {Thibierge}, \citenamefont {Alba-Simionesco},
  \citenamefont {Berthier}, \citenamefont {Biroli}, \citenamefont {Bouchaud},
  \citenamefont {Ladieu}, \citenamefont {L’H{\^o}te},\ and\ \citenamefont
  {Tarjus}}]{dalle2007spatial}%
  \BibitemOpen
  \bibfield  {author} {\bibinfo {author} {\bibfnamefont {C.}~\bibnamefont
  {Dalle-Ferrier}}, \bibinfo {author} {\bibfnamefont {C.}~\bibnamefont
  {Thibierge}}, \bibinfo {author} {\bibfnamefont {C.}~\bibnamefont
  {Alba-Simionesco}}, \bibinfo {author} {\bibfnamefont {L.}~\bibnamefont
  {Berthier}}, \bibinfo {author} {\bibfnamefont {G.}~\bibnamefont {Biroli}},
  \bibinfo {author} {\bibfnamefont {J.-P.}\ \bibnamefont {Bouchaud}}, \bibinfo
  {author} {\bibfnamefont {F.}~\bibnamefont {Ladieu}}, \bibinfo {author}
  {\bibfnamefont {D.}~\bibnamefont {L’H{\^o}te}}, \ and\ \bibinfo {author}
  {\bibfnamefont {G.}~\bibnamefont {Tarjus}},\ }\bibfield  {title} {\enquote
  {\bibinfo {title} {Spatial correlations in the dynamics of glassforming
  liquids: Experimental determination of their temperature dependence},}\
  }\href@noop {} {\bibfield  {journal} {\bibinfo  {journal} {Phys. Rev. E}\
  }\textbf {\bibinfo {volume} {76}},\ \bibinfo {pages} {041510} (\bibinfo
  {year} {2007})}\BibitemShut {NoStop}%
\bibitem [{\citenamefont {Karmakar}\ \emph {et~al.}(2014)\citenamefont
  {Karmakar}, \citenamefont {Dasgupta},\ and\ \citenamefont
  {Sastry}}]{karmakar2014growing}%
  \BibitemOpen
  \bibfield  {author} {\bibinfo {author} {\bibfnamefont {Smarajit}\
  \bibnamefont {Karmakar}}, \bibinfo {author} {\bibfnamefont {Chandan}\
  \bibnamefont {Dasgupta}}, \ and\ \bibinfo {author} {\bibfnamefont {Srikanth}\
  \bibnamefont {Sastry}},\ }\bibfield  {title} {\enquote {\bibinfo {title}
  {Growing length scales and their relation to timescales in glass-forming
  liquids},}\ }\href@noop {} {\bibfield  {journal} {\bibinfo  {journal} {Annu.
  Rev. Condens. Matter Phys.}\ }\textbf {\bibinfo {volume} {5}},\ \bibinfo
  {pages} {255--284} (\bibinfo {year} {2014})}\BibitemShut {NoStop}%
\bibitem [{\citenamefont {Kirkpatrick}\ \emph {et~al.}(1989)\citenamefont
  {Kirkpatrick}, \citenamefont {Thirumalai},\ and\ \citenamefont
  {Wolynes}}]{kirkpatrick1989scaling}%
  \BibitemOpen
  \bibfield  {author} {\bibinfo {author} {\bibfnamefont {Theodore~R}\
  \bibnamefont {Kirkpatrick}}, \bibinfo {author} {\bibfnamefont {Devarajan}\
  \bibnamefont {Thirumalai}}, \ and\ \bibinfo {author} {\bibfnamefont
  {Peter~G}\ \bibnamefont {Wolynes}},\ }\bibfield  {title} {\enquote {\bibinfo
  {title} {Scaling concepts for the dynamics of viscous liquids near an ideal
  glassy state},}\ }\href@noop {} {\bibfield  {journal} {\bibinfo  {journal}
  {Physical Review A}\ }\textbf {\bibinfo {volume} {40}},\ \bibinfo {pages}
  {1045} (\bibinfo {year} {1989})}\BibitemShut {NoStop}%
\bibitem [{\citenamefont {Lubchenko}\ and\ \citenamefont
  {Wolynes}(2007)}]{lubchenko2007theory}%
  \BibitemOpen
  \bibfield  {author} {\bibinfo {author} {\bibfnamefont {Vassiliy}\
  \bibnamefont {Lubchenko}}\ and\ \bibinfo {author} {\bibfnamefont {Peter~G}\
  \bibnamefont {Wolynes}},\ }\bibfield  {title} {\enquote {\bibinfo {title}
  {Theory of structural glasses and supercooled liquids},}\ }\href@noop {}
  {\bibfield  {journal} {\bibinfo  {journal} {Annu. Rev. Phys. Chem.}\ }\textbf
  {\bibinfo {volume} {58}},\ \bibinfo {pages} {235--266} (\bibinfo {year}
  {2007})}\BibitemShut {NoStop}%
\bibitem [{\citenamefont {Biroli}\ and\ \citenamefont
  {Bouchaud}(2012)}]{Biroli12}%
  \BibitemOpen
  \bibfield  {author} {\bibinfo {author} {\bibfnamefont {G.}~\bibnamefont
  {Biroli}}\ and\ \bibinfo {author} {\bibfnamefont {J.-P.}\ \bibnamefont
  {Bouchaud}},\ }\bibfield  {title} {\enquote {\bibinfo {title} {The random
  first-order transition theory of glasses: a critical assessment},}\
  }\href@noop {} {\bibfield  {journal} {\bibinfo  {journal} {Structural Glasses
  and Supercooled Liquids: Theory, Experiment, and Applications}\ ,\ \bibinfo
  {pages} {31--113}} (\bibinfo {year} {2012})}\BibitemShut {NoStop}%
\bibitem [{\citenamefont {Ritort}\ and\ \citenamefont
  {Sollich}(2003)}]{ritort2003glassy}%
  \BibitemOpen
  \bibfield  {author} {\bibinfo {author} {\bibfnamefont {Felix}\ \bibnamefont
  {Ritort}}\ and\ \bibinfo {author} {\bibfnamefont {Peter}\ \bibnamefont
  {Sollich}},\ }\bibfield  {title} {\enquote {\bibinfo {title} {Glassy dynamics
  of kinetically constrained models},}\ }\href@noop {} {\bibfield  {journal}
  {\bibinfo  {journal} {Advances in physics}\ }\textbf {\bibinfo {volume}
  {52}},\ \bibinfo {pages} {219--342} (\bibinfo {year} {2003})}\BibitemShut
  {NoStop}%
\bibitem [{\citenamefont {Berthier}\ \emph {et~al.}(2011)\citenamefont
  {Berthier}, \citenamefont {Biroli}, \citenamefont {Bouchaud}, \citenamefont
  {Cipelletti},\ and\ \citenamefont {van Saarloos}}]{berthier2011dynamical}%
  \BibitemOpen
  \bibfield  {author} {\bibinfo {author} {\bibfnamefont {Ludovic}\ \bibnamefont
  {Berthier}}, \bibinfo {author} {\bibfnamefont {Giulio}\ \bibnamefont
  {Biroli}}, \bibinfo {author} {\bibfnamefont {Jean-Philippe}\ \bibnamefont
  {Bouchaud}}, \bibinfo {author} {\bibfnamefont {Luca}\ \bibnamefont
  {Cipelletti}}, \ and\ \bibinfo {author} {\bibfnamefont {Wim}\ \bibnamefont
  {van Saarloos}},\ }\href@noop {} {\emph {\bibinfo {title} {Dynamical
  heterogeneities in glasses, colloids, and granular media}}},\ Vol.\ \bibinfo
  {volume} {150}\ (\bibinfo  {publisher} {OUP Oxford},\ \bibinfo {year}
  {2011})\BibitemShut {NoStop}%
\bibitem [{\citenamefont {Garrahan}\ and\ \citenamefont
  {Chandler}(2002)}]{garrahan2002geometrical}%
  \BibitemOpen
  \bibfield  {author} {\bibinfo {author} {\bibfnamefont {Juan~P}\ \bibnamefont
  {Garrahan}}\ and\ \bibinfo {author} {\bibfnamefont {David}\ \bibnamefont
  {Chandler}},\ }\bibfield  {title} {\enquote {\bibinfo {title} {Geometrical
  explanation and scaling of dynamical heterogeneities in glass forming
  systems},}\ }\href@noop {} {\bibfield  {journal} {\bibinfo  {journal}
  {Physical review letters}\ }\textbf {\bibinfo {volume} {89}},\ \bibinfo
  {pages} {035704} (\bibinfo {year} {2002})}\BibitemShut {NoStop}%
\bibitem [{\citenamefont {Garrahan}\ \emph {et~al.}(2007)\citenamefont
  {Garrahan}, \citenamefont {Jack}, \citenamefont {Lecomte}, \citenamefont
  {Pitard}, \citenamefont {van Duijvendijk},\ and\ \citenamefont {van
  Wijland}}]{garrahan2007dynamical}%
  \BibitemOpen
  \bibfield  {author} {\bibinfo {author} {\bibfnamefont {Juan~P}\ \bibnamefont
  {Garrahan}}, \bibinfo {author} {\bibfnamefont {Robert~L}\ \bibnamefont
  {Jack}}, \bibinfo {author} {\bibfnamefont {Vivien}\ \bibnamefont {Lecomte}},
  \bibinfo {author} {\bibfnamefont {Estelle}\ \bibnamefont {Pitard}}, \bibinfo
  {author} {\bibfnamefont {Kristina}\ \bibnamefont {van Duijvendijk}}, \ and\
  \bibinfo {author} {\bibfnamefont {Fr{\'e}d{\'e}ric}\ \bibnamefont {van
  Wijland}},\ }\bibfield  {title} {\enquote {\bibinfo {title} {Dynamical
  first-order phase transition in kinetically constrained models of glasses},}\
  }\href@noop {} {\bibfield  {journal} {\bibinfo  {journal} {Physical review
  letters}\ }\textbf {\bibinfo {volume} {98}},\ \bibinfo {pages} {195702}
  (\bibinfo {year} {2007})}\BibitemShut {NoStop}%
\bibitem [{\citenamefont {Hedges}\ \emph {et~al.}(2009)\citenamefont {Hedges},
  \citenamefont {Jack}, \citenamefont {Garrahan},\ and\ \citenamefont
  {Chandler}}]{hedges2009dynamic}%
  \BibitemOpen
  \bibfield  {author} {\bibinfo {author} {\bibfnamefont {L.O.}\ \bibnamefont
  {Hedges}}, \bibinfo {author} {\bibfnamefont {R.L.}\ \bibnamefont {Jack}},
  \bibinfo {author} {\bibfnamefont {J.P.}\ \bibnamefont {Garrahan}}, \ and\
  \bibinfo {author} {\bibfnamefont {D.}~\bibnamefont {Chandler}},\ }\bibfield
  {title} {\enquote {\bibinfo {title} {Dynamic order-disorder in atomistic
  models of structural glass formers},}\ }\href@noop {} {\bibfield  {journal}
  {\bibinfo  {journal} {Science}\ }\textbf {\bibinfo {volume} {323}},\ \bibinfo
  {pages} {1309--1313} (\bibinfo {year} {2009})}\BibitemShut {NoStop}%
\bibitem [{\citenamefont {Turnbull}\ and\ \citenamefont
  {Cohen}(1961)}]{turnbull1961free}%
  \BibitemOpen
  \bibfield  {author} {\bibinfo {author} {\bibfnamefont {David}\ \bibnamefont
  {Turnbull}}\ and\ \bibinfo {author} {\bibfnamefont {Morrel~H}\ \bibnamefont
  {Cohen}},\ }\bibfield  {title} {\enquote {\bibinfo {title} {Free-volume model
  of the amorphous phase: glass transition},}\ }\href@noop {} {\bibfield
  {journal} {\bibinfo  {journal} {The Journal of chemical physics}\ }\textbf
  {\bibinfo {volume} {34}},\ \bibinfo {pages} {120--125} (\bibinfo {year}
  {1961})}\BibitemShut {NoStop}%
\bibitem [{\citenamefont {Dyre}(2006)}]{dyre2006colloquium}%
  \BibitemOpen
  \bibfield  {author} {\bibinfo {author} {\bibfnamefont {J.C.}\ \bibnamefont
  {Dyre}},\ }\bibfield  {title} {\enquote {\bibinfo {title} {Colloquium: The
  glass transition and elastic models of glass-forming liquids},}\ }\href@noop
  {} {\bibfield  {journal} {\bibinfo  {journal} {Rev. Mod. Phys.}\ }\textbf
  {\bibinfo {volume} {78}},\ \bibinfo {pages} {953} (\bibinfo {year}
  {2006})}\BibitemShut {NoStop}%
\bibitem [{\citenamefont {Rainone}\ \emph {et~al.}(2020)\citenamefont
  {Rainone}, \citenamefont {Bouchbinder},\ and\ \citenamefont
  {Lerner}}]{rainone2020pinching}%
  \BibitemOpen
  \bibfield  {author} {\bibinfo {author} {\bibfnamefont {Corrado}\ \bibnamefont
  {Rainone}}, \bibinfo {author} {\bibfnamefont {Eran}\ \bibnamefont
  {Bouchbinder}}, \ and\ \bibinfo {author} {\bibfnamefont {Edan}\ \bibnamefont
  {Lerner}},\ }\bibfield  {title} {\enquote {\bibinfo {title} {Pinching a glass
  reveals key properties of its soft spots},}\ }\href@noop {} {\bibfield
  {journal} {\bibinfo  {journal} {Proceedings of the National Academy of
  Sciences}\ }\textbf {\bibinfo {volume} {117}},\ \bibinfo {pages} {5228--5234}
  (\bibinfo {year} {2020})}\BibitemShut {NoStop}%
\bibitem [{\citenamefont {Kapteijns}\ \emph {et~al.}(2021)\citenamefont
  {Kapteijns}, \citenamefont {Richard}, \citenamefont {Bouchbinder},
  \citenamefont {Schr{\o}der}, \citenamefont {Dyre},\ and\ \citenamefont
  {Lerner}}]{JeppeEdan}%
  \BibitemOpen
  \bibfield  {author} {\bibinfo {author} {\bibfnamefont {Geert}\ \bibnamefont
  {Kapteijns}}, \bibinfo {author} {\bibfnamefont {David}\ \bibnamefont
  {Richard}}, \bibinfo {author} {\bibfnamefont {Eran}\ \bibnamefont
  {Bouchbinder}}, \bibinfo {author} {\bibfnamefont {Thomas~B}\ \bibnamefont
  {Schr{\o}der}}, \bibinfo {author} {\bibfnamefont {Jeppe~C}\ \bibnamefont
  {Dyre}}, \ and\ \bibinfo {author} {\bibfnamefont {Edan}\ \bibnamefont
  {Lerner}},\ }\bibfield  {title} {\enquote {\bibinfo {title} {{Does mesoscopic
  elasticity control viscous slowing down in glassforming liquids?}}}\ }\href
  {\doibase 10.1063/5.0051193} {\bibfield  {journal} {\bibinfo  {journal} {The
  Journal of Chemical Physics}\ }\textbf {\bibinfo {volume} {155}},\ \bibinfo
  {pages} {74502} (\bibinfo {year} {2021})}\BibitemShut {NoStop}%
\bibitem [{\citenamefont {Ciamarra}\ \emph {et~al.}(2023)\citenamefont
  {Ciamarra}, \citenamefont {Ji},\ and\ \citenamefont {Wyart}}]{massimo}%
  \BibitemOpen
  \bibfield  {author} {\bibinfo {author} {\bibfnamefont {Massimo~Pica}\
  \bibnamefont {Ciamarra}}, \bibinfo {author} {\bibfnamefont {Wencheng}\
  \bibnamefont {Ji}}, \ and\ \bibinfo {author} {\bibfnamefont {Matthieu}\
  \bibnamefont {Wyart}},\ }\href {\doibase 10.48550/ARXIV.2302.05150} {\enquote
  {\bibinfo {title} {The energy cost of local rearrangements, not cooperative
  effects, makes glasses solid},}\ } (\bibinfo {year} {2023})\BibitemShut
  {NoStop}%
\bibitem [{\citenamefont {Schr{\o}der}\ and\ \citenamefont
  {Dyre}(2020)}]{schroder2020solid}%
  \BibitemOpen
  \bibfield  {author} {\bibinfo {author} {\bibfnamefont {Thomas~B}\
  \bibnamefont {Schr{\o}der}}\ and\ \bibinfo {author} {\bibfnamefont {Jeppe~C}\
  \bibnamefont {Dyre}},\ }\bibfield  {title} {\enquote {\bibinfo {title}
  {Solid-like mean-square displacement in glass-forming liquids},}\ }\href@noop
  {} {\bibfield  {journal} {\bibinfo  {journal} {The Journal of Chemical
  Physics}\ }\textbf {\bibinfo {volume} {152}},\ \bibinfo {pages} {141101}
  (\bibinfo {year} {2020})}\BibitemShut {NoStop}%
\bibitem [{\citenamefont {Lema{\^{i}}tre}(2014)}]{Lemaitre14}%
  \BibitemOpen
  \bibfield  {author} {\bibinfo {author} {\bibfnamefont {A.}~\bibnamefont
  {Lema{\^{i}}tre}},\ }\bibfield  {title} {\enquote {\bibinfo {title}
  {{Structural Relaxation is a Scale-Free Process}},}\ }\href@noop {}
  {\bibfield  {journal} {\bibinfo  {journal} {Phys. Rev. Lett.}\ }\textbf
  {\bibinfo {volume} {113}},\ \bibinfo {pages} {245702} (\bibinfo {year}
  {2014})}\BibitemShut {NoStop}%
\bibitem [{\citenamefont {Chowdhury}\ \emph {et~al.}(2016)\citenamefont
  {Chowdhury}, \citenamefont {Abraham}, \citenamefont {Hudson},\ and\
  \citenamefont {Harrowell}}]{chowdhury2016long}%
  \BibitemOpen
  \bibfield  {author} {\bibinfo {author} {\bibfnamefont {Sadrul}\ \bibnamefont
  {Chowdhury}}, \bibinfo {author} {\bibfnamefont {Sneha}\ \bibnamefont
  {Abraham}}, \bibinfo {author} {\bibfnamefont {Toby}\ \bibnamefont {Hudson}},
  \ and\ \bibinfo {author} {\bibfnamefont {Peter}\ \bibnamefont {Harrowell}},\
  }\bibfield  {title} {\enquote {\bibinfo {title} {Long range stress
  correlations in the inherent structures of liquids at rest},}\ }\href@noop {}
  {\bibfield  {journal} {\bibinfo  {journal} {The Journal of chemical physics}\
  }\textbf {\bibinfo {volume} {144}},\ \bibinfo {pages} {124508} (\bibinfo
  {year} {2016})}\BibitemShut {NoStop}%
\bibitem [{\citenamefont {Tong}\ \emph {et~al.}(2020)\citenamefont {Tong},
  \citenamefont {Sengupta},\ and\ \citenamefont {Tanaka}}]{tong2020emergent}%
  \BibitemOpen
  \bibfield  {author} {\bibinfo {author} {\bibfnamefont {Hua}\ \bibnamefont
  {Tong}}, \bibinfo {author} {\bibfnamefont {Shiladitya}\ \bibnamefont
  {Sengupta}}, \ and\ \bibinfo {author} {\bibfnamefont {Hajime}\ \bibnamefont
  {Tanaka}},\ }\bibfield  {title} {\enquote {\bibinfo {title} {Emergent
  solidity of amorphous materials as a consequence of mechanical
  self-organisation},}\ }\href@noop {} {\bibfield  {journal} {\bibinfo
  {journal} {Nature communications}\ }\textbf {\bibinfo {volume} {11}},\
  \bibinfo {pages} {1--10} (\bibinfo {year} {2020})}\BibitemShut {NoStop}%
\bibitem [{\citenamefont {Wu}\ \emph {et~al.}(2015)\citenamefont {Wu},
  \citenamefont {Iwashita},\ and\ \citenamefont {Egami}}]{wu2015anisotropic}%
  \BibitemOpen
  \bibfield  {author} {\bibinfo {author} {\bibfnamefont {Bin}\ \bibnamefont
  {Wu}}, \bibinfo {author} {\bibfnamefont {Takuya}\ \bibnamefont {Iwashita}}, \
  and\ \bibinfo {author} {\bibfnamefont {Takeshi}\ \bibnamefont {Egami}},\
  }\bibfield  {title} {\enquote {\bibinfo {title} {Anisotropic stress
  correlations in two-dimensional liquids},}\ }\href@noop {} {\bibfield
  {journal} {\bibinfo  {journal} {Physical Review E}\ }\textbf {\bibinfo
  {volume} {91}},\ \bibinfo {pages} {032301} (\bibinfo {year}
  {2015})}\BibitemShut {NoStop}%
\bibitem [{\citenamefont {Maier}\ \emph {et~al.}(2017)\citenamefont {Maier},
  \citenamefont {Zippelius},\ and\ \citenamefont {Fuchs}}]{maier2017emergence}%
  \BibitemOpen
  \bibfield  {author} {\bibinfo {author} {\bibfnamefont {Manuel}\ \bibnamefont
  {Maier}}, \bibinfo {author} {\bibfnamefont {Annette}\ \bibnamefont
  {Zippelius}}, \ and\ \bibinfo {author} {\bibfnamefont {Matthias}\
  \bibnamefont {Fuchs}},\ }\bibfield  {title} {\enquote {\bibinfo {title}
  {Emergence of long-ranged stress correlations at the liquid to glass
  transition},}\ }\href@noop {} {\bibfield  {journal} {\bibinfo  {journal}
  {Physical review letters}\ }\textbf {\bibinfo {volume} {119}},\ \bibinfo
  {pages} {265701} (\bibinfo {year} {2017})}\BibitemShut {NoStop}%
\bibitem [{\citenamefont {Steffen}\ \emph {et~al.}(2022)\citenamefont
  {Steffen}, \citenamefont {Schneider}, \citenamefont {M{\"u}ller},\ and\
  \citenamefont {Rottler}}]{steffen2022molecular}%
  \BibitemOpen
  \bibfield  {author} {\bibinfo {author} {\bibfnamefont {David}\ \bibnamefont
  {Steffen}}, \bibinfo {author} {\bibfnamefont {Ludwig}\ \bibnamefont
  {Schneider}}, \bibinfo {author} {\bibfnamefont {Marcus}\ \bibnamefont
  {M{\"u}ller}}, \ and\ \bibinfo {author} {\bibfnamefont {J{\"o}rg}\
  \bibnamefont {Rottler}},\ }\bibfield  {title} {\enquote {\bibinfo {title}
  {Molecular simulations and hydrodynamic theory of nonlocal shear-stress
  correlations in supercooled fluids},}\ }\href@noop {} {\bibfield  {journal}
  {\bibinfo  {journal} {The Journal of Chemical Physics}\ }\textbf {\bibinfo
  {volume} {157}},\ \bibinfo {pages} {064501} (\bibinfo {year}
  {2022})}\BibitemShut {NoStop}%
\bibitem [{\citenamefont {Flenner}\ and\ \citenamefont
  {Szamel}(2015)}]{flenner2015long}%
  \BibitemOpen
  \bibfield  {author} {\bibinfo {author} {\bibfnamefont {Elijah}\ \bibnamefont
  {Flenner}}\ and\ \bibinfo {author} {\bibfnamefont {Grzegorz}\ \bibnamefont
  {Szamel}},\ }\bibfield  {title} {\enquote {\bibinfo {title} {Long-range
  spatial correlations of particle displacements and the emergence of
  elasticity},}\ }\href@noop {} {\bibfield  {journal} {\bibinfo  {journal}
  {Physical Review Letters}\ }\textbf {\bibinfo {volume} {114}},\ \bibinfo
  {pages} {025501} (\bibinfo {year} {2015})}\BibitemShut {NoStop}%
\bibitem [{\citenamefont {Klochko}\ \emph {et~al.}(2022)\citenamefont
  {Klochko}, \citenamefont {Baschnagel}, \citenamefont {Wittmer}, \citenamefont
  {Meyer}, \citenamefont {Benzerara},\ and\ \citenamefont
  {Semenov}}]{klochko2022theory}%
  \BibitemOpen
  \bibfield  {author} {\bibinfo {author} {\bibfnamefont {L}~\bibnamefont
  {Klochko}}, \bibinfo {author} {\bibfnamefont {J}~\bibnamefont {Baschnagel}},
  \bibinfo {author} {\bibfnamefont {JP}~\bibnamefont {Wittmer}}, \bibinfo
  {author} {\bibfnamefont {H}~\bibnamefont {Meyer}}, \bibinfo {author}
  {\bibfnamefont {O}~\bibnamefont {Benzerara}}, \ and\ \bibinfo {author}
  {\bibfnamefont {AN}~\bibnamefont {Semenov}},\ }\bibfield  {title} {\enquote
  {\bibinfo {title} {Theory of length-scale dependent relaxation moduli and
  stress fluctuations in glass-forming and viscoelastic liquids},}\ }\href@noop
  {} {\bibfield  {journal} {\bibinfo  {journal} {The Journal of Chemical
  Physics}\ }\textbf {\bibinfo {volume} {156}},\ \bibinfo {pages} {164505}
  (\bibinfo {year} {2022})}\BibitemShut {NoStop}%
\bibitem [{\citenamefont {Chacko}\ \emph {et~al.}(2021)\citenamefont {Chacko},
  \citenamefont {Landes}, \citenamefont {Biroli}, \citenamefont {Dauchot},
  \citenamefont {Liu},\ and\ \citenamefont
  {Reichman}}]{chacko2021elastoplasticity}%
  \BibitemOpen
  \bibfield  {author} {\bibinfo {author} {\bibfnamefont {Rahul~N}\ \bibnamefont
  {Chacko}}, \bibinfo {author} {\bibfnamefont {Fran{\c{c}}ois~P}\ \bibnamefont
  {Landes}}, \bibinfo {author} {\bibfnamefont {Giulio}\ \bibnamefont {Biroli}},
  \bibinfo {author} {\bibfnamefont {Olivier}\ \bibnamefont {Dauchot}}, \bibinfo
  {author} {\bibfnamefont {Andrea~J}\ \bibnamefont {Liu}}, \ and\ \bibinfo
  {author} {\bibfnamefont {David~R}\ \bibnamefont {Reichman}},\ }\bibfield
  {title} {\enquote {\bibinfo {title} {Elastoplasticity mediates dynamical
  heterogeneity below the mode coupling temperature},}\ }\href@noop {}
  {\bibfield  {journal} {\bibinfo  {journal} {Physical Review Letters}\
  }\textbf {\bibinfo {volume} {127}},\ \bibinfo {pages} {048002} (\bibinfo
  {year} {2021})}\BibitemShut {NoStop}%
\bibitem [{\citenamefont {Lerbinger}\ \emph {et~al.}(2022)\citenamefont
  {Lerbinger}, \citenamefont {Barbot}, \citenamefont {Vandembroucq},\ and\
  \citenamefont {Patinet}}]{lerbinger2022relevance}%
  \BibitemOpen
  \bibfield  {author} {\bibinfo {author} {\bibfnamefont {Matthias}\
  \bibnamefont {Lerbinger}}, \bibinfo {author} {\bibfnamefont {Armand}\
  \bibnamefont {Barbot}}, \bibinfo {author} {\bibfnamefont {Damien}\
  \bibnamefont {Vandembroucq}}, \ and\ \bibinfo {author} {\bibfnamefont
  {Sylvain}\ \bibnamefont {Patinet}},\ }\bibfield  {title} {\enquote {\bibinfo
  {title} {Relevance of shear transformations in the relaxation of supercooled
  liquids},}\ }\href@noop {} {\bibfield  {journal} {\bibinfo  {journal}
  {Physical Review Letters}\ }\textbf {\bibinfo {volume} {129}},\ \bibinfo
  {pages} {195501} (\bibinfo {year} {2022})}\BibitemShut {NoStop}%
\bibitem [{\citenamefont {Falk}\ and\ \citenamefont
  {Langer}(1998)}]{falk1998dynamics}%
  \BibitemOpen
  \bibfield  {author} {\bibinfo {author} {\bibfnamefont {Michael~L}\
  \bibnamefont {Falk}}\ and\ \bibinfo {author} {\bibfnamefont {James~S}\
  \bibnamefont {Langer}},\ }\bibfield  {title} {\enquote {\bibinfo {title}
  {Dynamics of viscoplastic deformation in amorphous solids},}\ }\href@noop {}
  {\bibfield  {journal} {\bibinfo  {journal} {Physical Review E}\ }\textbf
  {\bibinfo {volume} {57}},\ \bibinfo {pages} {7192} (\bibinfo {year}
  {1998})}\BibitemShut {NoStop}%
\bibitem [{\citenamefont {Ozawa}\ and\ \citenamefont
  {Biroli}(2023)}]{ozawa2023elasticity}%
  \BibitemOpen
  \bibfield  {author} {\bibinfo {author} {\bibfnamefont {Misaki}\ \bibnamefont
  {Ozawa}}\ and\ \bibinfo {author} {\bibfnamefont {Giulio}\ \bibnamefont
  {Biroli}},\ }\bibfield  {title} {\enquote {\bibinfo {title} {Elasticity,
  facilitation, and dynamic heterogeneity in glass-forming liquids},}\
  }\href@noop {} {\bibfield  {journal} {\bibinfo  {journal} {Physical Review
  Letters}\ }\textbf {\bibinfo {volume} {130}},\ \bibinfo {pages} {138201}
  (\bibinfo {year} {2023})}\BibitemShut {NoStop}%
\bibitem [{\citenamefont {Picard}\ \emph {et~al.}(2004)\citenamefont {Picard},
  \citenamefont {Ajdari}, \citenamefont {Lequeux},\ and\ \citenamefont
  {Bocquet}}]{picard2004elastic}%
  \BibitemOpen
  \bibfield  {author} {\bibinfo {author} {\bibfnamefont {Guillemette}\
  \bibnamefont {Picard}}, \bibinfo {author} {\bibfnamefont {Armand}\
  \bibnamefont {Ajdari}}, \bibinfo {author} {\bibfnamefont {Fran{\c{c}}ois}\
  \bibnamefont {Lequeux}}, \ and\ \bibinfo {author} {\bibfnamefont
  {Lyd{\'e}ric}\ \bibnamefont {Bocquet}},\ }\bibfield  {title} {\enquote
  {\bibinfo {title} {Elastic consequences of a single plastic event: A step
  towards the microscopic modeling of the flow of yield stress fluids},}\
  }\href@noop {} {\bibfield  {journal} {\bibinfo  {journal} {The European
  Physical Journal E}\ }\textbf {\bibinfo {volume} {15}},\ \bibinfo {pages}
  {371--381} (\bibinfo {year} {2004})}\BibitemShut {NoStop}%
\bibitem [{\citenamefont {Nicolas}\ \emph {et~al.}(2018)\citenamefont
  {Nicolas}, \citenamefont {Ferrero}, \citenamefont {Martens},\ and\
  \citenamefont {Barrat}}]{Nicolas2018}%
  \BibitemOpen
  \bibfield  {author} {\bibinfo {author} {\bibfnamefont {A.}~\bibnamefont
  {Nicolas}}, \bibinfo {author} {\bibfnamefont {E.E.}\ \bibnamefont {Ferrero}},
  \bibinfo {author} {\bibfnamefont {K.}~\bibnamefont {Martens}}, \ and\
  \bibinfo {author} {\bibfnamefont {J.-L.}\ \bibnamefont {Barrat}},\ }\bibfield
   {title} {\enquote {\bibinfo {title} {Deformation and flow of amorphous
  solids: Insights from elastoplastic models},}\ }\href {\doibase
  10.1103/RevModPhys.90.045006} {\bibfield  {journal} {\bibinfo  {journal}
  {Rev. Mod. Phys.}\ }\textbf {\bibinfo {volume} {90}},\ \bibinfo {pages}
  {045006} (\bibinfo {year} {2018})}\BibitemShut {NoStop}%
\bibitem [{\citenamefont {Vandembroucq}\ \emph {et~al.}(2004)\citenamefont
  {Vandembroucq}, \citenamefont {Skoe},\ and\ \citenamefont
  {Roux}}]{Vandembroucq2004}%
  \BibitemOpen
  \bibfield  {author} {\bibinfo {author} {\bibfnamefont {D.}~\bibnamefont
  {Vandembroucq}}, \bibinfo {author} {\bibfnamefont {R.}~\bibnamefont {Skoe}},
  \ and\ \bibinfo {author} {\bibfnamefont {S.}~\bibnamefont {Roux}},\
  }\bibfield  {title} {\enquote {\bibinfo {title} {Universal depinning force
  fluctuations of an elastic line: Application to finite temperature
  behavior},}\ }\href {\doibase 10.1103/PhysRevE.70.051101} {\bibfield
  {journal} {\bibinfo  {journal} {Phys. Rev. E}\ }\textbf {\bibinfo {volume}
  {70}},\ \bibinfo {pages} {051101} (\bibinfo {year} {2004})}\BibitemShut
  {NoStop}%
\bibitem [{\citenamefont {Lin}\ \emph {et~al.}(2014{\natexlab{a}})\citenamefont
  {Lin}, \citenamefont {Lerner}, \citenamefont {Rosso},\ and\ \citenamefont
  {Wyart}}]{Lin2014b}%
  \BibitemOpen
  \bibfield  {author} {\bibinfo {author} {\bibfnamefont {J.}~\bibnamefont
  {Lin}}, \bibinfo {author} {\bibfnamefont {E.}~\bibnamefont {Lerner}},
  \bibinfo {author} {\bibfnamefont {A.}~\bibnamefont {Rosso}}, \ and\ \bibinfo
  {author} {\bibfnamefont {M.}~\bibnamefont {Wyart}},\ }\bibfield  {title}
  {\enquote {\bibinfo {title} {Scaling description of the yielding transition
  in soft amorphous solids at zero temperature},}\ }\href {\doibase
  10.1073/pnas.1406391111} {\bibfield  {journal} {\bibinfo  {journal} {Proc.
  Natl. Acad. Sci.}\ }\textbf {\bibinfo {volume} {111}},\ \bibinfo {pages}
  {14382–14387} (\bibinfo {year} {2014}{\natexlab{a}})}\BibitemShut {NoStop}%
\bibitem [{\citenamefont {Rossi}\ \emph {et~al.}(2022)\citenamefont {Rossi},
  \citenamefont {Biroli}, \citenamefont {Ozawa}, \citenamefont {Tarjus},\ and\
  \citenamefont {Zamponi}}]{rossi2022finite}%
  \BibitemOpen
  \bibfield  {author} {\bibinfo {author} {\bibfnamefont {Saverio}\ \bibnamefont
  {Rossi}}, \bibinfo {author} {\bibfnamefont {Giulio}\ \bibnamefont {Biroli}},
  \bibinfo {author} {\bibfnamefont {Misaki}\ \bibnamefont {Ozawa}}, \bibinfo
  {author} {\bibfnamefont {Gilles}\ \bibnamefont {Tarjus}}, \ and\ \bibinfo
  {author} {\bibfnamefont {Francesco}\ \bibnamefont {Zamponi}},\ }\bibfield
  {title} {\enquote {\bibinfo {title} {Finite-disorder critical point in the
  yielding transition of elastoplastic models},}\ }\href@noop {} {\bibfield
  {journal} {\bibinfo  {journal} {Physical Review Letters}\ }\textbf {\bibinfo
  {volume} {129}},\ \bibinfo {pages} {228002} (\bibinfo {year}
  {2022})}\BibitemShut {NoStop}%
\bibitem [{\citenamefont {Berthier}\ \emph {et~al.}(2005)\citenamefont
  {Berthier}, \citenamefont {Biroli}, \citenamefont {Bouchaud}, \citenamefont
  {Cipelletti}, \citenamefont {Masri}, \citenamefont {L'Hôte}, \citenamefont
  {Ladieu},\ and\ \citenamefont {Pierno}}]{berthier2005direct}%
  \BibitemOpen
  \bibfield  {author} {\bibinfo {author} {\bibfnamefont {Ludovic}\ \bibnamefont
  {Berthier}}, \bibinfo {author} {\bibfnamefont {Giulio}\ \bibnamefont
  {Biroli}}, \bibinfo {author} {\bibfnamefont {J-P}\ \bibnamefont {Bouchaud}},
  \bibinfo {author} {\bibfnamefont {Luca}\ \bibnamefont {Cipelletti}}, \bibinfo
  {author} {\bibfnamefont {D~El}\ \bibnamefont {Masri}}, \bibinfo {author}
  {\bibfnamefont {Denis}\ \bibnamefont {L'Hôte}}, \bibinfo {author}
  {\bibfnamefont {Francois}\ \bibnamefont {Ladieu}}, \ and\ \bibinfo {author}
  {\bibfnamefont {Matteo}\ \bibnamefont {Pierno}},\ }\bibfield  {title}
  {\enquote {\bibinfo {title} {Direct experimental evidence of a growing length
  scale accompanying the glass transition},}\ }\href@noop {} {\bibfield
  {journal} {\bibinfo  {journal} {Science}\ }\textbf {\bibinfo {volume}
  {310}},\ \bibinfo {pages} {1797--1800} (\bibinfo {year} {2005})}\BibitemShut
  {NoStop}%
\bibitem [{\citenamefont {Popovi{\'c}}\ \emph {et~al.}(2021)\citenamefont
  {Popovi{\'c}}, \citenamefont {de~Geus}, \citenamefont {Ji},\ and\
  \citenamefont {Wyart}}]{popovic2021thermally}%
  \BibitemOpen
  \bibfield  {author} {\bibinfo {author} {\bibfnamefont {M.}~\bibnamefont
  {Popovi{\'c}}}, \bibinfo {author} {\bibfnamefont {T.W.J.}\ \bibnamefont
  {de~Geus}}, \bibinfo {author} {\bibfnamefont {W.}~\bibnamefont {Ji}}, \ and\
  \bibinfo {author} {\bibfnamefont {M.}~\bibnamefont {Wyart}},\ }\bibfield
  {title} {\enquote {\bibinfo {title} {Thermally activated flow in models of
  amorphous solids},}\ }\href@noop {} {\bibfield  {journal} {\bibinfo
  {journal} {Phys. Rev. E}\ }\textbf {\bibinfo {volume} {104}},\ \bibinfo
  {pages} {025010} (\bibinfo {year} {2021})}\BibitemShut {NoStop}%
\bibitem [{\citenamefont {Candelier}\ \emph {et~al.}(2010)\citenamefont
  {Candelier}, \citenamefont {Widmer-Cooper}, \citenamefont {Kummerfeld},
  \citenamefont {Dauchot}, \citenamefont {Biroli}, \citenamefont {Harrowell},\
  and\ \citenamefont {Reichman}}]{candelier2010spatiotemporal}%
  \BibitemOpen
  \bibfield  {author} {\bibinfo {author} {\bibfnamefont {Rapha{\"e}l}\
  \bibnamefont {Candelier}}, \bibinfo {author} {\bibfnamefont {Asaph}\
  \bibnamefont {Widmer-Cooper}}, \bibinfo {author} {\bibfnamefont {Jonathan~K}\
  \bibnamefont {Kummerfeld}}, \bibinfo {author} {\bibfnamefont {Olivier}\
  \bibnamefont {Dauchot}}, \bibinfo {author} {\bibfnamefont {Giulio}\
  \bibnamefont {Biroli}}, \bibinfo {author} {\bibfnamefont {Peter}\
  \bibnamefont {Harrowell}}, \ and\ \bibinfo {author} {\bibfnamefont {David~R}\
  \bibnamefont {Reichman}},\ }\bibfield  {title} {\enquote {\bibinfo {title}
  {Spatiotemporal hierarchy of relaxation events, dynamical heterogeneities,
  and structural reorganization in a supercooled liquid},}\ }\href@noop {}
  {\bibfield  {journal} {\bibinfo  {journal} {Physical review letters}\
  }\textbf {\bibinfo {volume} {105}},\ \bibinfo {pages} {135702} (\bibinfo
  {year} {2010})}\BibitemShut {NoStop}%
\bibitem [{\citenamefont {Keys}\ \emph {et~al.}(2011)\citenamefont {Keys},
  \citenamefont {Hedges}, \citenamefont {Garrahan}, \citenamefont {Glotzer},\
  and\ \citenamefont {Chandler}}]{keys2011excitations}%
  \BibitemOpen
  \bibfield  {author} {\bibinfo {author} {\bibfnamefont {Aaron~S}\ \bibnamefont
  {Keys}}, \bibinfo {author} {\bibfnamefont {Lester~O}\ \bibnamefont {Hedges}},
  \bibinfo {author} {\bibfnamefont {Juan~P}\ \bibnamefont {Garrahan}}, \bibinfo
  {author} {\bibfnamefont {Sharon~C}\ \bibnamefont {Glotzer}}, \ and\ \bibinfo
  {author} {\bibfnamefont {David}\ \bibnamefont {Chandler}},\ }\bibfield
  {title} {\enquote {\bibinfo {title} {Excitations are localized and relaxation
  is hierarchical in glass-forming liquids},}\ }\href@noop {} {\bibfield
  {journal} {\bibinfo  {journal} {Physical Review X}\ }\textbf {\bibinfo
  {volume} {1}},\ \bibinfo {pages} {021013} (\bibinfo {year}
  {2011})}\BibitemShut {NoStop}%
\bibitem [{\citenamefont {Yanagishima}\ \emph {et~al.}(2017)\citenamefont
  {Yanagishima}, \citenamefont {Russo},\ and\ \citenamefont
  {Tanaka}}]{yanagishima2017common}%
  \BibitemOpen
  \bibfield  {author} {\bibinfo {author} {\bibfnamefont {Taiki}\ \bibnamefont
  {Yanagishima}}, \bibinfo {author} {\bibfnamefont {John}\ \bibnamefont
  {Russo}}, \ and\ \bibinfo {author} {\bibfnamefont {Hajime}\ \bibnamefont
  {Tanaka}},\ }\bibfield  {title} {\enquote {\bibinfo {title} {Common mechanism
  of thermodynamic and mechanical origin for ageing and crystallization of
  glasses},}\ }\href@noop {} {\bibfield  {journal} {\bibinfo  {journal} {Nature
  communications}\ }\textbf {\bibinfo {volume} {8}},\ \bibinfo {pages} {1--10}
  (\bibinfo {year} {2017})}\BibitemShut {NoStop}%
\bibitem [{\citenamefont {Sethna}\ \emph {et~al.}(2001)\citenamefont {Sethna},
  \citenamefont {Dahmen},\ and\ \citenamefont {Myers}}]{sethna2001crackling}%
  \BibitemOpen
  \bibfield  {author} {\bibinfo {author} {\bibfnamefont {James~P}\ \bibnamefont
  {Sethna}}, \bibinfo {author} {\bibfnamefont {Karin~A}\ \bibnamefont
  {Dahmen}}, \ and\ \bibinfo {author} {\bibfnamefont {Christopher~R}\
  \bibnamefont {Myers}},\ }\bibfield  {title} {\enquote {\bibinfo {title}
  {Crackling noise},}\ }\href@noop {} {\bibfield  {journal} {\bibinfo
  {journal} {Nature}\ }\textbf {\bibinfo {volume} {410}},\ \bibinfo {pages}
  {242--250} (\bibinfo {year} {2001})}\BibitemShut {NoStop}%
\bibitem [{\citenamefont {Rosso}\ \emph {et~al.}(2022)\citenamefont {Rosso},
  \citenamefont {Sethna},\ and\ \citenamefont {Wyart}}]{rosso2022avalanches}%
  \BibitemOpen
  \bibfield  {author} {\bibinfo {author} {\bibfnamefont {Alberto}\ \bibnamefont
  {Rosso}}, \bibinfo {author} {\bibfnamefont {James~P}\ \bibnamefont {Sethna}},
  \ and\ \bibinfo {author} {\bibfnamefont {Matthieu}\ \bibnamefont {Wyart}},\
  }\bibfield  {title} {\enquote {\bibinfo {title} {Avalanches and deformation
  in glasses and disordered systems},}\ }\href@noop {} {\bibfield  {journal}
  {\bibinfo  {journal} {arXiv preprint arXiv:2208.04090}\ } (\bibinfo {year}
  {2022})}\BibitemShut {NoStop}%
\bibitem [{\citenamefont {Purrello}\ \emph {et~al.}(2017)\citenamefont
  {Purrello}, \citenamefont {Iguain}, \citenamefont {Kolton},\ and\
  \citenamefont {Jagla}}]{purrello2017creep}%
  \BibitemOpen
  \bibfield  {author} {\bibinfo {author} {\bibfnamefont {V{\'\i}ctor~Hugo}\
  \bibnamefont {Purrello}}, \bibinfo {author} {\bibfnamefont {Jose~Luis}\
  \bibnamefont {Iguain}}, \bibinfo {author} {\bibfnamefont
  {Alejandro~Benedykt}\ \bibnamefont {Kolton}}, \ and\ \bibinfo {author}
  {\bibfnamefont {Eduardo~Alberto}\ \bibnamefont {Jagla}},\ }\bibfield  {title}
  {\enquote {\bibinfo {title} {Creep and thermal rounding close to the elastic
  depinning threshold},}\ }\href@noop {} {\bibfield  {journal} {\bibinfo
  {journal} {Physical Review E}\ }\textbf {\bibinfo {volume} {96}},\ \bibinfo
  {pages} {022112} (\bibinfo {year} {2017})}\BibitemShut {NoStop}%
\bibitem [{\citenamefont {Yao}\ and\ \citenamefont
  {Jack}(2023)}]{yao2023thermal}%
  \BibitemOpen
  \bibfield  {author} {\bibinfo {author} {\bibfnamefont {Liheng}\ \bibnamefont
  {Yao}}\ and\ \bibinfo {author} {\bibfnamefont {Robert~L}\ \bibnamefont
  {Jack}},\ }\bibfield  {title} {\enquote {\bibinfo {title} {Thermal vestiges
  of avalanches in the driven random field ising model},}\ }\href@noop {}
  {\bibfield  {journal} {\bibinfo  {journal} {Journal of Statistical Mechanics:
  Theory and Experiment}\ }\textbf {\bibinfo {volume} {2023}},\ \bibinfo
  {pages} {023303} (\bibinfo {year} {2023})}\BibitemShut {NoStop}%
\bibitem [{\citenamefont {Korchinski}\ and\ \citenamefont
  {Rottler}(2022)}]{korchinski2022dynamic}%
  \BibitemOpen
  \bibfield  {author} {\bibinfo {author} {\bibfnamefont {Daniel}\ \bibnamefont
  {Korchinski}}\ and\ \bibinfo {author} {\bibfnamefont {J{\"o}rg}\ \bibnamefont
  {Rottler}},\ }\bibfield  {title} {\enquote {\bibinfo {title} {Dynamic phase
  diagram of plastically deformed amorphous solids at finite temperature},}\
  }\href@noop {} {\bibfield  {journal} {\bibinfo  {journal} {Physical Review
  E}\ }\textbf {\bibinfo {volume} {106}},\ \bibinfo {pages} {034103} (\bibinfo
  {year} {2022})}\BibitemShut {NoStop}%
\bibitem [{\citenamefont {Scalliet}\ \emph {et~al.}(2021)\citenamefont
  {Scalliet}, \citenamefont {Guiselin},\ and\ \citenamefont
  {Berthier}}]{scalliet2021excess}%
  \BibitemOpen
  \bibfield  {author} {\bibinfo {author} {\bibfnamefont {Camille}\ \bibnamefont
  {Scalliet}}, \bibinfo {author} {\bibfnamefont {Benjamin}\ \bibnamefont
  {Guiselin}}, \ and\ \bibinfo {author} {\bibfnamefont {Ludovic}\ \bibnamefont
  {Berthier}},\ }\bibfield  {title} {\enquote {\bibinfo {title} {Excess wings
  and asymmetric relaxation spectra in a facilitated trap model},}\ }\href@noop
  {} {\bibfield  {journal} {\bibinfo  {journal} {The Journal of Chemical
  Physics}\ }\textbf {\bibinfo {volume} {155}},\ \bibinfo {pages} {064505}
  (\bibinfo {year} {2021})}\BibitemShut {NoStop}%
\bibitem [{\citenamefont {Scalliet}\ \emph {et~al.}(2022)\citenamefont
  {Scalliet}, \citenamefont {Guiselin},\ and\ \citenamefont
  {Berthier}}]{scalliet2022thirty}%
  \BibitemOpen
  \bibfield  {author} {\bibinfo {author} {\bibfnamefont {Camille}\ \bibnamefont
  {Scalliet}}, \bibinfo {author} {\bibfnamefont {Benjamin}\ \bibnamefont
  {Guiselin}}, \ and\ \bibinfo {author} {\bibfnamefont {Ludovic}\ \bibnamefont
  {Berthier}},\ }\bibfield  {title} {\enquote {\bibinfo {title} {Thirty
  milliseconds in the life of a supercooled liquid},}\ }\href@noop {}
  {\bibfield  {journal} {\bibinfo  {journal} {Physical Review X}\ }\textbf
  {\bibinfo {volume} {12}},\ \bibinfo {pages} {041028} (\bibinfo {year}
  {2022})}\BibitemShut {NoStop}%
\bibitem [{\citenamefont {Tarjus}\ and\ \citenamefont
  {Kivelson}(1995)}]{tarjus1995breakdown}%
  \BibitemOpen
  \bibfield  {author} {\bibinfo {author} {\bibfnamefont {Gilles}\ \bibnamefont
  {Tarjus}}\ and\ \bibinfo {author} {\bibfnamefont {Daniel}\ \bibnamefont
  {Kivelson}},\ }\bibfield  {title} {\enquote {\bibinfo {title} {Breakdown of
  the stokes--einstein relation in supercooled liquids},}\ }\href@noop {}
  {\bibfield  {journal} {\bibinfo  {journal} {The Journal of chemical physics}\
  }\textbf {\bibinfo {volume} {103}},\ \bibinfo {pages} {3071--3073} (\bibinfo
  {year} {1995})}\BibitemShut {NoStop}%
\bibitem [{\citenamefont {Ediger}(2000)}]{ediger2000spatially}%
  \BibitemOpen
  \bibfield  {author} {\bibinfo {author} {\bibfnamefont {Mark~D}\ \bibnamefont
  {Ediger}},\ }\bibfield  {title} {\enquote {\bibinfo {title} {Spatially
  heterogeneous dynamics in supercooled liquids},}\ }\href@noop {} {\bibfield
  {journal} {\bibinfo  {journal} {Annual review of physical chemistry}\
  }\textbf {\bibinfo {volume} {51}},\ \bibinfo {pages} {99--128} (\bibinfo
  {year} {2000})}\BibitemShut {NoStop}%
\bibitem [{\citenamefont {Sengupta}\ \emph {et~al.}(2013)\citenamefont
  {Sengupta}, \citenamefont {Karmakar}, \citenamefont {Dasgupta},\ and\
  \citenamefont {Sastry}}]{sengupta2013breakdown}%
  \BibitemOpen
  \bibfield  {author} {\bibinfo {author} {\bibfnamefont {Shiladitya}\
  \bibnamefont {Sengupta}}, \bibinfo {author} {\bibfnamefont {Smarajit}\
  \bibnamefont {Karmakar}}, \bibinfo {author} {\bibfnamefont {Chandan}\
  \bibnamefont {Dasgupta}}, \ and\ \bibinfo {author} {\bibfnamefont {Srikanth}\
  \bibnamefont {Sastry}},\ }\bibfield  {title} {\enquote {\bibinfo {title}
  {Breakdown of the stokes-einstein relation in two, three, and four
  dimensions},}\ }\href@noop {} {\bibfield  {journal} {\bibinfo  {journal} {The
  Journal of chemical physics}\ }\textbf {\bibinfo {volume} {138}},\ \bibinfo
  {pages} {12A548} (\bibinfo {year} {2013})}\BibitemShut {NoStop}%
\bibitem [{\citenamefont {Charbonneau}\ \emph {et~al.}(2014)\citenamefont
  {Charbonneau}, \citenamefont {Jin}, \citenamefont {Parisi},\ and\
  \citenamefont {Zamponi}}]{charbonneau2014hopping}%
  \BibitemOpen
  \bibfield  {author} {\bibinfo {author} {\bibfnamefont {Patrick}\ \bibnamefont
  {Charbonneau}}, \bibinfo {author} {\bibfnamefont {Yuliang}\ \bibnamefont
  {Jin}}, \bibinfo {author} {\bibfnamefont {Giorgio}\ \bibnamefont {Parisi}}, \
  and\ \bibinfo {author} {\bibfnamefont {Francesco}\ \bibnamefont {Zamponi}},\
  }\bibfield  {title} {\enquote {\bibinfo {title} {Hopping and the
  stokes--einstein relation breakdown in simple glass formers},}\ }\href@noop
  {} {\bibfield  {journal} {\bibinfo  {journal} {Proceedings of the National
  Academy of Sciences}\ }\textbf {\bibinfo {volume} {111}},\ \bibinfo {pages}
  {15025--15030} (\bibinfo {year} {2014})}\BibitemShut {NoStop}%
\bibitem [{\citenamefont {Kawasaki}\ and\ \citenamefont
  {Kim}(2017)}]{kawasaki2017identifying}%
  \BibitemOpen
  \bibfield  {author} {\bibinfo {author} {\bibfnamefont {Takeshi}\ \bibnamefont
  {Kawasaki}}\ and\ \bibinfo {author} {\bibfnamefont {Kang}\ \bibnamefont
  {Kim}},\ }\bibfield  {title} {\enquote {\bibinfo {title} {Identifying time
  scales for violation/preservation of stokes-einstein relation in supercooled
  water},}\ }\href@noop {} {\bibfield  {journal} {\bibinfo  {journal} {Science
  Advances}\ }\textbf {\bibinfo {volume} {3}},\ \bibinfo {pages} {e1700399}
  (\bibinfo {year} {2017})}\BibitemShut {NoStop}%
\bibitem [{\citenamefont {Jung}\ \emph {et~al.}(2004)\citenamefont {Jung},
  \citenamefont {Garrahan},\ and\ \citenamefont
  {Chandler}}]{jung2004excitation}%
  \BibitemOpen
  \bibfield  {author} {\bibinfo {author} {\bibfnamefont {YounJoon}\
  \bibnamefont {Jung}}, \bibinfo {author} {\bibfnamefont {Juan~P}\ \bibnamefont
  {Garrahan}}, \ and\ \bibinfo {author} {\bibfnamefont {David}\ \bibnamefont
  {Chandler}},\ }\bibfield  {title} {\enquote {\bibinfo {title} {Excitation
  lines and the breakdown of stokes-einstein relations in supercooled
  liquids},}\ }\href@noop {} {\bibfield  {journal} {\bibinfo  {journal}
  {Physical Review E}\ }\textbf {\bibinfo {volume} {69}},\ \bibinfo {pages}
  {061205} (\bibinfo {year} {2004})}\BibitemShut {NoStop}%
\bibitem [{\citenamefont {Berthier}\ \emph {et~al.}(2004)\citenamefont
  {Berthier}, \citenamefont {Chandler},\ and\ \citenamefont
  {Garrahan}}]{berthier2004length}%
  \BibitemOpen
  \bibfield  {author} {\bibinfo {author} {\bibfnamefont {Ludovic}\ \bibnamefont
  {Berthier}}, \bibinfo {author} {\bibfnamefont {David}\ \bibnamefont
  {Chandler}}, \ and\ \bibinfo {author} {\bibfnamefont {Juan~P}\ \bibnamefont
  {Garrahan}},\ }\bibfield  {title} {\enquote {\bibinfo {title} {Length scale
  for the onset of fickian diffusion in supercooled liquids},}\ }\href@noop {}
  {\bibfield  {journal} {\bibinfo  {journal} {Europhysics Letters}\ }\textbf
  {\bibinfo {volume} {69}},\ \bibinfo {pages} {320} (\bibinfo {year}
  {2004})}\BibitemShut {NoStop}%
\bibitem [{\citenamefont {Hedges}\ \emph {et~al.}(2007)\citenamefont {Hedges},
  \citenamefont {Maibaum}, \citenamefont {Chandler},\ and\ \citenamefont
  {Garrahan}}]{hedges2007decoupling}%
  \BibitemOpen
  \bibfield  {author} {\bibinfo {author} {\bibfnamefont {Lester~O}\
  \bibnamefont {Hedges}}, \bibinfo {author} {\bibfnamefont {Lutz}\ \bibnamefont
  {Maibaum}}, \bibinfo {author} {\bibfnamefont {David}\ \bibnamefont
  {Chandler}}, \ and\ \bibinfo {author} {\bibfnamefont {Juan~P}\ \bibnamefont
  {Garrahan}},\ }\bibfield  {title} {\enquote {\bibinfo {title} {Decoupling of
  exchange and persistence times in atomistic models of glass formers},}\
  }\href@noop {} {\bibfield  {journal} {\bibinfo  {journal} {The Journal of
  chemical physics}\ }\textbf {\bibinfo {volume} {127}},\ \bibinfo {pages}
  {211101} (\bibinfo {year} {2007})}\BibitemShut {NoStop}%
\bibitem [{\citenamefont {Chaudhuri}\ \emph {et~al.}(2007)\citenamefont
  {Chaudhuri}, \citenamefont {Berthier},\ and\ \citenamefont
  {Kob}}]{chaudhuri2007universal}%
  \BibitemOpen
  \bibfield  {author} {\bibinfo {author} {\bibfnamefont {Pinaki}\ \bibnamefont
  {Chaudhuri}}, \bibinfo {author} {\bibfnamefont {Ludovic}\ \bibnamefont
  {Berthier}}, \ and\ \bibinfo {author} {\bibfnamefont {Walter}\ \bibnamefont
  {Kob}},\ }\bibfield  {title} {\enquote {\bibinfo {title} {Universal nature of
  particle displacements close to glass and jamming transitions},}\ }\href@noop
  {} {\bibfield  {journal} {\bibinfo  {journal} {Physical review letters}\
  }\textbf {\bibinfo {volume} {99}},\ \bibinfo {pages} {060604} (\bibinfo
  {year} {2007})}\BibitemShut {NoStop}%
\bibitem [{\citenamefont {Pastore}\ \emph {et~al.}(2021)\citenamefont
  {Pastore}, \citenamefont {Kikutsuji}, \citenamefont {Rusciano}, \citenamefont
  {Matubayasi}, \citenamefont {Kim},\ and\ \citenamefont
  {Greco}}]{pastore2021breakdown}%
  \BibitemOpen
  \bibfield  {author} {\bibinfo {author} {\bibfnamefont {Raffaele}\
  \bibnamefont {Pastore}}, \bibinfo {author} {\bibfnamefont {Takuma}\
  \bibnamefont {Kikutsuji}}, \bibinfo {author} {\bibfnamefont {Francesco}\
  \bibnamefont {Rusciano}}, \bibinfo {author} {\bibfnamefont {Nobuyuki}\
  \bibnamefont {Matubayasi}}, \bibinfo {author} {\bibfnamefont {Kang}\
  \bibnamefont {Kim}}, \ and\ \bibinfo {author} {\bibfnamefont {Francesco}\
  \bibnamefont {Greco}},\ }\bibfield  {title} {\enquote {\bibinfo {title}
  {Breakdown of the stokes--einstein relation in supercooled liquids: A
  cage-jump perspective},}\ }\href@noop {} {\bibfield  {journal} {\bibinfo
  {journal} {The Journal of Chemical Physics}\ }\textbf {\bibinfo {volume}
  {155}},\ \bibinfo {pages} {114503} (\bibinfo {year} {2021})}\BibitemShut
  {NoStop}%
\bibitem [{\citenamefont {Bulatov}\ and\ \citenamefont
  {Argon}(1994)}]{bulatov1994stochastic}%
  \BibitemOpen
  \bibfield  {author} {\bibinfo {author} {\bibfnamefont {VV}~\bibnamefont
  {Bulatov}}\ and\ \bibinfo {author} {\bibfnamefont {AS}~\bibnamefont
  {Argon}},\ }\bibfield  {title} {\enquote {\bibinfo {title} {A stochastic
  model for continuum elasto-plastic behavior. ii. a study of the glass
  transition and structural relaxation},}\ }\href@noop {} {\bibfield  {journal}
  {\bibinfo  {journal} {Modelling and Simulation in Materials Science and
  Engineering}\ }\textbf {\bibinfo {volume} {2}},\ \bibinfo {pages} {185}
  (\bibinfo {year} {1994})}\BibitemShut {NoStop}%
\bibitem [{\citenamefont {Ferrero}\ \emph {et~al.}(2014)\citenamefont
  {Ferrero}, \citenamefont {Martens},\ and\ \citenamefont
  {Barrat}}]{ferrero2014relaxation}%
  \BibitemOpen
  \bibfield  {author} {\bibinfo {author} {\bibfnamefont {Ezequiel~E}\
  \bibnamefont {Ferrero}}, \bibinfo {author} {\bibfnamefont {Kirsten}\
  \bibnamefont {Martens}}, \ and\ \bibinfo {author} {\bibfnamefont
  {Jean-Louis}\ \bibnamefont {Barrat}},\ }\bibfield  {title} {\enquote
  {\bibinfo {title} {Relaxation in yield stress systems through elastically
  interacting activated events},}\ }\href@noop {} {\bibfield  {journal}
  {\bibinfo  {journal} {Physical review letters}\ }\textbf {\bibinfo {volume}
  {113}},\ \bibinfo {pages} {248301} (\bibinfo {year} {2014})}\BibitemShut
  {NoStop}%
\bibitem [{\citenamefont {Jagla}(2020)}]{jagla2020tensorial}%
  \BibitemOpen
  \bibfield  {author} {\bibinfo {author} {\bibfnamefont {Eduardo~Alberto}\
  \bibnamefont {Jagla}},\ }\bibfield  {title} {\enquote {\bibinfo {title}
  {Tensorial description of the plasticity of amorphous composites},}\
  }\href@noop {} {\bibfield  {journal} {\bibinfo  {journal} {Physical Review
  E}\ }\textbf {\bibinfo {volume} {101}},\ \bibinfo {pages} {043004} (\bibinfo
  {year} {2020})}\BibitemShut {NoStop}%
\bibitem [{\citenamefont {Ferrero}\ \emph {et~al.}(2021)\citenamefont
  {Ferrero}, \citenamefont {Kolton},\ and\ \citenamefont
  {Jagla}}]{ferrero2021yielding}%
  \BibitemOpen
  \bibfield  {author} {\bibinfo {author} {\bibfnamefont {Ezequiel~E}\
  \bibnamefont {Ferrero}}, \bibinfo {author} {\bibfnamefont {Alejandro~B}\
  \bibnamefont {Kolton}}, \ and\ \bibinfo {author} {\bibfnamefont {Eduardo~A}\
  \bibnamefont {Jagla}},\ }\bibfield  {title} {\enquote {\bibinfo {title}
  {Yielding of amorphous solids at finite temperatures},}\ }\href@noop {}
  {\bibfield  {journal} {\bibinfo  {journal} {Physical Review Materials}\
  }\textbf {\bibinfo {volume} {5}},\ \bibinfo {pages} {115602} (\bibinfo {year}
  {2021})}\BibitemShut {NoStop}%
\bibitem [{\citenamefont {Rodriguez-Lopez}\ \emph {et~al.}(2023)\citenamefont
  {Rodriguez-Lopez}, \citenamefont {Martens},\ and\ \citenamefont
  {Ferrero}}]{rodriguez2023temperature}%
  \BibitemOpen
  \bibfield  {author} {\bibinfo {author} {\bibfnamefont {Gieberth}\
  \bibnamefont {Rodriguez-Lopez}}, \bibinfo {author} {\bibfnamefont {Kirsten}\
  \bibnamefont {Martens}}, \ and\ \bibinfo {author} {\bibfnamefont
  {Ezequiel~E}\ \bibnamefont {Ferrero}},\ }\bibfield  {title} {\enquote
  {\bibinfo {title} {Temperature dependence of fast relaxation processes in
  amorphous materials},}\ }\href@noop {} {\bibfield  {journal} {\bibinfo
  {journal} {arXiv preprint arXiv:2302.06471}\ } (\bibinfo {year}
  {2023})}\BibitemShut {NoStop}%
\bibitem [{\citenamefont {Fan}\ \emph {et~al.}(2014)\citenamefont {Fan},
  \citenamefont {Iwashita},\ and\ \citenamefont {Egami}}]{fan2014thermally}%
  \BibitemOpen
  \bibfield  {author} {\bibinfo {author} {\bibfnamefont {Yue}\ \bibnamefont
  {Fan}}, \bibinfo {author} {\bibfnamefont {Takuya}\ \bibnamefont {Iwashita}},
  \ and\ \bibinfo {author} {\bibfnamefont {Takeshi}\ \bibnamefont {Egami}},\
  }\bibfield  {title} {\enquote {\bibinfo {title} {How thermally activated
  deformation starts in metallic glass},}\ }\href@noop {} {\bibfield  {journal}
  {\bibinfo  {journal} {Nature communications}\ }\textbf {\bibinfo {volume}
  {5}},\ \bibinfo {pages} {5083} (\bibinfo {year} {2014})}\BibitemShut
  {NoStop}%
\bibitem [{\citenamefont {Aguirre}\ and\ \citenamefont
  {Jagla}(2018)}]{aguirre2018critical}%
  \BibitemOpen
  \bibfield  {author} {\bibinfo {author} {\bibfnamefont {I~Fern{\'a}ndez}\
  \bibnamefont {Aguirre}}\ and\ \bibinfo {author} {\bibfnamefont
  {Eduardo~Alberto}\ \bibnamefont {Jagla}},\ }\bibfield  {title} {\enquote
  {\bibinfo {title} {Critical exponents of the yielding transition of amorphous
  solids},}\ }\href@noop {} {\bibfield  {journal} {\bibinfo  {journal}
  {Physical Review E}\ }\textbf {\bibinfo {volume} {98}},\ \bibinfo {pages}
  {013002} (\bibinfo {year} {2018})}\BibitemShut {NoStop}%
\bibitem [{\citenamefont {Maloney}\ and\ \citenamefont
  {Lemaitre}(2006)}]{maloney2006amorphous}%
  \BibitemOpen
  \bibfield  {author} {\bibinfo {author} {\bibfnamefont {Craig~E}\ \bibnamefont
  {Maloney}}\ and\ \bibinfo {author} {\bibfnamefont {Anael}\ \bibnamefont
  {Lemaitre}},\ }\bibfield  {title} {\enquote {\bibinfo {title} {Amorphous
  systems in athermal, quasistatic shear},}\ }\href@noop {} {\bibfield
  {journal} {\bibinfo  {journal} {Physical Review E}\ }\textbf {\bibinfo
  {volume} {74}},\ \bibinfo {pages} {016118} (\bibinfo {year}
  {2006})}\BibitemShut {NoStop}%
\bibitem [{Note1()}]{Note1}%
  \BibitemOpen
  \bibinfo {note} {Full isotropy is slightly broken by the choice of
  bi-periodic boundary conditions.}\BibitemShut {Stop}%
\bibitem [{\citenamefont {Donati}\ \emph {et~al.}(2002)\citenamefont {Donati},
  \citenamefont {Franz}, \citenamefont {Glotzer},\ and\ \citenamefont
  {Parisi}}]{donati2002theory}%
  \BibitemOpen
  \bibfield  {author} {\bibinfo {author} {\bibfnamefont {Claudio}\ \bibnamefont
  {Donati}}, \bibinfo {author} {\bibfnamefont {Silvio}\ \bibnamefont {Franz}},
  \bibinfo {author} {\bibfnamefont {Sharon~C}\ \bibnamefont {Glotzer}}, \ and\
  \bibinfo {author} {\bibfnamefont {Giorgio}\ \bibnamefont {Parisi}},\
  }\bibfield  {title} {\enquote {\bibinfo {title} {Theory of non-linear
  susceptibility and correlation length in glasses and liquids},}\ }\href@noop
  {} {\bibfield  {journal} {\bibinfo  {journal} {Journal of non-crystalline
  solids}\ }\textbf {\bibinfo {volume} {307}},\ \bibinfo {pages} {215--224}
  (\bibinfo {year} {2002})}\BibitemShut {NoStop}%
\bibitem [{\citenamefont {Capaccioli}\ \emph {et~al.}(2008)\citenamefont
  {Capaccioli}, \citenamefont {Ruocco},\ and\ \citenamefont
  {Zamponi}}]{capaccioli2008dynamically}%
  \BibitemOpen
  \bibfield  {author} {\bibinfo {author} {\bibfnamefont {Simone}\ \bibnamefont
  {Capaccioli}}, \bibinfo {author} {\bibfnamefont {Giancarlo}\ \bibnamefont
  {Ruocco}}, \ and\ \bibinfo {author} {\bibfnamefont {Francesco}\ \bibnamefont
  {Zamponi}},\ }\bibfield  {title} {\enquote {\bibinfo {title} {Dynamically
  correlated regions and configurational entropy in supercooled liquids},}\
  }\href@noop {} {\bibfield  {journal} {\bibinfo  {journal} {The Journal of
  Physical Chemistry B}\ }\textbf {\bibinfo {volume} {112}},\ \bibinfo {pages}
  {10652--10658} (\bibinfo {year} {2008})}\BibitemShut {NoStop}%
\bibitem [{\citenamefont {Dauchot}\ \emph {et~al.}(2022)\citenamefont
  {Dauchot}, \citenamefont {Ladieu},\ and\ \citenamefont
  {Royall}}]{dauchot2022glass}%
  \BibitemOpen
  \bibfield  {author} {\bibinfo {author} {\bibfnamefont {Olivier}\ \bibnamefont
  {Dauchot}}, \bibinfo {author} {\bibfnamefont {Fran{\c{c}}ois}\ \bibnamefont
  {Ladieu}}, \ and\ \bibinfo {author} {\bibfnamefont {C~Patrick}\ \bibnamefont
  {Royall}},\ }\bibfield  {title} {\enquote {\bibinfo {title} {The glass
  transition in molecules, colloids and grains: universality and
  specificity},}\ }\href@noop {} {\bibfield  {journal} {\bibinfo  {journal}
  {arXiv preprint arXiv:2211.03158}\ } (\bibinfo {year} {2022})}\BibitemShut
  {NoStop}%
\bibitem [{\citenamefont {Karmakar}\ \emph {et~al.}(2009)\citenamefont
  {Karmakar}, \citenamefont {Dasgupta},\ and\ \citenamefont
  {Sastry}}]{karmakar2009growing}%
  \BibitemOpen
  \bibfield  {author} {\bibinfo {author} {\bibfnamefont {Smarajit}\
  \bibnamefont {Karmakar}}, \bibinfo {author} {\bibfnamefont {Chandan}\
  \bibnamefont {Dasgupta}}, \ and\ \bibinfo {author} {\bibfnamefont {Srikanth}\
  \bibnamefont {Sastry}},\ }\bibfield  {title} {\enquote {\bibinfo {title}
  {Growing length and time scales in glass-forming liquids},}\ }\href@noop {}
  {\bibfield  {journal} {\bibinfo  {journal} {Proceedings of the National
  Academy of Sciences}\ }\textbf {\bibinfo {volume} {106}},\ \bibinfo {pages}
  {3675--3679} (\bibinfo {year} {2009})}\BibitemShut {NoStop}%
\bibitem [{\citenamefont {Coslovich}\ \emph {et~al.}(2018)\citenamefont
  {Coslovich}, \citenamefont {Ozawa},\ and\ \citenamefont
  {Kob}}]{coslovich2018dynamic}%
  \BibitemOpen
  \bibfield  {author} {\bibinfo {author} {\bibfnamefont {Daniele}\ \bibnamefont
  {Coslovich}}, \bibinfo {author} {\bibfnamefont {Misaki}\ \bibnamefont
  {Ozawa}}, \ and\ \bibinfo {author} {\bibfnamefont {Walter}\ \bibnamefont
  {Kob}},\ }\bibfield  {title} {\enquote {\bibinfo {title} {Dynamic and
  thermodynamic crossover scenarios in the kob-andersen mixture: Insights from
  multi-cpu and multi-gpu simulations},}\ }\href@noop {} {\bibfield  {journal}
  {\bibinfo  {journal} {The European Physical Journal E}\ }\textbf {\bibinfo
  {volume} {41}},\ \bibinfo {pages} {1--11} (\bibinfo {year}
  {2018})}\BibitemShut {NoStop}%
\bibitem [{\citenamefont {Chakrabarty}\ \emph {et~al.}(2017)\citenamefont
  {Chakrabarty}, \citenamefont {Tah}, \citenamefont {Karmakar},\ and\
  \citenamefont {Dasgupta}}]{chakrabarty2017block}%
  \BibitemOpen
  \bibfield  {author} {\bibinfo {author} {\bibfnamefont {Saurish}\ \bibnamefont
  {Chakrabarty}}, \bibinfo {author} {\bibfnamefont {Indrajit}\ \bibnamefont
  {Tah}}, \bibinfo {author} {\bibfnamefont {Smarajit}\ \bibnamefont
  {Karmakar}}, \ and\ \bibinfo {author} {\bibfnamefont {Chandan}\ \bibnamefont
  {Dasgupta}},\ }\bibfield  {title} {\enquote {\bibinfo {title} {Block analysis
  for the calculation of dynamic and static length scales in glass-forming
  liquids},}\ }\href@noop {} {\bibfield  {journal} {\bibinfo  {journal}
  {Physical Review Letters}\ }\textbf {\bibinfo {volume} {119}},\ \bibinfo
  {pages} {205502} (\bibinfo {year} {2017})}\BibitemShut {NoStop}%
\bibitem [{\citenamefont {La{\v{c}}evi{\'c}}\ \emph {et~al.}(2003)\citenamefont
  {La{\v{c}}evi{\'c}}, \citenamefont {Starr}, \citenamefont {Schr{\o}der},\
  and\ \citenamefont {Glotzer}}]{lavcevic2003spatially}%
  \BibitemOpen
  \bibfield  {author} {\bibinfo {author} {\bibfnamefont {N}~\bibnamefont
  {La{\v{c}}evi{\'c}}}, \bibinfo {author} {\bibfnamefont {Francis~W}\
  \bibnamefont {Starr}}, \bibinfo {author} {\bibfnamefont {TB}~\bibnamefont
  {Schr{\o}der}}, \ and\ \bibinfo {author} {\bibfnamefont {Sharon~C}\
  \bibnamefont {Glotzer}},\ }\bibfield  {title} {\enquote {\bibinfo {title}
  {Spatially heterogeneous dynamics investigated via a time-dependent
  four-point density correlation function},}\ }\href@noop {} {\bibfield
  {journal} {\bibinfo  {journal} {The Journal of chemical physics}\ }\textbf
  {\bibinfo {volume} {119}},\ \bibinfo {pages} {7372--7387} (\bibinfo {year}
  {2003})}\BibitemShut {NoStop}%
\bibitem [{Note2()}]{Note2}%
  \BibitemOpen
  \bibinfo {note} {We are considering either finite systems or the
  thermodynamic limit taken after the zero-temperature limit.}\BibitemShut
  {Stop}%
\bibitem [{\citenamefont {Paczuski}\ \emph {et~al.}(1996)\citenamefont
  {Paczuski}, \citenamefont {Maslov},\ and\ \citenamefont
  {Bak}}]{paczuski1996avalanche}%
  \BibitemOpen
  \bibfield  {author} {\bibinfo {author} {\bibfnamefont {Maya}\ \bibnamefont
  {Paczuski}}, \bibinfo {author} {\bibfnamefont {Sergei}\ \bibnamefont
  {Maslov}}, \ and\ \bibinfo {author} {\bibfnamefont {Per}\ \bibnamefont
  {Bak}},\ }\bibfield  {title} {\enquote {\bibinfo {title} {Avalanche dynamics
  in evolution, growth, and depinning models},}\ }\href@noop {} {\bibfield
  {journal} {\bibinfo  {journal} {Physical Review E}\ }\textbf {\bibinfo
  {volume} {53}},\ \bibinfo {pages} {414} (\bibinfo {year} {1996})}\BibitemShut
  {NoStop}%
\bibitem [{\citenamefont {Baret}\ \emph {et~al.}(2002)\citenamefont {Baret},
  \citenamefont {Vandembroucq},\ and\ \citenamefont {Roux}}]{Baret2002}%
  \BibitemOpen
  \bibfield  {author} {\bibinfo {author} {\bibfnamefont {J.-C.}\ \bibnamefont
  {Baret}}, \bibinfo {author} {\bibfnamefont {D.}~\bibnamefont {Vandembroucq}},
  \ and\ \bibinfo {author} {\bibfnamefont {S.}~\bibnamefont {Roux}},\
  }\bibfield  {title} {\enquote {\bibinfo {title} {Extremal model for amorphous
  media plasticity},}\ }\href {\doibase 10.1103/PhysRevLett.89.195506}
  {\bibfield  {journal} {\bibinfo  {journal} {Phys. Rev. Lett.}\ }\textbf
  {\bibinfo {volume} {89}},\ \bibinfo {pages} {195506} (\bibinfo {year}
  {2002})}\BibitemShut {NoStop}%
\bibitem [{\citenamefont {Kumar}\ \emph {et~al.}(2022)\citenamefont {Kumar},
  \citenamefont {Patinet}, \citenamefont {Maloney}, \citenamefont {Regev},
  \citenamefont {Vandembroucq},\ and\ \citenamefont
  {Mungan}}]{kumar2022mapping}%
  \BibitemOpen
  \bibfield  {author} {\bibinfo {author} {\bibfnamefont {Dheeraj}\ \bibnamefont
  {Kumar}}, \bibinfo {author} {\bibfnamefont {Sylvain}\ \bibnamefont
  {Patinet}}, \bibinfo {author} {\bibfnamefont {Craig~E}\ \bibnamefont
  {Maloney}}, \bibinfo {author} {\bibfnamefont {Ido}\ \bibnamefont {Regev}},
  \bibinfo {author} {\bibfnamefont {Damien}\ \bibnamefont {Vandembroucq}}, \
  and\ \bibinfo {author} {\bibfnamefont {Muhittin}\ \bibnamefont {Mungan}},\
  }\bibfield  {title} {\enquote {\bibinfo {title} {Mapping out the glassy
  landscape of a mesoscopic elastoplastic model},}\ }\href@noop {} {\bibfield
  {journal} {\bibinfo  {journal} {The Journal of Chemical Physics}\ }\textbf
  {\bibinfo {volume} {157}},\ \bibinfo {pages} {174504} (\bibinfo {year}
  {2022})}\BibitemShut {NoStop}%
\bibitem [{\citenamefont {Han}\ \emph {et~al.}(2018)\citenamefont {Han},
  \citenamefont {Li},\ and\ \citenamefont {Deng}}]{han2018critical}%
  \BibitemOpen
  \bibfield  {author} {\bibinfo {author} {\bibfnamefont {Jihui}\ \bibnamefont
  {Han}}, \bibinfo {author} {\bibfnamefont {Wei}\ \bibnamefont {Li}}, \ and\
  \bibinfo {author} {\bibfnamefont {Weibing}\ \bibnamefont {Deng}},\ }\bibfield
   {title} {\enquote {\bibinfo {title} {Critical behavior of a generalized
  bak-sneppen model},}\ }in\ \href@noop {} {\emph {\bibinfo {booktitle}
  {Journal of Physics: Conference Series}}},\ Vol.\ \bibinfo {volume} {1113}\
  (\bibinfo {organization} {IOP Publishing},\ \bibinfo {year} {2018})\ p.\
  \bibinfo {pages} {012011}\BibitemShut {NoStop}%
\bibitem [{\citenamefont {Lin}\ \emph {et~al.}(2014{\natexlab{b}})\citenamefont
  {Lin}, \citenamefont {Saade}, \citenamefont {Lerner}, \citenamefont {Rosso},\
  and\ \citenamefont {Wyart}}]{Lin14a}%
  \BibitemOpen
  \bibfield  {author} {\bibinfo {author} {\bibfnamefont {J.}~\bibnamefont
  {Lin}}, \bibinfo {author} {\bibfnamefont {A.}~\bibnamefont {Saade}}, \bibinfo
  {author} {\bibfnamefont {E.}~\bibnamefont {Lerner}}, \bibinfo {author}
  {\bibfnamefont {A.}~\bibnamefont {Rosso}}, \ and\ \bibinfo {author}
  {\bibfnamefont {M.}~\bibnamefont {Wyart}},\ }\bibfield  {title} {\enquote
  {\bibinfo {title} {On the density of shear transformations in amorphous
  solids},}\ }\href {\doibase 10.1209/0295-5075/105/26003} {\bibfield
  {journal} {\bibinfo  {journal} {Europhys. Lett.}\ }\textbf {\bibinfo {volume}
  {105}},\ \bibinfo {pages} {26003} (\bibinfo {year}
  {2014}{\natexlab{b}})}\BibitemShut {NoStop}%
\bibitem [{\citenamefont {Karmakar}\ \emph {et~al.}(2010)\citenamefont
  {Karmakar}, \citenamefont {Lerner},\ and\ \citenamefont
  {Procaccia}}]{Karmakar2010}%
  \BibitemOpen
  \bibfield  {author} {\bibinfo {author} {\bibfnamefont {S.}~\bibnamefont
  {Karmakar}}, \bibinfo {author} {\bibfnamefont {E.}~\bibnamefont {Lerner}}, \
  and\ \bibinfo {author} {\bibfnamefont {I.}~\bibnamefont {Procaccia}},\
  }\bibfield  {title} {\enquote {\bibinfo {title} {Statistical physics of the
  yielding transition in amorphous solids},}\ }\href {\doibase
  10.1103/PhysRevE.82.055103} {\bibfield  {journal} {\bibinfo  {journal} {Phys.
  Rev. E}\ }\textbf {\bibinfo {volume} {82}},\ \bibinfo {pages} {055103}
  (\bibinfo {year} {2010})}\BibitemShut {NoStop}%
\bibitem [{\citenamefont {M{\"u}ller}\ and\ \citenamefont
  {Wyart}(2015)}]{muller2015marginal}%
  \BibitemOpen
  \bibfield  {author} {\bibinfo {author} {\bibfnamefont {Markus}\ \bibnamefont
  {M{\"u}ller}}\ and\ \bibinfo {author} {\bibfnamefont {Matthieu}\ \bibnamefont
  {Wyart}},\ }\bibfield  {title} {\enquote {\bibinfo {title} {Marginal
  stability in structural, spin, and electron glasses},}\ }\href@noop {}
  {\bibfield  {journal} {\bibinfo  {journal} {Annu. Rev. Condens. Matter
  Phys.}\ }\textbf {\bibinfo {volume} {6}},\ \bibinfo {pages} {177--200}
  (\bibinfo {year} {2015})}\BibitemShut {NoStop}%
\bibitem [{\citenamefont {Lema{\^{i}}tre}\ and\ \citenamefont
  {Caroli}(2007)}]{Lemaitre07}%
  \BibitemOpen
  \bibfield  {author} {\bibinfo {author} {\bibfnamefont {A.}~\bibnamefont
  {Lema{\^{i}}tre}}\ and\ \bibinfo {author} {\bibfnamefont {C.}~\bibnamefont
  {Caroli}},\ }\bibfield  {title} {\enquote {\bibinfo {title} {{Plastic
  Response of a 2D Amorphous Solid to Quasi-Static Shear : II - Dynamical Noise
  and Avalanches in a Mean Field Model}},}\ }\href {\doibase
  10.48550/arXiv.0705.3122} {\bibfield  {journal} {\bibinfo  {journal} {arXiv
  preprint 0705.3122}\ } (\bibinfo {year} {2007}),\
  10.48550/arXiv.0705.3122}\BibitemShut {NoStop}%
\bibitem [{\citenamefont {Lin}\ and\ \citenamefont
  {Wyart}(2016)}]{lin2016mean}%
  \BibitemOpen
  \bibfield  {author} {\bibinfo {author} {\bibfnamefont {Jie}\ \bibnamefont
  {Lin}}\ and\ \bibinfo {author} {\bibfnamefont {Matthieu}\ \bibnamefont
  {Wyart}},\ }\bibfield  {title} {\enquote {\bibinfo {title} {Mean-field
  description of plastic flow in amorphous solids},}\ }\href@noop {} {\bibfield
   {journal} {\bibinfo  {journal} {Physical review X}\ }\textbf {\bibinfo
  {volume} {6}},\ \bibinfo {pages} {011005} (\bibinfo {year}
  {2016})}\BibitemShut {NoStop}%
\bibitem [{\citenamefont {Ferrero}\ and\ \citenamefont
  {Jagla}(2021)}]{ferrero2021properties}%
  \BibitemOpen
  \bibfield  {author} {\bibinfo {author} {\bibfnamefont {Ezequiel~E}\
  \bibnamefont {Ferrero}}\ and\ \bibinfo {author} {\bibfnamefont {Eduardo~A}\
  \bibnamefont {Jagla}},\ }\bibfield  {title} {\enquote {\bibinfo {title}
  {Properties of the density of shear transformations in driven amorphous
  solids},}\ }\href@noop {} {\bibfield  {journal} {\bibinfo  {journal} {Journal
  of Physics: Condensed Matter}\ }\textbf {\bibinfo {volume} {33}},\ \bibinfo
  {pages} {124001} (\bibinfo {year} {2021})}\BibitemShut {NoStop}%
\bibitem [{\citenamefont {Korchinski}\ \emph {et~al.}(2021)\citenamefont
  {Korchinski}, \citenamefont {Ruscher},\ and\ \citenamefont
  {Rottler}}]{korchinski2021signatures}%
  \BibitemOpen
  \bibfield  {author} {\bibinfo {author} {\bibfnamefont {Daniel}\ \bibnamefont
  {Korchinski}}, \bibinfo {author} {\bibfnamefont {C{\'e}line}\ \bibnamefont
  {Ruscher}}, \ and\ \bibinfo {author} {\bibfnamefont {J{\"o}rg}\ \bibnamefont
  {Rottler}},\ }\bibfield  {title} {\enquote {\bibinfo {title} {Signatures of
  the spatial extent of plastic events in the yielding transition in amorphous
  solids},}\ }\href@noop {} {\bibfield  {journal} {\bibinfo  {journal}
  {Physical Review E}\ }\textbf {\bibinfo {volume} {104}},\ \bibinfo {pages}
  {034603} (\bibinfo {year} {2021})}\BibitemShut {NoStop}%
\bibitem [{\citenamefont {Ciamarra}\ \emph {et~al.}(2016)\citenamefont
  {Ciamarra}, \citenamefont {Pastore},\ and\ \citenamefont
  {Coniglio}}]{ciamarra2016particle}%
  \BibitemOpen
  \bibfield  {author} {\bibinfo {author} {\bibfnamefont {Massimo~Pica}\
  \bibnamefont {Ciamarra}}, \bibinfo {author} {\bibfnamefont {Raffaele}\
  \bibnamefont {Pastore}}, \ and\ \bibinfo {author} {\bibfnamefont {Antonio}\
  \bibnamefont {Coniglio}},\ }\bibfield  {title} {\enquote {\bibinfo {title}
  {Particle jumps in structural glasses},}\ }\href@noop {} {\bibfield
  {journal} {\bibinfo  {journal} {Soft matter}\ }\textbf {\bibinfo {volume}
  {12}},\ \bibinfo {pages} {358--366} (\bibinfo {year} {2016})}\BibitemShut
  {NoStop}%
\bibitem [{\citenamefont {Swallen}\ \emph {et~al.}(2003)\citenamefont
  {Swallen}, \citenamefont {Bonvallet}, \citenamefont {McMahon},\ and\
  \citenamefont {Ediger}}]{swallen2003self}%
  \BibitemOpen
  \bibfield  {author} {\bibinfo {author} {\bibfnamefont {Stephen~F}\
  \bibnamefont {Swallen}}, \bibinfo {author} {\bibfnamefont {Paul~A}\
  \bibnamefont {Bonvallet}}, \bibinfo {author} {\bibfnamefont {Robert~J}\
  \bibnamefont {McMahon}}, \ and\ \bibinfo {author} {\bibfnamefont
  {MD}~\bibnamefont {Ediger}},\ }\bibfield  {title} {\enquote {\bibinfo {title}
  {Self-diffusion of tris-naphthylbenzene near the glass transition
  temperature},}\ }\href@noop {} {\bibfield  {journal} {\bibinfo  {journal}
  {Physical review letters}\ }\textbf {\bibinfo {volume} {90}},\ \bibinfo
  {pages} {015901} (\bibinfo {year} {2003})}\BibitemShut {NoStop}%
\bibitem [{\citenamefont {Mallamace}\ \emph {et~al.}(2010)\citenamefont
  {Mallamace}, \citenamefont {Branca}, \citenamefont {Corsaro}, \citenamefont
  {Leone}, \citenamefont {Spooren}, \citenamefont {Chen},\ and\ \citenamefont
  {Stanley}}]{mallamace2010transport}%
  \BibitemOpen
  \bibfield  {author} {\bibinfo {author} {\bibfnamefont {Francesco}\
  \bibnamefont {Mallamace}}, \bibinfo {author} {\bibfnamefont {Caterina}\
  \bibnamefont {Branca}}, \bibinfo {author} {\bibfnamefont {Carmelo}\
  \bibnamefont {Corsaro}}, \bibinfo {author} {\bibfnamefont {Nancy}\
  \bibnamefont {Leone}}, \bibinfo {author} {\bibfnamefont {Jeroen}\
  \bibnamefont {Spooren}}, \bibinfo {author} {\bibfnamefont {Sow-Hsin}\
  \bibnamefont {Chen}}, \ and\ \bibinfo {author} {\bibfnamefont {H~Eugene}\
  \bibnamefont {Stanley}},\ }\bibfield  {title} {\enquote {\bibinfo {title}
  {Transport properties of glass-forming liquids suggest that dynamic crossover
  temperature is as important as the glass transition temperature},}\
  }\href@noop {} {\bibfield  {journal} {\bibinfo  {journal} {Proceedings of the
  National Academy of Sciences}\ }\textbf {\bibinfo {volume} {107}},\ \bibinfo
  {pages} {22457--22462} (\bibinfo {year} {2010})}\BibitemShut {NoStop}%
\bibitem [{Note3()}]{Note3}%
  \BibitemOpen
  \bibinfo {note} {In fact, in some kinetically constrained models~\cite
  {garrahan2011kinetically}, the fractional Stokes-Einstein violation (with a
  similar value of $\zeta $ with experiments) was conjectured based on
  numerical data and physical arguments. Recent rigorous results~\cite
  {blondel2014there} showed that this is not what happens, but there is likely
  a violation as the one we find. This suggests that the logarithmic
  Stokes-Einstein violation can be easily misinterpreted as a fractional one
  with a small exponent $1-\zeta $.}\BibitemShut {Stop}%
\bibitem [{\citenamefont {Ozawa}\ \emph {et~al.}(2016)\citenamefont {Ozawa},
  \citenamefont {Kim},\ and\ \citenamefont {Miyazaki}}]{ozawa2016tuning}%
  \BibitemOpen
  \bibfield  {author} {\bibinfo {author} {\bibfnamefont {Misaki}\ \bibnamefont
  {Ozawa}}, \bibinfo {author} {\bibfnamefont {Kang}\ \bibnamefont {Kim}}, \
  and\ \bibinfo {author} {\bibfnamefont {Kunimasa}\ \bibnamefont {Miyazaki}},\
  }\bibfield  {title} {\enquote {\bibinfo {title} {Tuning pairwise potential
  can control the fragility of glass-forming liquids: From a tetrahedral
  network to isotropic soft sphere models},}\ }\href@noop {} {\bibfield
  {journal} {\bibinfo  {journal} {Journal of Statistical Mechanics: Theory and
  Experiment}\ }\textbf {\bibinfo {volume} {2016}},\ \bibinfo {pages} {074002}
  (\bibinfo {year} {2016})}\BibitemShut {NoStop}%
\bibitem [{\citenamefont {Glarum}(1960)}]{glarum1960dielectric}%
  \BibitemOpen
  \bibfield  {author} {\bibinfo {author} {\bibfnamefont {Sivert~H}\
  \bibnamefont {Glarum}},\ }\bibfield  {title} {\enquote {\bibinfo {title}
  {Dielectric relaxation of isoamyl bromide},}\ }\href@noop {} {\bibfield
  {journal} {\bibinfo  {journal} {The Journal of Chemical Physics}\ }\textbf
  {\bibinfo {volume} {33}},\ \bibinfo {pages} {639--643} (\bibinfo {year}
  {1960})}\BibitemShut {NoStop}%
\bibitem [{\citenamefont {Hasyim}\ and\ \citenamefont
  {Mandadapu}(2021)}]{hasyim2021theory}%
  \BibitemOpen
  \bibfield  {author} {\bibinfo {author} {\bibfnamefont {Muhammad~R}\
  \bibnamefont {Hasyim}}\ and\ \bibinfo {author} {\bibfnamefont {Kranthi~K}\
  \bibnamefont {Mandadapu}},\ }\bibfield  {title} {\enquote {\bibinfo {title}
  {A theory of localized excitations in supercooled liquids},}\ }\href@noop {}
  {\bibfield  {journal} {\bibinfo  {journal} {The Journal of chemical physics}\
  }\textbf {\bibinfo {volume} {155}},\ \bibinfo {pages} {044504} (\bibinfo
  {year} {2021})}\BibitemShut {NoStop}%
\bibitem [{\citenamefont {Isobe}\ \emph {et~al.}(2016)\citenamefont {Isobe},
  \citenamefont {Keys}, \citenamefont {Chandler},\ and\ \citenamefont
  {Garrahan}}]{isobe2016applicability}%
  \BibitemOpen
  \bibfield  {author} {\bibinfo {author} {\bibfnamefont {Masaharu}\
  \bibnamefont {Isobe}}, \bibinfo {author} {\bibfnamefont {Aaron~S}\
  \bibnamefont {Keys}}, \bibinfo {author} {\bibfnamefont {David}\ \bibnamefont
  {Chandler}}, \ and\ \bibinfo {author} {\bibfnamefont {Juan~P}\ \bibnamefont
  {Garrahan}},\ }\bibfield  {title} {\enquote {\bibinfo {title} {Applicability
  of dynamic facilitation theory to binary hard disk systems},}\ }\href@noop {}
  {\bibfield  {journal} {\bibinfo  {journal} {Physical review letters}\
  }\textbf {\bibinfo {volume} {117}},\ \bibinfo {pages} {145701} (\bibinfo
  {year} {2016})}\BibitemShut {NoStop}%
\bibitem [{\citenamefont {Elmatad}\ \emph {et~al.}(2010)\citenamefont
  {Elmatad}, \citenamefont {Jack}, \citenamefont {Chandler},\ and\
  \citenamefont {Garrahan}}]{elmatad2010finite}%
  \BibitemOpen
  \bibfield  {author} {\bibinfo {author} {\bibfnamefont {Yael~S}\ \bibnamefont
  {Elmatad}}, \bibinfo {author} {\bibfnamefont {Robert~L}\ \bibnamefont
  {Jack}}, \bibinfo {author} {\bibfnamefont {David}\ \bibnamefont {Chandler}},
  \ and\ \bibinfo {author} {\bibfnamefont {Juan~P}\ \bibnamefont {Garrahan}},\
  }\bibfield  {title} {\enquote {\bibinfo {title} {Finite-temperature critical
  point of a glass transition},}\ }\href@noop {} {\bibfield  {journal}
  {\bibinfo  {journal} {Proceedings of the National Academy of Sciences}\
  }\textbf {\bibinfo {volume} {107}},\ \bibinfo {pages} {12793--12798}
  (\bibinfo {year} {2010})}\BibitemShut {NoStop}%
\bibitem [{\citenamefont {Turci}\ \emph {et~al.}(2017)\citenamefont {Turci},
  \citenamefont {Royall},\ and\ \citenamefont
  {Speck}}]{turci2017nonequilibrium}%
  \BibitemOpen
  \bibfield  {author} {\bibinfo {author} {\bibfnamefont {Francesco}\
  \bibnamefont {Turci}}, \bibinfo {author} {\bibfnamefont {C~Patrick}\
  \bibnamefont {Royall}}, \ and\ \bibinfo {author} {\bibfnamefont {Thomas}\
  \bibnamefont {Speck}},\ }\bibfield  {title} {\enquote {\bibinfo {title}
  {Nonequilibrium phase transition in an atomistic glassformer: The connection
  to thermodynamics},}\ }\href@noop {} {\bibfield  {journal} {\bibinfo
  {journal} {Physical Review X}\ }\textbf {\bibinfo {volume} {7}},\ \bibinfo
  {pages} {031028} (\bibinfo {year} {2017})}\BibitemShut {NoStop}%
\bibitem [{\citenamefont {Agoritsas}\ \emph {et~al.}(2015)\citenamefont
  {Agoritsas}, \citenamefont {Bertin}, \citenamefont {Martens},\ and\
  \citenamefont {Barrat}}]{agoritsas2015relevance}%
  \BibitemOpen
  \bibfield  {author} {\bibinfo {author} {\bibfnamefont {Elisabeth}\
  \bibnamefont {Agoritsas}}, \bibinfo {author} {\bibfnamefont {Eric}\
  \bibnamefont {Bertin}}, \bibinfo {author} {\bibfnamefont {Kirsten}\
  \bibnamefont {Martens}}, \ and\ \bibinfo {author} {\bibfnamefont
  {Jean-Louis}\ \bibnamefont {Barrat}},\ }\bibfield  {title} {\enquote
  {\bibinfo {title} {On the relevance of disorder in athermal amorphous
  materials under shear},}\ }\href@noop {} {\bibfield  {journal} {\bibinfo
  {journal} {The European Physical Journal E}\ }\textbf {\bibinfo {volume}
  {38}},\ \bibinfo {pages} {1--22} (\bibinfo {year} {2015})}\BibitemShut
  {NoStop}%
\bibitem [{\citenamefont {Tanaka}\ \emph {et~al.}(2010)\citenamefont {Tanaka},
  \citenamefont {Kawasaki}, \citenamefont {Shintani},\ and\ \citenamefont
  {Watanabe}}]{tanaka2010critical}%
  \BibitemOpen
  \bibfield  {author} {\bibinfo {author} {\bibfnamefont {H.}~\bibnamefont
  {Tanaka}}, \bibinfo {author} {\bibfnamefont {T.}~\bibnamefont {Kawasaki}},
  \bibinfo {author} {\bibfnamefont {H.}~\bibnamefont {Shintani}}, \ and\
  \bibinfo {author} {\bibfnamefont {K.}~\bibnamefont {Watanabe}},\ }\bibfield
  {title} {\enquote {\bibinfo {title} {Critical-like behaviour of glass-forming
  liquids},}\ }\href {\doibase 10.1038/nmat2634} {\bibfield  {journal}
  {\bibinfo  {journal} {Nat. Mater.}\ }\textbf {\bibinfo {volume} {9}},\
  \bibinfo {pages} {324--331} (\bibinfo {year} {2010})}\BibitemShut {NoStop}%
\bibitem [{\citenamefont {Royall}\ and\ \citenamefont
  {Williams}(2015)}]{royall2015role}%
  \BibitemOpen
  \bibfield  {author} {\bibinfo {author} {\bibfnamefont {C~Patrick}\
  \bibnamefont {Royall}}\ and\ \bibinfo {author} {\bibfnamefont {Stephen~R}\
  \bibnamefont {Williams}},\ }\bibfield  {title} {\enquote {\bibinfo {title}
  {The role of local structure in dynamical arrest},}\ }\href@noop {}
  {\bibfield  {journal} {\bibinfo  {journal} {Physics Reports}\ }\textbf
  {\bibinfo {volume} {560}},\ \bibinfo {pages} {1--75} (\bibinfo {year}
  {2015})}\BibitemShut {NoStop}%
\bibitem [{\citenamefont {Tanaka}\ \emph {et~al.}(2019)\citenamefont {Tanaka},
  \citenamefont {Tong}, \citenamefont {Shi},\ and\ \citenamefont
  {Russo}}]{tanaka2019revealing}%
  \BibitemOpen
  \bibfield  {author} {\bibinfo {author} {\bibfnamefont {Hajime}\ \bibnamefont
  {Tanaka}}, \bibinfo {author} {\bibfnamefont {Hua}\ \bibnamefont {Tong}},
  \bibinfo {author} {\bibfnamefont {Rui}\ \bibnamefont {Shi}}, \ and\ \bibinfo
  {author} {\bibfnamefont {John}\ \bibnamefont {Russo}},\ }\bibfield  {title}
  {\enquote {\bibinfo {title} {Revealing key structural features hidden in
  liquids and glasses},}\ }\href@noop {} {\bibfield  {journal} {\bibinfo
  {journal} {Nature Reviews Physics}\ }\textbf {\bibinfo {volume} {1}},\
  \bibinfo {pages} {333--348} (\bibinfo {year} {2019})}\BibitemShut {NoStop}%
\bibitem [{\citenamefont {Paret}\ \emph {et~al.}(2020)\citenamefont {Paret},
  \citenamefont {Jack},\ and\ \citenamefont {Coslovich}}]{paret2020assessing}%
  \BibitemOpen
  \bibfield  {author} {\bibinfo {author} {\bibfnamefont {Joris}\ \bibnamefont
  {Paret}}, \bibinfo {author} {\bibfnamefont {Robert~L}\ \bibnamefont {Jack}},
  \ and\ \bibinfo {author} {\bibfnamefont {Daniele}\ \bibnamefont
  {Coslovich}},\ }\bibfield  {title} {\enquote {\bibinfo {title} {Assessing the
  structural heterogeneity of supercooled liquids through community
  inference},}\ }\href@noop {} {\bibfield  {journal} {\bibinfo  {journal} {The
  Journal of chemical physics}\ }\textbf {\bibinfo {volume} {152}},\ \bibinfo
  {pages} {144502} (\bibinfo {year} {2020})}\BibitemShut {NoStop}%
\bibitem [{\citenamefont {Widmer-Cooper}\ \emph {et~al.}(2004)\citenamefont
  {Widmer-Cooper}, \citenamefont {Harrowell},\ and\ \citenamefont
  {Fynewever}}]{widmer2004reproducible}%
  \BibitemOpen
  \bibfield  {author} {\bibinfo {author} {\bibfnamefont {Asaph}\ \bibnamefont
  {Widmer-Cooper}}, \bibinfo {author} {\bibfnamefont {Peter}\ \bibnamefont
  {Harrowell}}, \ and\ \bibinfo {author} {\bibfnamefont {H}~\bibnamefont
  {Fynewever}},\ }\bibfield  {title} {\enquote {\bibinfo {title} {How
  reproducible are dynamic heterogeneities in a supercooled liquid?}}\
  }\href@noop {} {\bibfield  {journal} {\bibinfo  {journal} {Physical review
  letters}\ }\textbf {\bibinfo {volume} {93}},\ \bibinfo {pages} {135701}
  (\bibinfo {year} {2004})}\BibitemShut {NoStop}%
\bibitem [{\citenamefont {Widmer-Cooper}\ \emph {et~al.}(2008)\citenamefont
  {Widmer-Cooper}, \citenamefont {Perry}, \citenamefont {Harrowell},\ and\
  \citenamefont {Reichman}}]{widmer2008irreversible}%
  \BibitemOpen
  \bibfield  {author} {\bibinfo {author} {\bibfnamefont {A.}~\bibnamefont
  {Widmer-Cooper}}, \bibinfo {author} {\bibfnamefont {H.}~\bibnamefont
  {Perry}}, \bibinfo {author} {\bibfnamefont {P.}~\bibnamefont {Harrowell}}, \
  and\ \bibinfo {author} {\bibfnamefont {D.R.}\ \bibnamefont {Reichman}},\
  }\bibfield  {title} {\enquote {\bibinfo {title} {Irreversible reorganization
  in a supercooled liquid originates from localized soft modes},}\ }\href
  {\doibase 10.1038/nphys1025} {\bibfield  {journal} {\bibinfo  {journal} {Nat.
  Phys.}\ }\textbf {\bibinfo {volume} {4}},\ \bibinfo {pages} {711--715}
  (\bibinfo {year} {2008})}\BibitemShut {NoStop}%
\bibitem [{\citenamefont {Hocky}\ \emph {et~al.}(2014)\citenamefont {Hocky},
  \citenamefont {Coslovich}, \citenamefont {Ikeda},\ and\ \citenamefont
  {Reichman}}]{hocky2014correlation}%
  \BibitemOpen
  \bibfield  {author} {\bibinfo {author} {\bibfnamefont {Glen~M}\ \bibnamefont
  {Hocky}}, \bibinfo {author} {\bibfnamefont {Daniele}\ \bibnamefont
  {Coslovich}}, \bibinfo {author} {\bibfnamefont {Atsushi}\ \bibnamefont
  {Ikeda}}, \ and\ \bibinfo {author} {\bibfnamefont {David~R}\ \bibnamefont
  {Reichman}},\ }\bibfield  {title} {\enquote {\bibinfo {title} {Correlation of
  local order with particle mobility in supercooled liquids is highly system
  dependent},}\ }\href@noop {} {\bibfield  {journal} {\bibinfo  {journal}
  {Physical review letters}\ }\textbf {\bibinfo {volume} {113}},\ \bibinfo
  {pages} {157801} (\bibinfo {year} {2014})}\BibitemShut {NoStop}%
\bibitem [{\citenamefont {Lema{\^\i}tre}\ \emph {et~al.}(2021)\citenamefont
  {Lema{\^\i}tre}, \citenamefont {Mondal}, \citenamefont {Moshe}, \citenamefont
  {Procaccia}, \citenamefont {Roy},\ and\ \citenamefont
  {Screiber-Re'em}}]{lemaitre2021anomalous}%
  \BibitemOpen
  \bibfield  {author} {\bibinfo {author} {\bibfnamefont {Ana{\"e}l}\
  \bibnamefont {Lema{\^\i}tre}}, \bibinfo {author} {\bibfnamefont {Chandana}\
  \bibnamefont {Mondal}}, \bibinfo {author} {\bibfnamefont {Michael}\
  \bibnamefont {Moshe}}, \bibinfo {author} {\bibfnamefont {Itamar}\
  \bibnamefont {Procaccia}}, \bibinfo {author} {\bibfnamefont {Saikat}\
  \bibnamefont {Roy}}, \ and\ \bibinfo {author} {\bibfnamefont {Keren}\
  \bibnamefont {Screiber-Re'em}},\ }\bibfield  {title} {\enquote {\bibinfo
  {title} {Anomalous elasticity and plastic screening in amorphous solids},}\
  }\href@noop {} {\bibfield  {journal} {\bibinfo  {journal} {Physical Review
  E}\ }\textbf {\bibinfo {volume} {104}},\ \bibinfo {pages} {024904} (\bibinfo
  {year} {2021})}\BibitemShut {NoStop}%
\bibitem [{\citenamefont {Lerner}\ \emph {et~al.}(2014)\citenamefont {Lerner},
  \citenamefont {DeGiuli}, \citenamefont {D{\"{u}}ring},\ and\ \citenamefont
  {Wyart}}]{Lerner14}%
  \BibitemOpen
  \bibfield  {author} {\bibinfo {author} {\bibfnamefont {E.}~\bibnamefont
  {Lerner}}, \bibinfo {author} {\bibfnamefont {E.}~\bibnamefont {DeGiuli}},
  \bibinfo {author} {\bibfnamefont {G.}~\bibnamefont {D{\"{u}}ring}}, \ and\
  \bibinfo {author} {\bibfnamefont {M.}~\bibnamefont {Wyart}},\ }\bibfield
  {title} {\enquote {\bibinfo {title} {{Breakdown of continuum elasticity in
  amorphous solids}},}\ }\href {\doibase 10.1039/C4SM00311J} {\bibfield
  {journal} {\bibinfo  {journal} {Soft Matter}\ }\textbf {\bibinfo {volume}
  {10}},\ \bibinfo {pages} {5085} (\bibinfo {year} {2014})}\BibitemShut
  {NoStop}%
\bibitem [{\citenamefont {Liu}\ \emph {et~al.}(2021)\citenamefont {Liu},
  \citenamefont {Dutta}, \citenamefont {Chaudhuri},\ and\ \citenamefont
  {Martens}}]{liu2021elastoplastic}%
  \BibitemOpen
  \bibfield  {author} {\bibinfo {author} {\bibfnamefont {Chen}\ \bibnamefont
  {Liu}}, \bibinfo {author} {\bibfnamefont {Suman}\ \bibnamefont {Dutta}},
  \bibinfo {author} {\bibfnamefont {Pinaki}\ \bibnamefont {Chaudhuri}}, \ and\
  \bibinfo {author} {\bibfnamefont {Kirsten}\ \bibnamefont {Martens}},\
  }\bibfield  {title} {\enquote {\bibinfo {title} {Elastoplastic approach based
  on microscopic insights for the steady state and transient dynamics of
  sheared disordered solids},}\ }\href@noop {} {\bibfield  {journal} {\bibinfo
  {journal} {Physical Review Letters}\ }\textbf {\bibinfo {volume} {126}},\
  \bibinfo {pages} {138005} (\bibinfo {year} {2021})}\BibitemShut {NoStop}%
\bibitem [{\citenamefont {Castellanos}\ \emph {et~al.}(2021)\citenamefont
  {Castellanos}, \citenamefont {Roux},\ and\ \citenamefont
  {Patinet}}]{castellanos2021insights}%
  \BibitemOpen
  \bibfield  {author} {\bibinfo {author} {\bibfnamefont {David~Fern{\'a}ndez}\
  \bibnamefont {Castellanos}}, \bibinfo {author} {\bibfnamefont {St{\'e}phane}\
  \bibnamefont {Roux}}, \ and\ \bibinfo {author} {\bibfnamefont {Sylvain}\
  \bibnamefont {Patinet}},\ }\bibfield  {title} {\enquote {\bibinfo {title}
  {Insights from the quantitative calibration of an elasto-plastic model from a
  lennard-jones atomic glass},}\ }\href@noop {} {\bibfield  {journal} {\bibinfo
   {journal} {Comptes Rendus. Physique}\ }\textbf {\bibinfo {volume} {22}},\
  \bibinfo {pages} {1--28} (\bibinfo {year} {2021})}\BibitemShut {NoStop}%
\bibitem [{\citenamefont {Castellanos}\ \emph {et~al.}(2022)\citenamefont
  {Castellanos}, \citenamefont {Roux},\ and\ \citenamefont
  {Patinet}}]{castellanos2022history}%
  \BibitemOpen
  \bibfield  {author} {\bibinfo {author} {\bibfnamefont {David~F}\ \bibnamefont
  {Castellanos}}, \bibinfo {author} {\bibfnamefont {St{\'e}phane}\ \bibnamefont
  {Roux}}, \ and\ \bibinfo {author} {\bibfnamefont {Sylvain}\ \bibnamefont
  {Patinet}},\ }\bibfield  {title} {\enquote {\bibinfo {title} {History
  dependent plasticity of glass: A mapping between atomistic and elasto-plastic
  models},}\ }\href@noop {} {\bibfield  {journal} {\bibinfo  {journal} {Acta
  Materialia}\ }\textbf {\bibinfo {volume} {241}},\ \bibinfo {pages} {118405}
  (\bibinfo {year} {2022})}\BibitemShut {NoStop}%
\bibitem [{\citenamefont {Tah}\ \emph {et~al.}(2022)\citenamefont {Tah},
  \citenamefont {Ridout},\ and\ \citenamefont {Liu}}]{tah2022fragility}%
  \BibitemOpen
  \bibfield  {author} {\bibinfo {author} {\bibfnamefont {Indrajit}\
  \bibnamefont {Tah}}, \bibinfo {author} {\bibfnamefont {Sean~A}\ \bibnamefont
  {Ridout}}, \ and\ \bibinfo {author} {\bibfnamefont {Andrea~J}\ \bibnamefont
  {Liu}},\ }\bibfield  {title} {\enquote {\bibinfo {title} {Fragility in glassy
  liquids: A structural approach based on machine learning},}\ }\href@noop {}
  {\bibfield  {journal} {\bibinfo  {journal} {The Journal of Chemical Physics}\
  }\textbf {\bibinfo {volume} {157}},\ \bibinfo {pages} {124501} (\bibinfo
  {year} {2022})}\BibitemShut {NoStop}%
\bibitem [{\citenamefont {Xiao}\ \emph {et~al.}(2023)\citenamefont {Xiao},
  \citenamefont {Zhang}, \citenamefont {Yang}, \citenamefont {Ivancic},
  \citenamefont {Ridout}, \citenamefont {Riggleman}, \citenamefont {Durian},\
  and\ \citenamefont {Liu}}]{xiao2023machine}%
  \BibitemOpen
  \bibfield  {author} {\bibinfo {author} {\bibfnamefont {Hongyi}\ \bibnamefont
  {Xiao}}, \bibinfo {author} {\bibfnamefont {Ge}~\bibnamefont {Zhang}},
  \bibinfo {author} {\bibfnamefont {Entao}\ \bibnamefont {Yang}}, \bibinfo
  {author} {\bibfnamefont {Robert~JS}\ \bibnamefont {Ivancic}}, \bibinfo
  {author} {\bibfnamefont {Sean~A}\ \bibnamefont {Ridout}}, \bibinfo {author}
  {\bibfnamefont {Robert}\ \bibnamefont {Riggleman}}, \bibinfo {author}
  {\bibfnamefont {Douglas~J}\ \bibnamefont {Durian}}, \ and\ \bibinfo {author}
  {\bibfnamefont {Andrea~J}\ \bibnamefont {Liu}},\ }\bibfield  {title}
  {\enquote {\bibinfo {title} {Machine learning-informed
  structuro-elastoplasticity predicts ductility of disordered solids},}\
  }\href@noop {} {\bibfield  {journal} {\bibinfo  {journal} {arXiv preprint
  arXiv:2303.12486}\ } (\bibinfo {year} {2023})}\BibitemShut {NoStop}%
\bibitem [{\citenamefont {Hecksher}\ and\ \citenamefont
  {Dyre}(2015)}]{hecksher2015review}%
  \BibitemOpen
  \bibfield  {author} {\bibinfo {author} {\bibfnamefont {Tina}\ \bibnamefont
  {Hecksher}}\ and\ \bibinfo {author} {\bibfnamefont {Jeppe~C}\ \bibnamefont
  {Dyre}},\ }\bibfield  {title} {\enquote {\bibinfo {title} {A review of
  experiments testing the shoving model},}\ }\href@noop {} {\bibfield
  {journal} {\bibinfo  {journal} {Journal of Non-Crystalline Solids}\ }\textbf
  {\bibinfo {volume} {407}},\ \bibinfo {pages} {14--22} (\bibinfo {year}
  {2015})}\BibitemShut {NoStop}%
\bibitem [{\citenamefont {Ji}\ \emph {et~al.}(2022)\citenamefont {Ji},
  \citenamefont {de~Geus}, \citenamefont {Agoritsas},\ and\ \citenamefont
  {Wyart}}]{Wencheng22}%
  \BibitemOpen
  \bibfield  {author} {\bibinfo {author} {\bibfnamefont {Wencheng}\
  \bibnamefont {Ji}}, \bibinfo {author} {\bibfnamefont {Tom~WJ}\ \bibnamefont
  {de~Geus}}, \bibinfo {author} {\bibfnamefont {Elisabeth}\ \bibnamefont
  {Agoritsas}}, \ and\ \bibinfo {author} {\bibfnamefont {Matthieu}\
  \bibnamefont {Wyart}},\ }\bibfield  {title} {\enquote {\bibinfo {title}
  {Mean-field description for the architecture of low-energy excitations in
  glasses},}\ }\href@noop {} {\bibfield  {journal} {\bibinfo  {journal}
  {Physical Review E}\ }\textbf {\bibinfo {volume} {105}},\ \bibinfo {pages}
  {044601} (\bibinfo {year} {2022})}\BibitemShut {NoStop}%
\bibitem [{\citenamefont {Scalliet}\ \emph {et~al.}(2019)\citenamefont
  {Scalliet}, \citenamefont {Berthier},\ and\ \citenamefont
  {Zamponi}}]{scalliet2019nature}%
  \BibitemOpen
  \bibfield  {author} {\bibinfo {author} {\bibfnamefont {Camille}\ \bibnamefont
  {Scalliet}}, \bibinfo {author} {\bibfnamefont {Ludovic}\ \bibnamefont
  {Berthier}}, \ and\ \bibinfo {author} {\bibfnamefont {Francesco}\
  \bibnamefont {Zamponi}},\ }\bibfield  {title} {\enquote {\bibinfo {title}
  {Nature of excitations and defects in structural glasses},}\ }\href@noop {}
  {\bibfield  {journal} {\bibinfo  {journal} {Nature communications}\ }\textbf
  {\bibinfo {volume} {10}},\ \bibinfo {pages} {5102} (\bibinfo {year}
  {2019})}\BibitemShut {NoStop}%
\bibitem [{\citenamefont {Mizuno}\ \emph {et~al.}(2020)\citenamefont {Mizuno},
  \citenamefont {Tong}, \citenamefont {Ikeda},\ and\ \citenamefont
  {Mossa}}]{mizuno2020intermittent}%
  \BibitemOpen
  \bibfield  {author} {\bibinfo {author} {\bibfnamefont {Hideyuki}\
  \bibnamefont {Mizuno}}, \bibinfo {author} {\bibfnamefont {Hua}\ \bibnamefont
  {Tong}}, \bibinfo {author} {\bibfnamefont {Atsushi}\ \bibnamefont {Ikeda}}, \
  and\ \bibinfo {author} {\bibfnamefont {Stefano}\ \bibnamefont {Mossa}},\
  }\bibfield  {title} {\enquote {\bibinfo {title} {Intermittent rearrangements
  accompanying thermal fluctuations distinguish glasses from crystals},}\
  }\href@noop {} {\bibfield  {journal} {\bibinfo  {journal} {The Journal of
  Chemical Physics}\ }\textbf {\bibinfo {volume} {153}},\ \bibinfo {pages}
  {154501} (\bibinfo {year} {2020})}\BibitemShut {NoStop}%
\bibitem [{\citenamefont {Richard}\ \emph {et~al.}(2023)\citenamefont
  {Richard}, \citenamefont {Kapteijns},\ and\ \citenamefont
  {Lerner}}]{richard2023detecting}%
  \BibitemOpen
  \bibfield  {author} {\bibinfo {author} {\bibfnamefont {David}\ \bibnamefont
  {Richard}}, \bibinfo {author} {\bibfnamefont {Geert}\ \bibnamefont
  {Kapteijns}}, \ and\ \bibinfo {author} {\bibfnamefont {Edan}\ \bibnamefont
  {Lerner}},\ }\bibfield  {title} {\enquote {\bibinfo {title} {Detecting
  low-energy quasilocalized excitations in computer glasses},}\ }\href@noop {}
  {\bibfield  {journal} {\bibinfo  {journal} {arXiv preprint arXiv:2303.12887}\
  } (\bibinfo {year} {2023})}\BibitemShut {NoStop}%
\bibitem [{\citenamefont {Coslovich}\ and\ \citenamefont
  {Pastore}(2007)}]{coslovich2007understanding}%
  \BibitemOpen
  \bibfield  {author} {\bibinfo {author} {\bibfnamefont {Daniele}\ \bibnamefont
  {Coslovich}}\ and\ \bibinfo {author} {\bibfnamefont {Giorgio}\ \bibnamefont
  {Pastore}},\ }\bibfield  {title} {\enquote {\bibinfo {title} {Understanding
  fragility in supercooled lennard-jones mixtures. i. locally preferred
  structures},}\ }\href@noop {} {\bibfield  {journal} {\bibinfo  {journal} {The
  Journal of chemical physics}\ }\textbf {\bibinfo {volume} {127}},\ \bibinfo
  {pages} {124504} (\bibinfo {year} {2007})}\BibitemShut {NoStop}%
\bibitem [{\citenamefont {Biroli}\ and\ \citenamefont
  {Bouchaud}(2022)}]{biroli2022rfot}%
  \BibitemOpen
  \bibfield  {author} {\bibinfo {author} {\bibfnamefont {Giulio}\ \bibnamefont
  {Biroli}}\ and\ \bibinfo {author} {\bibfnamefont {Jean-Philippe}\
  \bibnamefont {Bouchaud}},\ }\bibfield  {title} {\enquote {\bibinfo {title}
  {The rfot theory of glasses: Recent progress and open issues},}\ }\href@noop
  {} {\bibfield  {journal} {\bibinfo  {journal} {arXiv preprint
  arXiv:2208.05866}\ } (\bibinfo {year} {2022})}\BibitemShut {NoStop}%
\bibitem [{\citenamefont {Castellanos}\ and\ \citenamefont
  {Zaiser}(2018)}]{castellanos2018avalanche}%
  \BibitemOpen
  \bibfield  {author} {\bibinfo {author} {\bibfnamefont {David~Fernandez}\
  \bibnamefont {Castellanos}}\ and\ \bibinfo {author} {\bibfnamefont {Michael}\
  \bibnamefont {Zaiser}},\ }\bibfield  {title} {\enquote {\bibinfo {title}
  {Avalanche behavior in creep failure of disordered materials},}\ }\href@noop
  {} {\bibfield  {journal} {\bibinfo  {journal} {Physical review letters}\
  }\textbf {\bibinfo {volume} {121}},\ \bibinfo {pages} {125501} (\bibinfo
  {year} {2018})}\BibitemShut {NoStop}%
\bibitem [{\citenamefont {Bauer}\ \emph {et~al.}(2006)\citenamefont {Bauer},
  \citenamefont {Oberdisse},\ and\ \citenamefont {Ramos}}]{Bauer2006}%
  \BibitemOpen
  \bibfield  {author} {\bibinfo {author} {\bibfnamefont {T.}~\bibnamefont
  {Bauer}}, \bibinfo {author} {\bibfnamefont {J.}~\bibnamefont {Oberdisse}}, \
  and\ \bibinfo {author} {\bibfnamefont {L.}~\bibnamefont {Ramos}},\ }\bibfield
   {title} {\enquote {\bibinfo {title} {Collective rearrangement at the onset
  of flow of a polycrystalline hexagonal columnar phase},}\ }\href {\doibase
  10.1103/PhysRevLett.97.258303} {\bibfield  {journal} {\bibinfo  {journal}
  {Phys. Rev. Lett.}\ }\textbf {\bibinfo {volume} {97}} (\bibinfo {year}
  {2006}),\ 10.1103/PhysRevLett.97.258303}\BibitemShut {NoStop}%
\bibitem [{\citenamefont {Caton}\ and\ \citenamefont
  {Baravian}(2008)}]{Caton2008}%
  \BibitemOpen
  \bibfield  {author} {\bibinfo {author} {\bibfnamefont {F.}~\bibnamefont
  {Caton}}\ and\ \bibinfo {author} {\bibfnamefont {C.}~\bibnamefont
  {Baravian}},\ }\bibfield  {title} {\enquote {\bibinfo {title} {Plastic
  behavior of some yield stress fluids: from creep to long-time yield},}\
  }\href {\doibase 10.1007/s00397-008-0267-2} {\bibfield  {journal} {\bibinfo
  {journal} {Rheol. Acta}\ }\textbf {\bibinfo {volume} {47}},\ \bibinfo {pages}
  {601–607} (\bibinfo {year} {2008})}\BibitemShut {NoStop}%
\bibitem [{\citenamefont {Divoux}\ \emph {et~al.}(2011)\citenamefont {Divoux},
  \citenamefont {Barentin},\ and\ \citenamefont {Manneville}}]{Divoux2011}%
  \BibitemOpen
  \bibfield  {author} {\bibinfo {author} {\bibfnamefont {T.}~\bibnamefont
  {Divoux}}, \bibinfo {author} {\bibfnamefont {C.}~\bibnamefont {Barentin}}, \
  and\ \bibinfo {author} {\bibfnamefont {S.}~\bibnamefont {Manneville}},\
  }\bibfield  {title} {\enquote {\bibinfo {title} {From stress-induced
  fluidization processes to herschel-bulkley behaviour in simple yield stress
  fluids},}\ }\href {\doibase 10.1039/c1sm05607g} {\bibfield  {journal}
  {\bibinfo  {journal} {Soft Matter}\ }\textbf {\bibinfo {volume} {7}},\
  \bibinfo {pages} {8409} (\bibinfo {year} {2011})}\BibitemShut {NoStop}%
\bibitem [{\citenamefont {Siebenb{\" u}rger}\ \emph {et~al.}(2012)\citenamefont
  {Siebenb{\" u}rger}, \citenamefont {Ballauff},\ and\ \citenamefont
  {Voigtmann}}]{Siebenbuerger2012}%
  \BibitemOpen
  \bibfield  {author} {\bibinfo {author} {\bibfnamefont {M.}~\bibnamefont
  {Siebenb{\" u}rger}}, \bibinfo {author} {\bibfnamefont {M.}~\bibnamefont
  {Ballauff}}, \ and\ \bibinfo {author} {\bibfnamefont {T.}~\bibnamefont
  {Voigtmann}},\ }\bibfield  {title} {\enquote {\bibinfo {title} {Creep in
  colloidal glasses},}\ }\href {\doibase 10.1103/PhysRevLett.108.255701}
  {\bibfield  {journal} {\bibinfo  {journal} {Phys. Rev. Lett.}\ }\textbf
  {\bibinfo {volume} {108}},\ \bibinfo {pages} {255701} (\bibinfo {year}
  {2012})}\BibitemShut {NoStop}%
\bibitem [{\citenamefont {Grenard}\ \emph {et~al.}(2014)\citenamefont
  {Grenard}, \citenamefont {Divoux}, \citenamefont {Taberlet},\ and\
  \citenamefont {Manneville}}]{Grenard2014}%
  \BibitemOpen
  \bibfield  {author} {\bibinfo {author} {\bibfnamefont {V.}~\bibnamefont
  {Grenard}}, \bibinfo {author} {\bibfnamefont {T.}~\bibnamefont {Divoux}},
  \bibinfo {author} {\bibfnamefont {N.}~\bibnamefont {Taberlet}}, \ and\
  \bibinfo {author} {\bibfnamefont {S.}~\bibnamefont {Manneville}},\ }\bibfield
   {title} {\enquote {\bibinfo {title} {Timescales in creep and yielding of
  attractive gels},}\ }\href@noop {} {\bibfield  {journal} {\bibinfo  {journal}
  {Soft Matter}\ ,\ \bibinfo {pages} {17}} (\bibinfo {year}
  {2014})}\BibitemShut {NoStop}%
\bibitem [{\citenamefont {Leocmach}\ \emph {et~al.}(2014)\citenamefont
  {Leocmach}, \citenamefont {Perge}, \citenamefont {Divoux},\ and\
  \citenamefont {Manneville}}]{Leocmach2014}%
  \BibitemOpen
  \bibfield  {author} {\bibinfo {author} {\bibfnamefont {M.}~\bibnamefont
  {Leocmach}}, \bibinfo {author} {\bibfnamefont {C.}~\bibnamefont {Perge}},
  \bibinfo {author} {\bibfnamefont {T.}~\bibnamefont {Divoux}}, \ and\ \bibinfo
  {author} {\bibfnamefont {S.}~\bibnamefont {Manneville}},\ }\bibfield  {title}
  {\enquote {\bibinfo {title} {Creep and fracture of a protein gel under
  stress},}\ }\href {\doibase 10.1103/PhysRevLett.113.038303} {\bibfield
  {journal} {\bibinfo  {journal} {Phys. Rev. Lett.}\ }\textbf {\bibinfo
  {volume} {113}} (\bibinfo {year} {2014}),\
  10.1103/PhysRevLett.113.038303}\BibitemShut {NoStop}%
\bibitem [{\citenamefont {Bustingorry}\ \emph {et~al.}(2007)\citenamefont
  {Bustingorry}, \citenamefont {Kolton},\ and\ \citenamefont
  {Giamarchi}}]{bustingorry2007thermal}%
  \BibitemOpen
  \bibfield  {author} {\bibinfo {author} {\bibfnamefont {Sebastian}\
  \bibnamefont {Bustingorry}}, \bibinfo {author} {\bibfnamefont
  {AB}~\bibnamefont {Kolton}}, \ and\ \bibinfo {author} {\bibfnamefont
  {Thierry}\ \bibnamefont {Giamarchi}},\ }\bibfield  {title} {\enquote
  {\bibinfo {title} {Thermal rounding of the depinning transition},}\
  }\href@noop {} {\bibfield  {journal} {\bibinfo  {journal} {Europhysics
  Letters}\ }\textbf {\bibinfo {volume} {81}},\ \bibinfo {pages} {26005}
  (\bibinfo {year} {2007})}\BibitemShut {NoStop}%
\bibitem [{\citenamefont {Kolton}\ \emph {et~al.}(2005)\citenamefont {Kolton},
  \citenamefont {Rosso},\ and\ \citenamefont {Giamarchi}}]{kolton2005creep}%
  \BibitemOpen
  \bibfield  {author} {\bibinfo {author} {\bibfnamefont {Alejandro~B}\
  \bibnamefont {Kolton}}, \bibinfo {author} {\bibfnamefont {Alberto}\
  \bibnamefont {Rosso}}, \ and\ \bibinfo {author} {\bibfnamefont {Thierry}\
  \bibnamefont {Giamarchi}},\ }\bibfield  {title} {\enquote {\bibinfo {title}
  {Creep motion of an elastic string in a random potential},}\ }\href@noop {}
  {\bibfield  {journal} {\bibinfo  {journal} {Physical review letters}\
  }\textbf {\bibinfo {volume} {94}},\ \bibinfo {pages} {047002} (\bibinfo
  {year} {2005})}\BibitemShut {NoStop}%
\bibitem [{\citenamefont {Berthier}\ and\ \citenamefont
  {Kob}(2007)}]{berthier2007monte}%
  \BibitemOpen
  \bibfield  {author} {\bibinfo {author} {\bibfnamefont {Ludovic}\ \bibnamefont
  {Berthier}}\ and\ \bibinfo {author} {\bibfnamefont {Walter}\ \bibnamefont
  {Kob}},\ }\bibfield  {title} {\enquote {\bibinfo {title} {The monte carlo
  dynamics of a binary lennard-jones glass-forming mixture},}\ }\href@noop {}
  {\bibfield  {journal} {\bibinfo  {journal} {Journal of Physics: Condensed
  Matter}\ }\textbf {\bibinfo {volume} {19}},\ \bibinfo {pages} {205130}
  (\bibinfo {year} {2007})}\BibitemShut {NoStop}%
\bibitem [{\citenamefont {Maloney}\ and\ \citenamefont
  {Lacks}(2006)}]{maloney2006energy}%
  \BibitemOpen
  \bibfield  {author} {\bibinfo {author} {\bibfnamefont {Craig~E}\ \bibnamefont
  {Maloney}}\ and\ \bibinfo {author} {\bibfnamefont {Daniel~J}\ \bibnamefont
  {Lacks}},\ }\bibfield  {title} {\enquote {\bibinfo {title} {Energy barrier
  scalings in driven systems},}\ }\href@noop {} {\bibfield  {journal} {\bibinfo
   {journal} {Physical Review E}\ }\textbf {\bibinfo {volume} {73}},\ \bibinfo
  {pages} {061106} (\bibinfo {year} {2006})}\BibitemShut {NoStop}%
\bibitem [{\citenamefont {Barbot}\ \emph {et~al.}(2018)\citenamefont {Barbot},
  \citenamefont {Lerbinger}, \citenamefont {Hernandez-Garcia}, \citenamefont
  {Garc{\'\i}a-Garc{\'\i}a}, \citenamefont {Falk}, \citenamefont
  {Vandembroucq},\ and\ \citenamefont {Patinet}}]{barbot2018local}%
  \BibitemOpen
  \bibfield  {author} {\bibinfo {author} {\bibfnamefont {Armand}\ \bibnamefont
  {Barbot}}, \bibinfo {author} {\bibfnamefont {Matthias}\ \bibnamefont
  {Lerbinger}}, \bibinfo {author} {\bibfnamefont {Anier}\ \bibnamefont
  {Hernandez-Garcia}}, \bibinfo {author} {\bibfnamefont {Reinaldo}\
  \bibnamefont {Garc{\'\i}a-Garc{\'\i}a}}, \bibinfo {author} {\bibfnamefont
  {Michael~L}\ \bibnamefont {Falk}}, \bibinfo {author} {\bibfnamefont {Damien}\
  \bibnamefont {Vandembroucq}}, \ and\ \bibinfo {author} {\bibfnamefont
  {Sylvain}\ \bibnamefont {Patinet}},\ }\bibfield  {title} {\enquote {\bibinfo
  {title} {Local yield stress statistics in model amorphous solids},}\
  }\href@noop {} {\bibfield  {journal} {\bibinfo  {journal} {Physical Review
  E}\ }\textbf {\bibinfo {volume} {97}},\ \bibinfo {pages} {033001} (\bibinfo
  {year} {2018})}\BibitemShut {NoStop}%
\bibitem [{\citenamefont {Popovi{\'c}}\ \emph {et~al.}(2018)\citenamefont
  {Popovi{\'c}}, \citenamefont {de~Geus},\ and\ \citenamefont
  {Wyart}}]{popovic2018elastoplastic}%
  \BibitemOpen
  \bibfield  {author} {\bibinfo {author} {\bibfnamefont {Marko}\ \bibnamefont
  {Popovi{\'c}}}, \bibinfo {author} {\bibfnamefont {Tom~WJ}\ \bibnamefont
  {de~Geus}}, \ and\ \bibinfo {author} {\bibfnamefont {Matthieu}\ \bibnamefont
  {Wyart}},\ }\bibfield  {title} {\enquote {\bibinfo {title} {Elastoplastic
  description of sudden failure in athermal amorphous materials during
  quasistatic loading},}\ }\href@noop {} {\bibfield  {journal} {\bibinfo
  {journal} {Physical Review E}\ }\textbf {\bibinfo {volume} {98}},\ \bibinfo
  {pages} {040901} (\bibinfo {year} {2018})}\BibitemShut {NoStop}%
\bibitem [{\citenamefont {Pollard}\ and\ \citenamefont
  {Fielding}(2022)}]{pollard2022yielding}%
  \BibitemOpen
  \bibfield  {author} {\bibinfo {author} {\bibfnamefont {Joseph}\ \bibnamefont
  {Pollard}}\ and\ \bibinfo {author} {\bibfnamefont {Suzanne~M}\ \bibnamefont
  {Fielding}},\ }\bibfield  {title} {\enquote {\bibinfo {title} {Yielding,
  shear banding, and brittle failure of amorphous materials},}\ }\href@noop {}
  {\bibfield  {journal} {\bibinfo  {journal} {Physical Review Research}\
  }\textbf {\bibinfo {volume} {4}},\ \bibinfo {pages} {043037} (\bibinfo {year}
  {2022})}\BibitemShut {NoStop}%
\bibitem [{\citenamefont {Flenner}\ \emph {et~al.}(2014)\citenamefont
  {Flenner}, \citenamefont {Staley},\ and\ \citenamefont
  {Szamel}}]{flenner2014universal}%
  \BibitemOpen
  \bibfield  {author} {\bibinfo {author} {\bibfnamefont {Elijah}\ \bibnamefont
  {Flenner}}, \bibinfo {author} {\bibfnamefont {Hannah}\ \bibnamefont
  {Staley}}, \ and\ \bibinfo {author} {\bibfnamefont {Grzegorz}\ \bibnamefont
  {Szamel}},\ }\bibfield  {title} {\enquote {\bibinfo {title} {Universal
  features of dynamic heterogeneity in supercooled liquids},}\ }\href@noop {}
  {\bibfield  {journal} {\bibinfo  {journal} {Physical review letters}\
  }\textbf {\bibinfo {volume} {112}},\ \bibinfo {pages} {097801} (\bibinfo
  {year} {2014})}\BibitemShut {NoStop}%
\bibitem [{\citenamefont {Kim}\ and\ \citenamefont
  {Saito}(2013)}]{kim2013multiple}%
  \BibitemOpen
  \bibfield  {author} {\bibinfo {author} {\bibfnamefont {Kang}\ \bibnamefont
  {Kim}}\ and\ \bibinfo {author} {\bibfnamefont {Shinji}\ \bibnamefont
  {Saito}},\ }\bibfield  {title} {\enquote {\bibinfo {title} {Multiple length
  and time scales of dynamic heterogeneities in model glass-forming liquids: A
  systematic analysis of multi-point and multi-time correlations},}\
  }\href@noop {} {\bibfield  {journal} {\bibinfo  {journal} {The Journal of
  chemical physics}\ }\textbf {\bibinfo {volume} {138}},\ \bibinfo {pages}
  {12A506} (\bibinfo {year} {2013})}\BibitemShut {NoStop}%
\bibitem [{\citenamefont {Biroli}\ \emph {et~al.}(2022)\citenamefont {Biroli},
  \citenamefont {Miyazaki},\ and\ \citenamefont
  {Reichman}}]{biroli2022dynamical}%
  \BibitemOpen
  \bibfield  {author} {\bibinfo {author} {\bibfnamefont {Giulio}\ \bibnamefont
  {Biroli}}, \bibinfo {author} {\bibfnamefont {Kunimasa}\ \bibnamefont
  {Miyazaki}}, \ and\ \bibinfo {author} {\bibfnamefont {David~R}\ \bibnamefont
  {Reichman}},\ }\bibfield  {title} {\enquote {\bibinfo {title} {Dynamical
  heterogeneity in glass-forming liquids},}\ }\href@noop {} {\bibfield
  {journal} {\bibinfo  {journal} {arXiv preprint arXiv:2209.02825}\ } (\bibinfo
  {year} {2022})}\BibitemShut {NoStop}%
\bibitem [{\citenamefont {Flenner}\ and\ \citenamefont
  {Szamel}(2016)}]{flenner2016dynamic}%
  \BibitemOpen
  \bibfield  {author} {\bibinfo {author} {\bibfnamefont {Elijah}\ \bibnamefont
  {Flenner}}\ and\ \bibinfo {author} {\bibfnamefont {Grzegorz}\ \bibnamefont
  {Szamel}},\ }\bibfield  {title} {\enquote {\bibinfo {title} {Dynamic
  heterogeneity in two-dimensional supercooled liquids: Comparison of
  bond-breaking and bond-orientational correlations},}\ }\href@noop {}
  {\bibfield  {journal} {\bibinfo  {journal} {Journal of Statistical Mechanics:
  Theory and Experiment}\ }\textbf {\bibinfo {volume} {2016}},\ \bibinfo
  {pages} {074008} (\bibinfo {year} {2016})}\BibitemShut {NoStop}%
\bibitem [{\citenamefont {Garrahan}\ \emph {et~al.}(2011)\citenamefont
  {Garrahan}, \citenamefont {Sollich},\ and\ \citenamefont
  {Toninelli}}]{garrahan2011kinetically}%
  \BibitemOpen
  \bibfield  {author} {\bibinfo {author} {\bibfnamefont {Juan~P}\ \bibnamefont
  {Garrahan}}, \bibinfo {author} {\bibfnamefont {Peter}\ \bibnamefont
  {Sollich}}, \ and\ \bibinfo {author} {\bibfnamefont {Cristina}\ \bibnamefont
  {Toninelli}},\ }\bibfield  {title} {\enquote {\bibinfo {title} {Kinetically
  constrained models},}\ }\href@noop {} {\bibfield  {journal} {\bibinfo
  {journal} {Dynamical heterogeneities in glasses, colloids, and granular
  media}\ }\textbf {\bibinfo {volume} {150}},\ \bibinfo {pages} {111--137}
  (\bibinfo {year} {2011})}\BibitemShut {NoStop}%
\bibitem [{\citenamefont {Blondel}\ and\ \citenamefont
  {Toninelli}(2014)}]{blondel2014there}%
  \BibitemOpen
  \bibfield  {author} {\bibinfo {author} {\bibfnamefont {Oriane}\ \bibnamefont
  {Blondel}}\ and\ \bibinfo {author} {\bibfnamefont {Cristina}\ \bibnamefont
  {Toninelli}},\ }\bibfield  {title} {\enquote {\bibinfo {title} {Is there a
  fractional breakdown of the stokes-einstein relation in kinetically
  constrained models at low temperature?}}\ }\href@noop {} {\bibfield
  {journal} {\bibinfo  {journal} {Europhysics Letters}\ }\textbf {\bibinfo
  {volume} {107}},\ \bibinfo {pages} {26005} (\bibinfo {year}
  {2014})}\BibitemShut {NoStop}%
\end{thebibliography}%

\end{document}